\documentclass[utf8,12pt]{mystyle} 
\usepackage{url,lineno,microtype,subcaption,datetime, fmtcount, etoolbox, fcprefix}
\usepackage{booktabs}
\usepackage{multirow}
\usepackage{rotating}
\usepackage{makecell}
\usepackage{adjustbox}
\usepackage[flushleft]{threeparttable}

\usepackage{float}
\usepackage[onehalfspacing]{setspace}
\usepackage[acronym,nonumberlist]{glossaries}
\usepackage{amsmath}
\usepackage{hyperref}
\hypersetup{
    colorlinks = false,
    hidelinks=true    
}

\makeglossaries
\newacronym{WHO}{WHO}{World Health Organization}
\newacronym{FDA}{FDA}{United States Food and Drug Administration}
 \newacronym{LASC 2018}{LASC'18}{left atrium segmentation challenge 2018}
  \newacronym{CETUS}{CETUS}{Challenge on Endocardial Three-dimensional Ultrasound Segmentation}
 \newacronym{MICCAI}{MICCAI}{ International Conference on
Medical Image Computing and Computer-assisted Intervention}
\newacronym{ACDC}{ACDC}{Automated Cardiac Diagnosis Challenge}
\newacronym{MM-WHS}{MM-WHS}{Multi-Modality Whole Heart Segmentation}
\newacronym{WHS}{WHS}{whole heart segmentation}
\newacronym{CVD}{CVD}{cardiovascular diseases}
\newacronym{CAD}{CAD}{coronary artery disease}
\newacronym{CHD}{CHD}{congenital heart disease}
\newacronym{HCM}{HCM}{hypertrophic cardiomyopathy}
\newacronym{DCM}{DCM}{dilated cardiomyopathy}
\newacronym{MI}{MI}{myocardial infarction}
\newacronym{AS}{AS}{aortic stenosis}
\newacronym{HF}{HF}{heart failure}
\newacronym{IC}{IC}{ischemic cardiomyopathy}

\newacronym{LA}{LA}{left atrium}
\newacronym{AF}{AF}{atrial fibrillation}
\newacronym{RA}{RA}{right atrium}
\newacronym{LAD}{LAD}{left atrial descending coronary artery}
\newacronym{LV}{LV}{left ventricle}
\newacronym{MYO}{MYO}{myocardium}
\newacronym{IS}{IS}{interventricular septum}
\newacronym{AO}{AO}{aorta}
\newacronym{PA}{PA}{pulmonary artery}
\newacronym{PV}{PV}{pulmonary vein}
\newacronym{RV}{RV}{right ventricle}

\newacronym{CMR}{CMR}{cardiovascular magnetic resonance}
\newacronym{MRI}{MRI}{magnetic resonance imaging}
\newacronym{fMRI}{fMRI}{functional magnetic resonance imaging}
\newacronym{CT}{CT}{computed tomography}
\newacronym{CTA}{CTA}{computed tomography angiography}
\newacronym{MR}{MR}{magnetic resonance}
\newacronym{MRA}{MRA}{magnetic resonance angiography}
\newacronym{SPECT}{SPECT}{single photon emission computed tomography}
\newacronym{XA}{XA}{x-ray angiography}
\newacronym{GE}{GE}{gadolinium-enhanced}
\newacronym{LGE}{LGE}{late gadolinium enhancement}
\newacronym{US}{US}{ultrasound}
\newacronym{PET}{PET}{positron emission tomography}
\newacronym{VEC-MRI}{VEC-MRI}{velocity-encoded cine MR imaging}
\newacronym{MPR}{MPR}{multi-planar reformatted}
\newacronym{bSSFP}{bSSFP}{balanced steady state free precession}
\newacronym{T}{T}{Tesla}
\newacronym{RF}{RF}{radio frequency}
\newacronym{SNR}{SNR}{signal-to-noise ratio}
\newacronym{SAX}{SAX}{short-axis}
\newacronym{LAX}{LAX}{long-axis}
\newacronym{2CH}{2CH}{2-chamber}
\newacronym{3CH}{3CH}{3-chamber}
\newacronym{4CH}{4CH}{4-chamber}

\newacronym{SOP}{SOP}{standard operating procedures}
\newacronym{FOV}{FOV}{field-of-view}
\newacronym{ECG}{ECG}{electrocardiogram}
\newacronym{FFR}{FFR}{fractional flow reserve}

\newacronym{RFCA}{RFCA}{radio-frequency catheter ablation}

\newacronym{ED}{ED}{end-diastole}
\newacronym{ES}{ES}{end-systole}
\newacronym{LVOT}{LVOT}{left ventricular outflow tract}
\newacronym{EF}{EF}{ejection fraction}
\newacronym{LAV}{lAV}{left atrial volume}
\newacronym{RAV}{RAV}{right atrial volume}
\newacronym{LVEDV}{LVEDV}{left ventricular end-diastolic volume}
\newacronym{LVESV}{LVESV}{left ventricular end-systolic volume}
\newacronym{RVEDV}{RVEDV}{right ventricular end-diastolic volume}
\newacronym{RVESV}{RVESV}{right ventricular end-systolic volume}
\newacronym{RVEF}{RVEF}{right ventricular ejection fraction}
\newacronym{SV}{SV}{stroke volume}
\newacronym{NCP}{NCP}{non-calcified plaque}
\newacronym{MCP}{MCP}{mixed-calcified plaque}
\newacronym{CAC}{CAC}{coronary artery calcium}

\newacronym{CNN}{CNN}{convolutional neural network}
\newacronym{CONV}{CONV}{Convolutional}

\newacronym{FC}{FC}{fully connected}
\newacronym{DNN}{DNN}{deep neural network}
\newacronym{GRU}{GRU}{gated recurrent unit}
\newacronym{AE}{AE}{autoencoder}
\newacronym{FCN}{FCN}{fully convolutional neural network}
\newacronym{RNN}{RNN}{recurrent neural network}
\newacronym{GAN}{GAN}{generative adversarial network}
\newacronym{BN}{BN}{batch normalization}
\newacronym{DL}{DL}{deep learning}
\newacronym{ROI}{ROI}{region-of-interest}
\newacronym{CRF}{CRF}{conditional random field}
\newacronym{EM}{EM}{expectation maximization}
\newacronym{DBN}{DBN}{deep belief networks}
\newacronym{MSL}{MSL}{marginal space learning}
\newacronym{MRF}{MRF}{markov random field}
\newacronym{LSTM}{LSTM}{long-short term memory}
\newacronym{SVM}{SVM}{support vector machine}
\newacronym{ReLU}{ReLU}{rectified linear unit}
\newacronym{SMC}{SMC}{sequential monte carlo}
\newacronym{ASM}{ASM}{active shape model}
\newacronym{SRF}{SRF}{structured random forest}

\newacronym{nu}{nu}{not used}

\newacronym{MSE}{MSE}{mean squared error}
\newacronym{GDPR}{GDPR}{The General Data Protection Regulation}
\newacronym{GPU}{GPU}{graphical processing units}

\newglossaryentry{formula}
{
        name=formula,
        description={A mathematical expression}
}
 
 \def\keyFont{\fontsize{8}{11}\helveticabold }
\def\firstAuthorLast{Chen Chen {et~al.}} 

\def\Authors{Chen Chen\,$^{1,*}$, Chen Qin\,$^{1,*}$, Huaqi Qiu\,$^{1,*}$, Giacomo Tarroni$^{1,2}$, Jinming Duan$^{3}$, Wenjia Bai$^{4,5}$, and Daniel Rueckert\,$^{1}$}
\def\Address{$^{1}$ Biomedical Image Analysis Group, Department of Computing, Imperial College London, London, UK;\\
$^{2}$ Department of Computer Science,
City, University of London, London, UK;\\
$^{3}$ School of Computer Science,
University of Birmingham, Birmingham, UK;\\
$^{4}$ Data Science Institute, Imperial College London, London, UK;\\
$^{5}$ Division of Brain Sciences, Department of Medicine, Imperial College London, London, UK}
\def\corrAuthor{Chen Chen}
\def\corrEmail{chen.chen15@imperial.ac.uk}

\begin{document}

\twocolumn
\firstpage{1}
\title[Deep learning for cardiac image segmentation: A review]{Deep learning for cardiac image segmentation: A review} 
\author[\firstAuthorLast ]{\Authors} 
\address{} 
\correspondance{} 
\extraAuth{}
\maketitle

\begin{abstract}
Deep learning has become the most widely used approach for cardiac image segmentation in recent years. In this paper, we provide a review of over 100 cardiac image segmentation papers using deep learning, which covers common imaging modalities including magnetic resonance imaging (MRI), computed tomography (CT), and ultrasound (US) and major anatomical structures of interest (ventricles, atria and vessels). In addition, a summary of publicly available cardiac image datasets and code repositories are included to provide a base for encouraging reproducible research. Finally, we discuss the challenges and limitations with current deep learning-based approaches (scarcity of labels, model generalizability across different domains, interpretability) and suggest potential directions for future research.   
\tiny
 \keyFont{\section{Keywords:} Artificial intelligence, deep learning, neural networks, cardiac image segmentation, cardiac image analysis, MRI, CT, US} 
\end{abstract}
 
\section*{Article types}
Review

\begin{figure*}[!ht]
    \centering
    \begin{minipage}{\textwidth}
    \includegraphics[width=\textwidth]{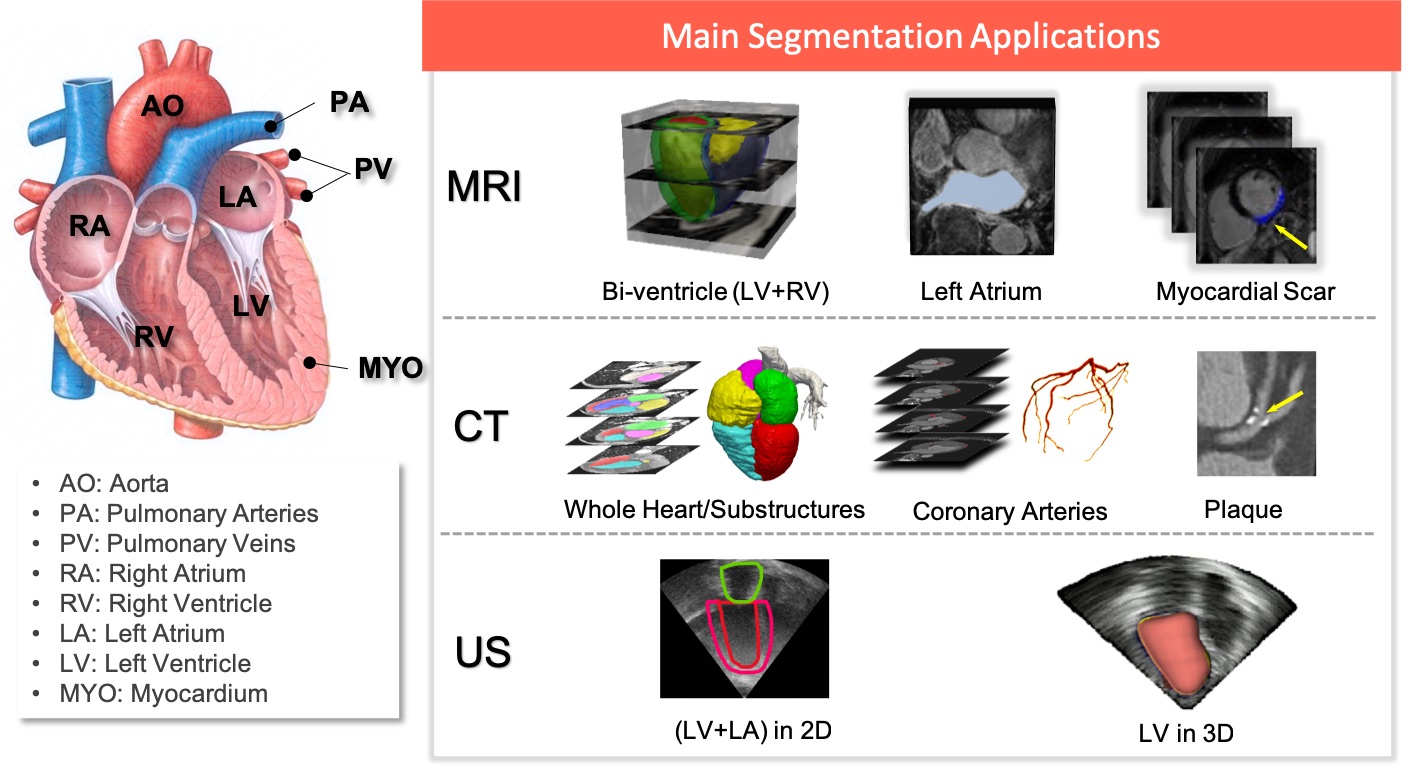}
    \caption[xx]{\textbf{Overview of cardiac image segmentation tasks for different imaging modalities.} For better understanding, we provide the anatomy of the heart on the left (image source: clipart-library.com). Of note, for simplicity, we list the tasks for which deep learning techniques have been applied, which will be discussed in Section~\ref{sec: dl for cardiac image segmentation}. }
    \label{fig:overview}
     \end{minipage}
\end{figure*}
\section{Introduction}
\label{SEC:Introduction}

\Glspl{CVD} are the leading cause of death
globally according to \Gls{WHO}. About 17.9 million people died from CVDs in 2016, from \Gls{CVD}, mainly from heart disease and stroke\footnote{\url{https://www.who.int/cardiovascular_diseases/about_cvd/en/}}. The number is still increasing annually. In recent decades, major advances have been made in cardiovascular research and practice aiming to improve diagnosis and treatment of cardiac diseases as well as reducing the mortality of \gls{CVD}. Modern medical imaging techniques such as \gls{MRI}, \gls{CT} and \gls{US} are now widely used, which enable non-invasive qualitative and quantitative assessment of cardiac anatomical structures and functions and provide support for diagnosis, disease monitoring, treatment planning and prognosis. 

Of particular interest, cardiac image segmentation is an important first step in numerous applications. It  partitions the image into a number of semantically (i.e. anatomically) meaningful regions, based on which quantitative measures can be extracted, such as the myocardial mass, wall thickness, \gls{LV} and \gls{RV} volume as well as \gls{EF} etc. Typically, the anatomical structures of interest for cardiac image segmentation include the \gls{LV}, \gls{RV}, \gls{LA}, \gls{RA}, and coronary arteries. An overview of typical tasks related to cardiac image segmentation is presented in Fig.~\ref{fig:overview}, where applications for the three most commonly used modalities, i.e., \gls{MRI}, \gls{CT} and \gls{US}, are shown.

\begin{figure*}[t]
    \begin{center}
    \includegraphics[width=\textwidth]{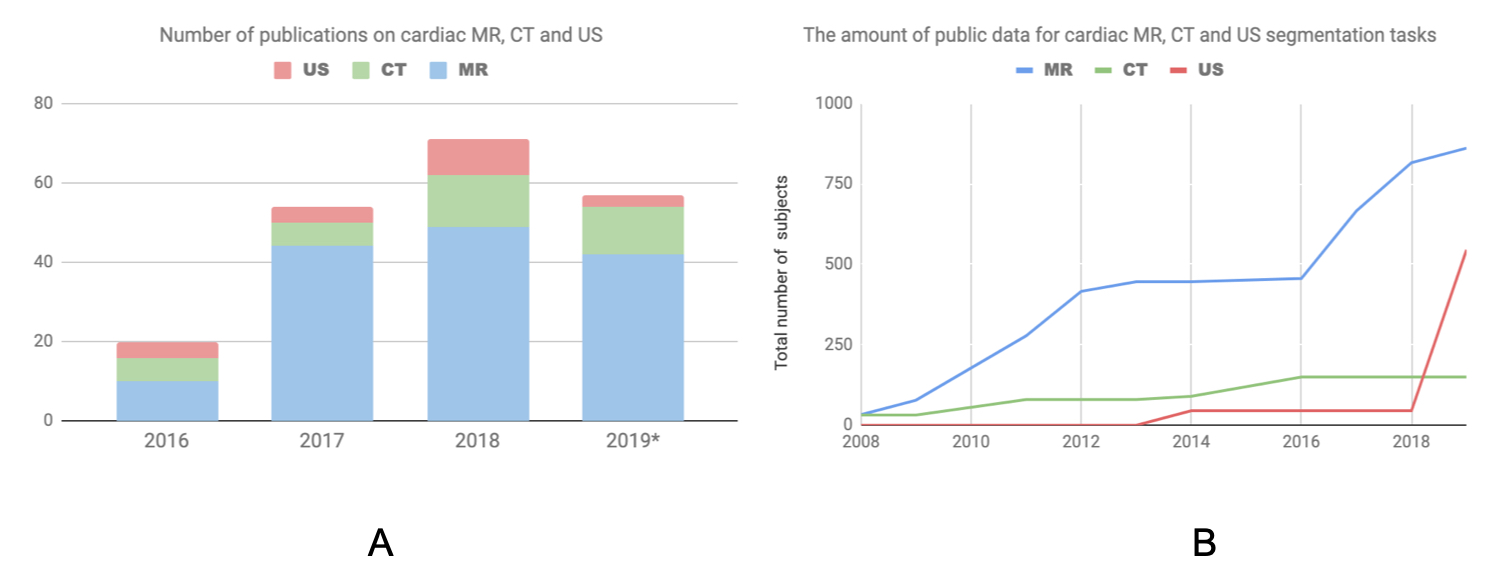}
    \end{center}
    \caption{\textbf{(A)} Overview of numbers of papers published from 1st January 2016 to 1st August 2019 regarding deep learning-based methods for cardiac image segmentation reviewed in this work. \textbf{(B)} The increase of public data for cardiac image segmentation in the past ten years. CT: computed tomography, MR: magnetic resonance, US: ultrasound.}
    \label{fig:number_of_publications}
\end{figure*}

Before the rise of deep learning, traditional machine learning techniques such as model-based methods (e.g. active shape and appearance models) and atlas-based methods had been shown to achieve good performance in cardiac image segmentation \citep{Petitjean_2015_MedIA,Peng_2016_MAGMA,Tavakoli_2013_CVIU,Lesage_2009_MedIA}. However, they often require significant feature engineering or prior knowledge to achieve satisfactory accuracy. In contrast, \gls{DL}-based algorithms are good at \emph{automatically} discovering intricate features from data for object detection and segmentation. These features are directly learned from data using a general-purpose learning procedure and in end-to-end fashion. This makes \gls{DL}-based algorithms easy to apply to other image analysis applications. Benefiting from advanced computer hardware (e.g. graphical processing units (GPUs) and tensor processing units (TPUs)) as well as increased available data for training, \gls{DL}-based segmentation algorithms have gradually outperformed previous state-of-the-art traditional methods, gaining more popularity in research. This trend can be observed in Fig.~\ref{fig:number_of_publications}A, which shows how the number of DL-based papers for cardiac image segmentation has increased strongly in the last years. In particular, the number of the publications for MR image segmentation is significantly higher than the numbers of the other two domains, especially in 2017. One reason, which can be observed in Fig.~\ref{fig:number_of_publications}B, is that the publicly available data for MR segmentation has increased remarkably since 2016.  

In this paper, we provide an overview of state-of-the-art deep learning techniques for cardiac image segmentation in the three most commonly used modalities (i.e. \gls{MRI}, \gls{CT}, \gls{US}) in clinical practice and discuss the advantages and remaining limitations of current deep learning-based segmentation methods that hinder widespread clinical deployment. To our knowledge, there have been several review papers that presented overviews about applications of \gls{DL}-based methods for general medical image analysis~\citep{Greenspan_2016_TMI,Shen_2017_Review,LITJENS_2017_MedIA}, as well as some surveys dedicated to applications designed for cardiovascular image analysis ~\citep{Gandhi_2018_Echocardiography,Mazurowski_2019_JMRI}. However, none of them has provided a systematic overview focused on \emph{cardiac segmentation applications}. This review paper aims at providing a comprehensive overview from the debut to the state-of-the-art of deep learning algorithms, focusing on a variety of cardiac image segmentation tasks (e.g. the \gls{LV}, \gls{RV}, and vessel segmentation) (Sec. \ref{sec: dl for cardiac image segmentation}). Particularly, we aim to cover most influential \gls{DL}-related works in this field published until 1st August 2019 and categorized these publications in terms of specific methodology. Besides, in addition to the basics of deep learning introduced in Sec.\ref{SEC:basics}, we also provide a summary of public datasets (see Table~\ref{tab:public datasets}) as well as public code (see Table~\ref{tab:public code}), aiming to present a good reading basis for newcomers to the topic and encourage future contributions. More importantly, we provide insightful discussions about the current research situations~(Sec.\ref{sec: Dicussion}) as well as challenges and potential directions for future work (Sec. \ref{SEC:Challenges}). 

\textbf{Search criterion}
To identify related contributions, search engines like Scopus and PubMed were queried for papers containing (“convolutional” OR “deep learning”) and (“cardiac") and ("image segmentation") in title or abstract. Additionally, conference proceedings for MICCAI, ISBI and EMBC were searched based on the titles of papers. Papers which do not primarily focus on segmentation problems were excluded. The last update to the included papers was on Aug 1, 2019. 

\begin{figure*}[!ht]
    \centering
    \includegraphics[width=0.9\textwidth]{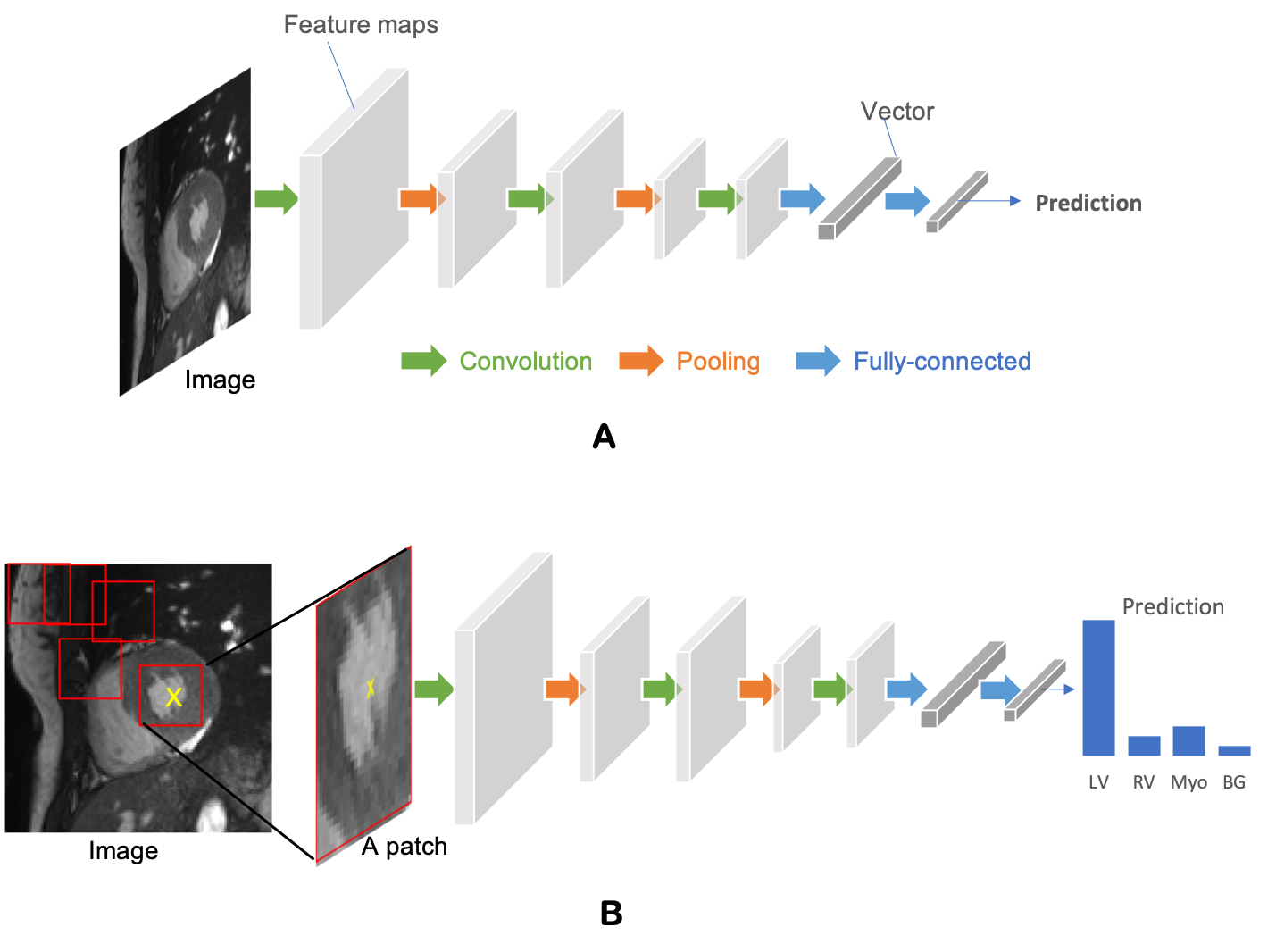}
    \caption{\textbf{(A) Generic architecture of convolutional neural networks (CNN).} A CNN takes a cardiac MR image as input, learning hierarchical features through a stack of convolutions and pooling operations. These spatial feature maps are then flattened and reduced into a vector through fully connected layers. This vector can be in many forms, depending on the specific task. It can be probabilities for a set of classes (image classification) or coordinates of a bounding box (object localization) or a predicted label for the center pixel of the input (patch-based segmentation) or a real value for regression tasks (e.g. left ventricular volume estimation).
    \textbf{(B) Patch-based segmentation method based on a CNN classifier.}
    The CNN takes a patch as input and outputs the probabilities for four classes where the class with the highest score is the prediction for the center pixel (see the yellow cross) in this patch. By repeatedly forwarding patches located at different locations into the CNN for classification, one can finally get a pixel-wise segmentation map for the whole image. LV:left ventricle; RV: right ventricle; BG: Background; Myo: left ventricular myocardium.}
    \label{fig:CNN}
\end{figure*}
\section{Fundamentals of Deep Learning}
\label{SEC:basics}
Deep learning models are deep artificial neural networks. Each neural network consists of an input layer, an output layer, and multiple hidden layers.  In the following section, we will review several deep learning networks and key techniques that have been commonly used in state-of-the-art segmentation algorithms. For a more detailed and thorough illustration of the mathematical background and fundamentals of deep learning we refer the interested reader to~\cite{Goodfellow_2016_MIT}.
\newline
\medskip\subsection{Neural Networks}
In this section, we first introduce basic neural network architectures and then briefly introduce building blocks which are commonly used to boost the ability of the networks to learn features that are useful for image segmentation.
\\

\smallskip\subsubsection{Convolutional Neural Networks (CNNs)}
In this part, we will introduce \gls{CNN}, which is the most common type of deep neural networks for image analysis. \gls{CNN} have been successfully applied to advance the state-of-the-art on many image classification, object detection and segmentation tasks. 

As shown in Fig.~\ref{fig:CNN}A, a standard \gls{CNN} consists of an input layer, an output layer and a stack of functional layers in between that transform an input into an output in a specific form (e.g. vectors). These functional layers often contains convolutional layers, pooling layers and/or fully-connected layers. In general, each convolution uses a $n \times n$ kernel (for 2D input) or $n \times n \times n$ kernel (for 3D input) followed by batch normalization~\citep{Ioffe_2015_ICML} after which the output is passed through a nonlinear activation function (e.g. \gls{ReLU}), which is used to extract feature maps from an image. These feature maps are then downsampled by pooling layers, typically by a factor of 2, which removes redundant features to improve the statistical efficiency and model generalization. After that, fully connected layers are applied to reduce the dimension of features and find the most task-relevant features for inference. The output of the network is a fix-sized vector where each element can be a probabilistic score for each category (for image classification), a real value for a regression task (e.g. the left ventricular volume estimation) or a set of values (e.g. the coordinates of a bounding box for object detection and localization). 

In general, the size of convolution kernel $n$ is chosen to be small in general, e.g. $n=3$, in order to reduce computational costs. While the kernels are small, one can increase the receptive field (the area of the input image that potentially impacts the activation of a particular convolutional kernel/neuron) by increasing the number of convolutional layers. For example, a convolutonal layer with large $7\times7$ kernels can be replaced by three layers with small $3\times3$ kernels. The number of parameters is reduced by a factor of $7^2/(3\times(3^2))\approx 2$ while the receptive field remains the same ($7\times7$). An online resource~\footnote{https://fomoro.com/research/article/receptive-field-calculator} is referred here, which illustrates and visualizes the change of receptive field by varying the number of hidden layers and the size of kernels. In general, increasing the depth of convolution neural networks (the number of hidden layers) to enlarge the receptive field can lead to improved model performance, e.g. classification accuracy~\citep{Simonyan_2015_ICLR}. 

CNNs for image classification can also be employed for image segmentation applications without major adaptations to the network architecture    ~\citep{Ciresan_2012_NIPS}, as shown in Fig.~\ref{fig:CNN}B. However, this requires to divide each image into patches and then train a CNN to predict the class label of the center pixel for every patch. One major disadvantage of this patch-based approach is that, at inference time, the network has to be deployed for every patch individually despite the fact that there is a lot of redundancy due to multiple overlapping patches in the image. As a result of this inefficiency, the main application of CNNs with fully connected layers is object localization, which aims to estimate the bounding box of the object of interest in an image. This bounding box is then used to crop the image, forming an image pre-processing step to reduce the computational cost for segmentation ~\citep{Avendi_2016_MedIA}. For efficient, end-to-end pixel-wise segmentation, a variant of CNNs called \gls{FCN} is more commonly used, which will be discussed in the next section.

\begin{figure*}[!ht]
\begin{center}
\includegraphics[width=\textwidth]{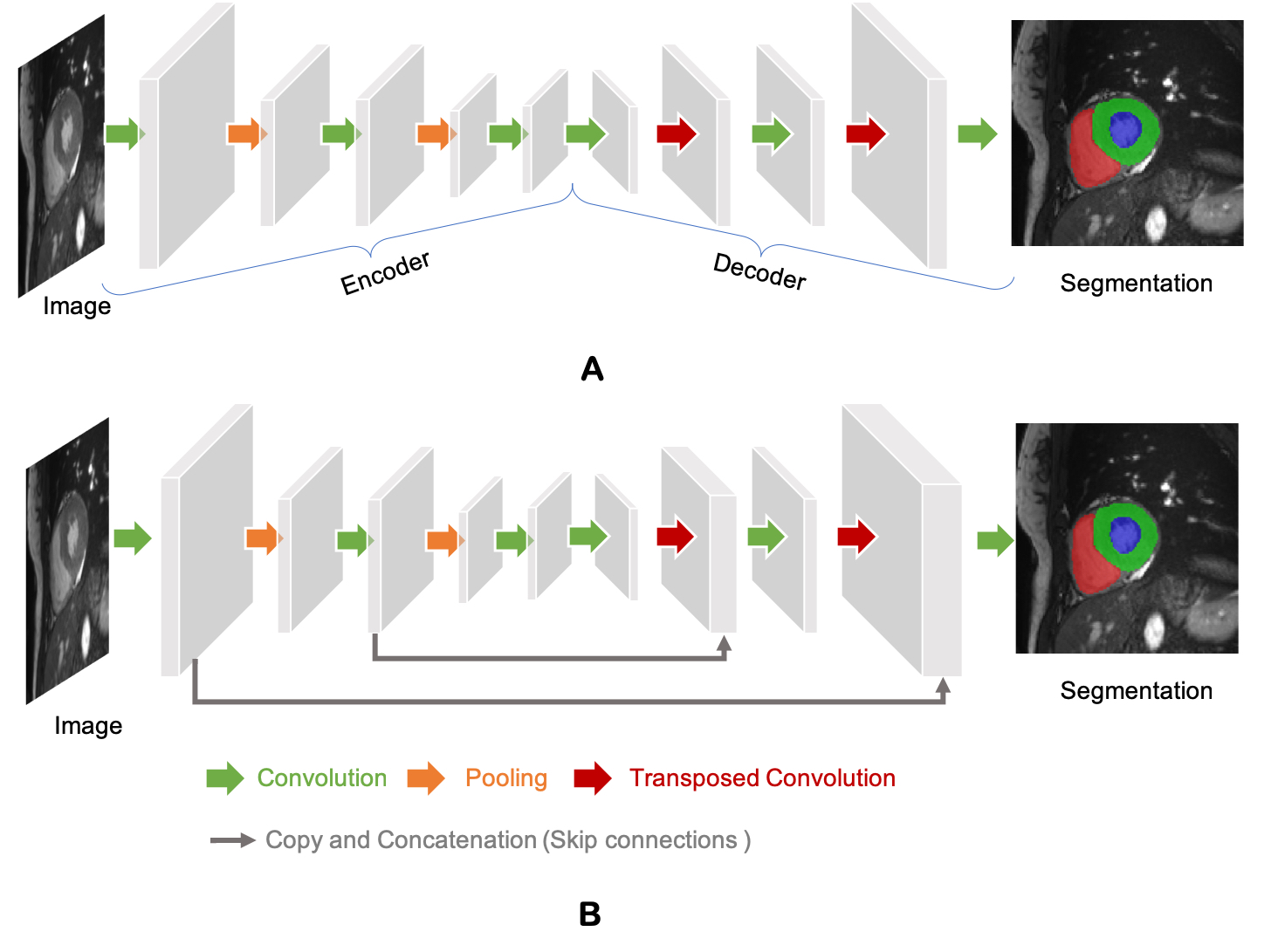}
\end{center}
\caption{\textbf{(A) Generic architecture of fully convolutional neural networks (FCN) for segmentation.} The FCN first takes the whole image as input, learns deep image features though the encoder, gradually recovers the spatial dimension by a series of transposed convolution layers in the decoder and finally predicts a pixel-wise image segmentation for the left ventricle cavity (the blue region), the left ventricular myocardium (the green region) and the right ventricle (the red region). One use case of this FCN-based cardiac segmentation can be found in~\cite{Tran_2016_Arxiv}.
\textbf{(B) A schematic drawing of {U-net}}. On the basis of the basic structure of FCN, U-net employs `skip connections' ( the gray arrows) to aggregate feature maps from coarse to fine. Of note, for simplicity, we reduce the number of downsampling and upsampling blocks. For detailed information, we recommend readers to the original paper~\citep{Ronneberger_2015_MICCAI}.
}
\label{fig:FCNvsUnet}
\end{figure*}

\smallskip\subsubsection{Fully Convolutional Neural Networks (FCNs)}
The idea of \gls{FCN} was first introduced by~\cite{Long_2014_CVPR} for image segmentation. FCNs are a special type of \glspl{CNN} that do not have any fully connected layers. In general, as shown in Fig.~\ref{fig:FCNvsUnet}A, \glspl{FCN} are designed to have an encoder-decoder structure such that they can take input of arbitrary size and produce the output with the same size. Given an input image, the encoder first transforms the input into high-level feature representation whereas the decoder interprets the feature maps and recovers spatial details back to the image space for pixel-wise prediction through a series of transposed convolution and convolution operations. Here, transposed convolutions are used for up-scaling the feature maps, typically by a factor of 2. These transposed convolutions can also be replaced by unpooling layers and upsampling layers. Compared to a patch-based CNN for segmentation, \gls{FCN} is trained and applied to the entire images, removing the need for patch selection~\citep{Shelhamer_2017_TPAMI}.

\Gls{FCN} with the simple encoder-decoder structure in Fig.~\ref{fig:FCNvsUnet}A may be limited to capture detailed context information in an image for precise segmentation as some features may be eliminated by the pooling layers in the encoder. Several variants of \glspl{FCN} have been proposed to propagate features from the encoder to the decoder, in order to boost the segmentation accuracy. The most well-known and most popular variant of \glspl{FCN} for biomedical image segmentation is the U-net~\citep{Ronneberger_2015_MICCAI}. On the basis of the vanilla FCN~\citep{Long_2014_CVPR}, the U-net employs skip connections between the encoder and decoder to recover spatial context loss in the down-sampling path, yielding more precise segmentation (see Fig.~\ref{fig:FCNvsUnet}B). Several state-of-the-art cardiac image segmentation methods have adopted the U-net or its 3D variants, the 3D U-net~\citep{Cicek_2016_MICCAI} and the 3D V-net~\citep{Milletari_2016_3DV}, as their backbone networks, achieving promising segmentation accuracy for a number of cardiac segmentation tasks~\citep{Tao_2019_Radiology,Isensee_2017_STACOM, Xia_2018_STACOM}.

\subsubsection{Recurrent Neural Networks (RNNs)}

\begin{figure}[t]
	\begin{center}
	\includegraphics[width=0.5\textwidth]{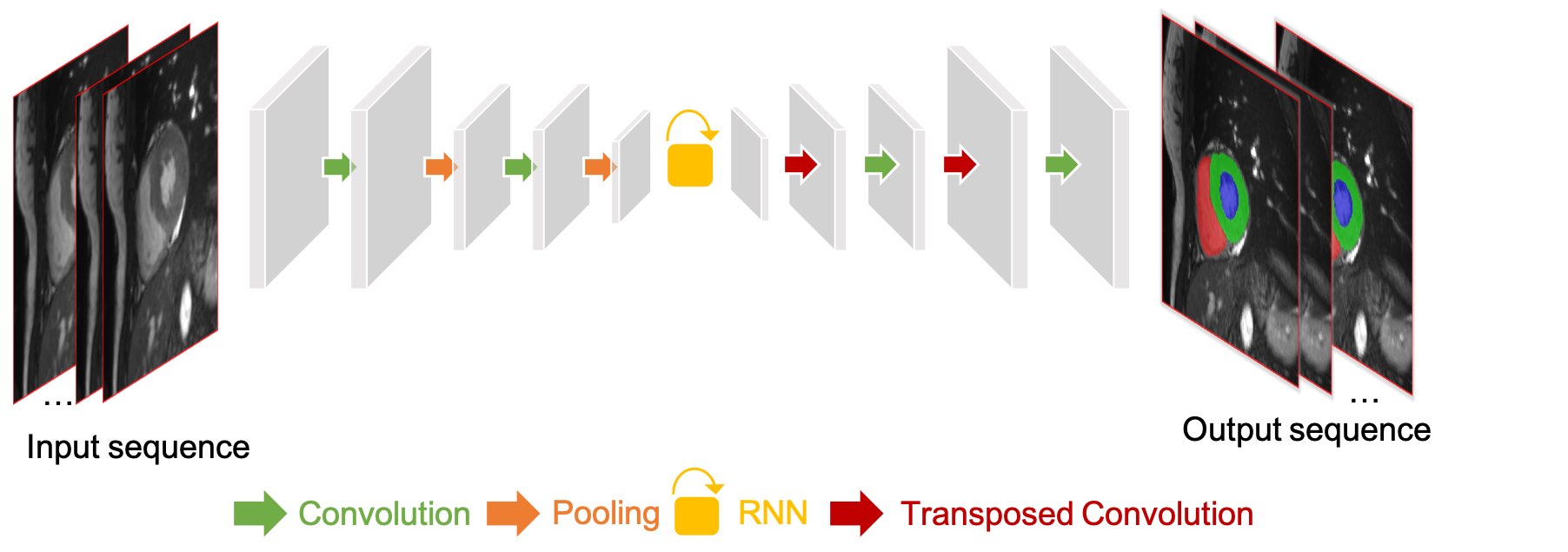}
    \end{center}
	\caption{\textbf{An example of {RNN} for cardiac image segmentation}. The yellow block with a curved arrow represents a RNN module, which can memorize the past and use the knowledge learned from the past to make its present decision. This type of network is ideal for sequential data such as cine MR images and ultrasound movies, as well as volumetric data. In this example, the network is used to segment cardiac ventricles from a stack of 2D cardiac MR slice, which allows to propagate
		contextual information from adjacent slices in the z-direction for better inter-slice coherence~\citep{Poudel_2016_HVSCMR}.}
	\label{fig:RNN}
\end{figure}

\Glspl{RNN} are another type of artificial neural networks which are used for sequential data, such as cine \gls{MRI} and ultrasound image sequences. An RNN can `remember' the past and use the knowledge learned from the past to make its present decision, see Fig~\ref{fig:RNN}. For example, given a sequence of images, an RNN takes the first image as input, captures the information to make a prediction and then memorize this information which is then utilized to make a prediction for the next image. The two most widely used architectures in the family of \glspl{RNN} are LSTM~\citep{Hochreiter_1997_NC} and \gls{GRU}~\citep{Cho_2014_EMNLP}, which are capable of modeling long-term memory. A use case for cardiac segmentation is to combine an RNN with a 2D FCN so that the combined network is capable of capturing information from adjacent slices to improve the inter-slice coherence of segmentation results~\citep{Poudel_2016_HVSCMR}.

\subsubsection{Autoencoders (AE)}

\begin{figure}[t]
	\centering
	\includegraphics[width=0.5\textwidth]{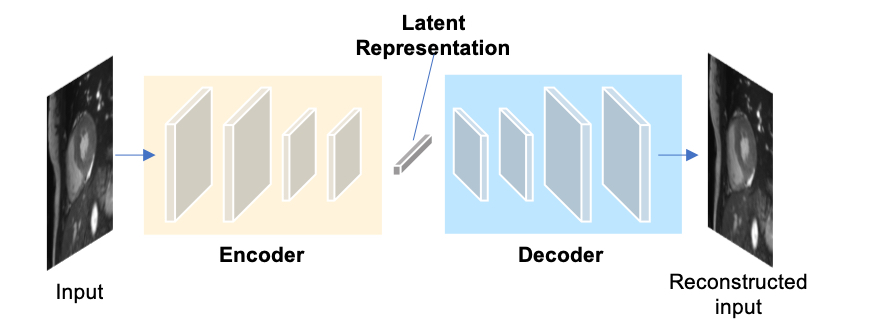}
	\caption{\textbf{A generic architecture of an autoencoder.} An autoencocer employs an encoder-decoder structure, where the encoder maps the input data to a low-dimensional latent representation and the decoder interprets the code and reconstructs the input.}.
	\label{fig:AE}
\end{figure}

\Glspl{AE} are a type of neural networks that are designed to learn compact latent representations from data without supervision. A typical architecture of an autoencoder consists of two networks: an encoder network and a decoder network for the reconstruction of the input, see Fig.~\ref{fig:AE}. Since the learned representations contain generally useful information in the original data, many researchers have employed autoencoders to extract general semantic features or shape information from input images or labels and then use those features to guide the cardiac image segmentation~\citep{Oktay_2016_MICCAI,Schlemper_2018_MICCAI,Yue_2019_MICCAI}.

\subsubsection{Generative Adversarial Networks (GAN)}
\begin{figure}[t]
	\centering
	\includegraphics[width=0.5\textwidth]{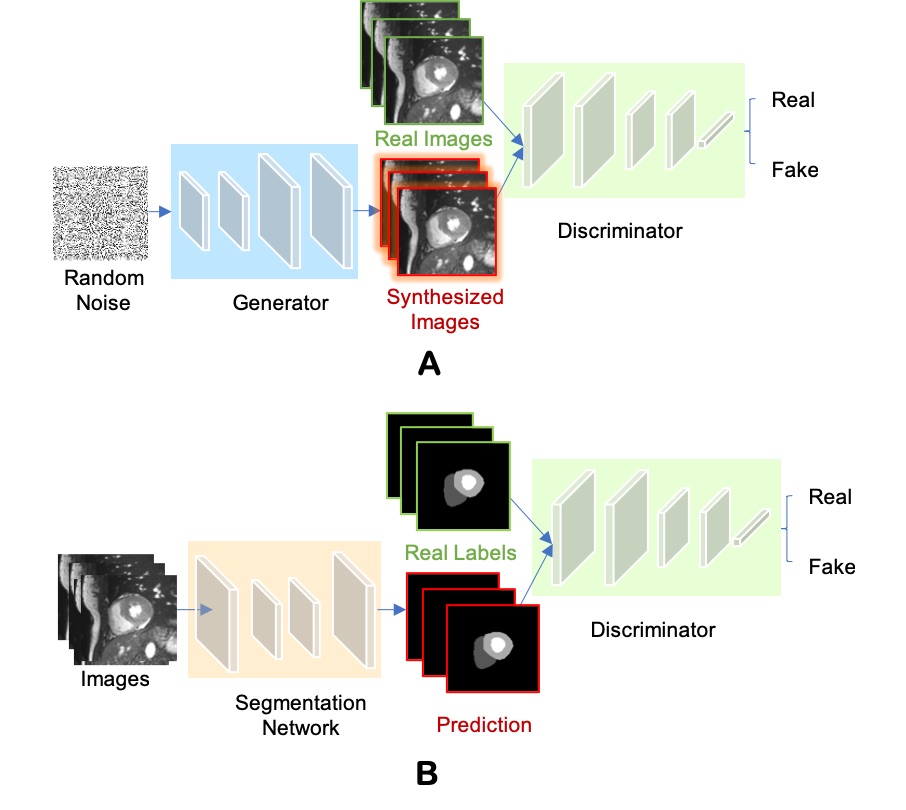}
	\caption{\textbf{(A)} Overview of GAN for image synthesis; \textbf{(B)} Overview of adversarial training for image segmentation.}
	\label{fig:GAN vs Adversarial Training}
\end{figure}

The concept of \Gls{GAN} was proposed by \cite{Goodfellow_2014_NIPS} for image synthesis from noise. GANs are a type of generative models that learn to model the data distribution of real data and thus are able to create new image examples. As shown in Fig.~\ref{fig:GAN vs Adversarial Training}A, a GAN consists of two networks: a generator network and a discriminator network. During training, the two networks are trained to compete against each other: the generator produces fake images aimed at fooling the discriminator, whereas the discriminator tries to identify real images from fake ones. This type of training is referred to as `adversarial training', since the two models are both set to win the competition.  This training scheme can also be used for training a segmentation network. As shown in Fig.~\ref{fig:GAN vs Adversarial Training}B, the generator is replaced by a segmentation network and the discriminator is required to distinguish the generated segmentation maps from the ground truth ones (the target segmentation maps). In this way, the segmentation network is encouraged to produce more anatomically plausible segmentation maps~\citep{Luc_2016_NIPS_workshop,Savioli_2018_Arxiv}.

\label{SEC: advanced blocks}

\subsubsection{Advanced building blocks for improved segmentation}

\begin{figure}[t]
	\centering
	\includegraphics[width=0.5\textwidth]{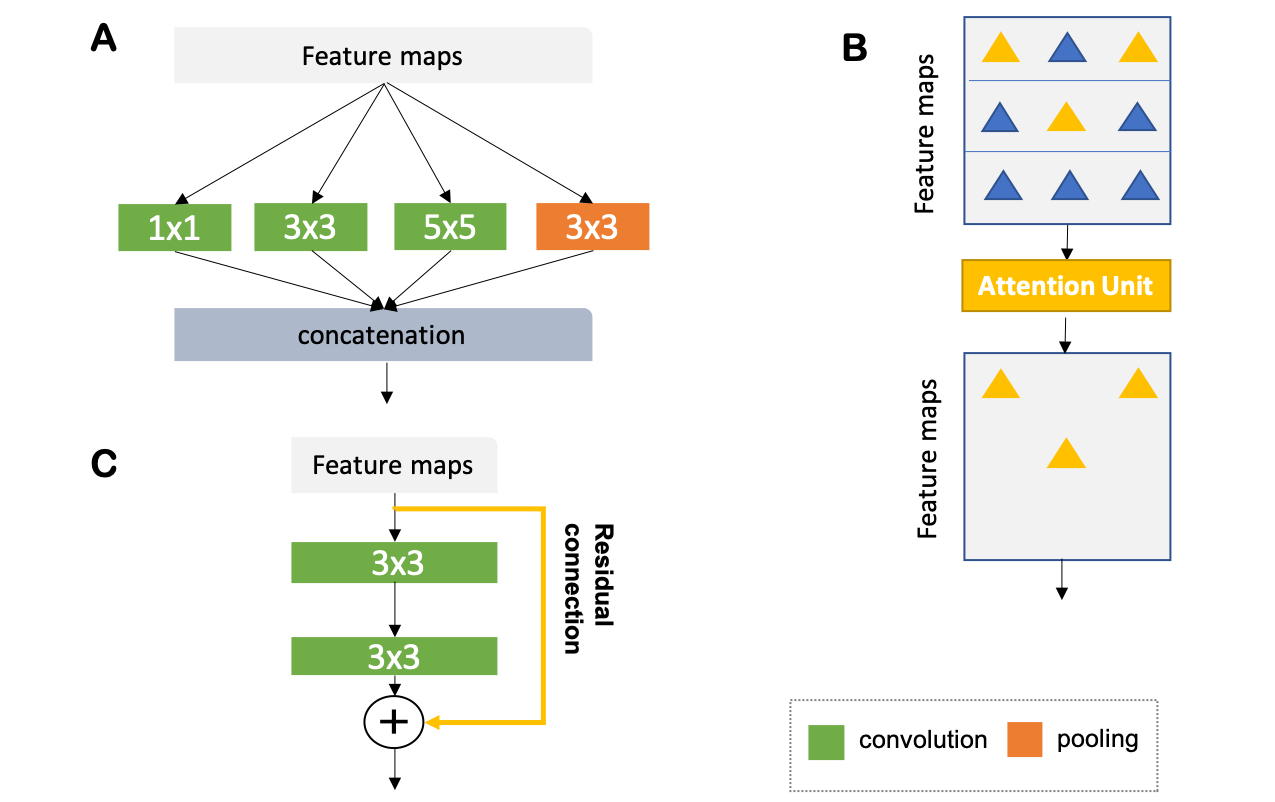}
	\caption{\textbf{(A) Naive version of the inception module~\citep{Szegedy_2015_CVPR}.} In this module, convolutional kernels with varying sizes are applied to the same input for multi-scale feature fusion. \textbf{(B) Schematic diagram of the attention module~\citep{Vaswani_2017_NIPS,Oktay_2018_MIDL}}. The attention module teaches the network to pay attention to important features (e.g. features relevant to anatomy) and ignore redundant features. \textbf{(C) Schematic diagram of a residual unit~\citep{He_2016_CVPR}.} The yellow arrow represents a residual connection which is applied to reusing the features from the previous layer. The numbers in the green and orange blocks denote the sizes of corresponding convolutional or pooling kernels. Here, for simplicity, all diagrams have been reproduced based on the illustration in the original papers.}
	\label{fig:advanced_blocks}
\end{figure}

Medical image segmentation, as an important step for quantitative analysis and clinical research, requires a pixel-wise accuracy. Over the past years, many researchers have developed advanced building blocks to learn robust, representative features for precise segmentation. These techniques have been widely applied to state-of-the-art neural networks (e.g. U-net) to improve cardiac image segmentation performance. Therefore, we identified several important techniques reported in the literature to this end and present them with corresponding references for further reading. These techniques are: 
\begin{enumerate}
\item Advanced convolutional modules for multi-scale feature aggregation in the hidden layers:
\begin{itemize}
\item {Inception modules}~\citep{Szegedy_2015_CVPR}, see Fig.~\ref{fig:advanced_blocks}A;
\item {Dilated convolutional kernels}~\citep{Yu_2015_ICLR};
\item {Deep supervision}~\citep{Lee_2015_AIS};
\item {Atrous spatial pyramid pooling}~\citep{Chen_2017_Arxiv};
\end{itemize}
\item Adaptive convolutional kernels designed to pay attention to important features:
\begin{itemize}
\item {Attention units}~\citep{Vaswani_2017_NIPS}, see Fig.~\ref{fig:advanced_blocks}B;
\item {Squeeze-and-excitation blocks}~\citep{Hu_2018_CVPR};
\end{itemize}
\item Interlayer connections designed to reuse features from previous layers:
\begin{itemize}
\item {Residual connections}~\citep{He_2016_CVPR}, see Fig.~\ref{fig:advanced_blocks}C;
\item {Dense connections}~\citep{Huang_2017_CVPR}.
\end{itemize}
\end{enumerate}

\subsection{Training Neural Networks}
Before being able to perform inference, neural networks must be trained. This training process requires a dataset that contains paired images and labels $\{\boldsymbol{x},\boldsymbol{y}\}$ for training and testing, an optimizer (e.g. stochastic gradient descent, Adam) and a loss function to update the model parameters. This function accounts for the error of the network prediction in each iteration during training, providing signals for the optimizer to update the network parameters through  backpropagation. The goal of training is to find proper values of the network parameters to minimize the loss function.
\smallskip\subsubsection{Common Loss Functions}
For regression tasks (e.g. heart localization, calcium scoring, landmark detection, image reconstruction), the simplest loss function is the \gls{MSE}:
\begin{equation}
\label{Eq: MSE Loss}
\mathcal{L}_{\text{MSE}} ={\frac {1}{n}}\sum _{i=1}^{n}{(\boldsymbol{y}_{i}-{\hat {\boldsymbol{y}_{i}}})^{2}},
\end{equation}
 where $\boldsymbol{y}_{i}$ is the vector of target values and $\hat{\boldsymbol{y}_{i}}$ is the vector of the predicted values; $n$ is the number of data samples.

Cross-entropy is the most common loss for both image classification and segmentation tasks. In particular, the cross-entropy loss for segmentation summarizes the pixel-wise probability errors between the predicted probabilistic output $\boldsymbol{p}$ and its corresponding target segmentation map $\boldsymbol{y}$  for each class $c$: 
\begin{equation}
\label{Eq: CE Loss}
\mathcal{L}_{\text{CE}} =-{\frac {1}{n}}\sum _{i=1}^{n}\sum_{c=1}^{C} {\boldsymbol{y}_i^c}{\log(\boldsymbol{p}_i^c)},    
\end{equation}
where $C$ is the number of all classes. Another loss function which is specifically designed for object segmentation is called soft-Dice loss function~\citep{Milletari_2016_3DV}, which penalizes the mismatch between a predicted segmentation map and its target map at pixel-level:
\begin{equation}
\label{Eq: Dice Loss}
\mathcal{L}_{Dice}=1-\frac{2\sum_{i=1}^{n} {\sum_{c=1}^{C} \boldsymbol{y}_{i} ^c} \boldsymbol{p}_{i} ^{c}} {\sum_{i=1}^{n}\sum_{c=1}^{C} (\boldsymbol{y}_{i}^{ c}+{\boldsymbol{p}_{i}^{c}})}.    
\end{equation}

In addition, there are several variants of the cross-entropy or soft-Dice loss such as the weighted cross-entropy loss~\citep{Jang_2017_STACOM,Baumgartner_2017_STACOM} and weighted soft-Dice loss~\citep{Yang_2017_STACOM,Khened_2019_MedIA} that are used to address potential class imbalance problem in medical image segmentation tasks where the loss term is weighted to account for rare classes or small objects.

\smallskip
\subsubsection{Reduce over-fitting}
The biggest challenge of training deep networks for medical image analysis is over-fitting, due to the fact that there is often a limited number of training images in comparison with the number of learnable parameters in a deep network. A number of techniques have been developed to alleviate this problem. Some of the techniques are the following ones:
\begin{itemize}
    \item Weight initialization~\citep{He_2015_ICCV} and weight regularization (i.e. L1/L2 regularization)
    \item Dropout~\citep{Srivastava_2014_JMLR}
    \item Ensemble learning~\citep{Kamnitsas_2017_Arxiv}
    \item Data augmentation by artificially generating training samples via affine transformations
    \item Transfer learning with a model pre-trained on existing large datasets.
\end{itemize}
\medskip\subsection{Evaluation Metrics}
To quantitatively evaluate the performance of automated segmentation algorithms, three types of metrics are commonly used: a) volume-based metrics (e.g. Dice metric, Jaccard similarity index); b) surface distance-based metrics (e.g. mean contour distance, Hausdorff distance); c) clinical performance metrics (e.g. ventricular volume and mass). For a detailed illustration of common used clinical indices in cardiac image analysis, we recommend the review paper by~\cite{Peng_2016_MAGMA}. In our paper, we mainly report the accuracy of methods in terms of the Dice metric for ease of comparison. The Dice score measures the ratio of overlap between two results (e.g. automatic segmentation vs manual segmentation), ranging from 0 (mismatch) to 1 (perfect match).

 \section{Deep Learning for Cardiac Image Segmentation}
\label{sec: dl for cardiac image segmentation}

\begin{table*}[!ht]
\centering
  \caption{\textbf{A summary of representative deep learning methods on cardiac MRI segmentation.} SAX: short-axis view; 2CH: 2-chamber view; 4CH: 4-chamber view; ED: end-diastolic; ES: end-systolic. }
\label{tab:MR}
\resizebox{\textwidth}{!}{
\begin{tabular}{@{}clccr@{}}
\toprule
 Application & Selected works & Description & Type of Images & Structure(s) \\ \midrule
\multirow{27}{*}{\begin{tabular}[c]{@{}c@{}}Ventricle \\ Segmentation\end{tabular}} 
 & \multicolumn{2}{l}{\textbf{FCN-based}} &  &  \\
 & \cite{Tran_2016_Arxiv} & 2D FCN & SAX & Bi-ventricle \\
 & \cite{Lieman-Sifry_2017_FIMH} & A lightweight FCN (E-Net) & SAX & Bi-ventricle \\
 & \cite{Isensee_2017_STACOM} & 2D U-net +3D U-net  (ensemble) & SAX & Bi-ventricle \\
 & \cite{Jang_2017_STACOM} & 2D M-Net with weighted cross entropy loss & SAX & Bi-ventricle \\
 & \cite{Baumgartner_2017_STACOM} & 2D U-net with cross entropy & SAX & Bi-ventricle \\
 & \cite{Bai_2018_JCMR} & \begin{tabular}[c]{@{}c@{}}2D FCN trained and verified on a large dataset ($\sim 5000$ subjects);\end{tabular} & SAX, 2CH, 4CH & Four chambers \\
 & \cite{Tao_2019_Radiology} & \begin{tabular}[c]{@{}c@{}}2D U-net trained and verified on a multi-vendor, multi-scanner dataset\end{tabular} & SAX & LV \\
 & \cite{Khened_2019_MedIA} & 2D Dense U-net with inception module & SAX & Bi-ventricle \\
  & \cite{Fahmy_2019_JCMR} & 2D FCN & SAX & LV\\
 \cmidrule(l){2-5}
 & \multicolumn{2}{l}{\textbf{Introducing spatial or temporal context}} &  &  \\
 & \cite{Poudel_2016_HVSCMR} & 2D FCN with RNN to model inter-slice coherency & SAX & Bi-ventricle \\
 & \cite{Patravali_2017_STACOM} & 2D multi-channel FCN to aggregate inter-slice information & SAX & Bi-ventricle \\
 & \cite{Wolterink_2017_STACOM} & Dilated U-net to segment ED and ES simultaneously & SAX & Bi-ventricle \\
 \cmidrule(l){2-5} 
 & \textbf{Applying anatomical constraints} &  &  &  \\
 & \cite{Oktay_2018_TMI} & FCN trained with additional anatomical shape-based regularization & SAX;US & LV \\
 \cmidrule(l){2-5} 
 & \multicolumn{2}{l}{\textbf{Multi-stage networks}} &  &  \\
 & \cite{Tan_2017_MedIA} & \begin{tabular}[c]{@{}c@{}}Semi-automated method; CNN (localization) \\ followed by another CNN to derive contour parameters\end{tabular} & SAX & LV \\
 & \cite{Zheng_2018_TMI} & \begin{tabular}[c]{@{}c@{}}FCN (localization) + FCN (segmentation);\\ Propagate labels from adjacent slices\end{tabular} & SAX & Bi-ventricle \\
 & \cite{Vigneault_2018_MedIA} & \begin{tabular}[c]{@{}c@{}}U-net (initial segmentation) + CNN (localization and transformation) \\ + Cascaded U-net (segmentation)\end{tabular} & SAX, 2CH, 4CH & Four chambers \\
 \cmidrule(l){2-5}
&\multicolumn{4}{l}{\textbf{Hybrid segmentation methods}} \\
 & \cite{Avendi_2016_MedIA, Avendi_2017_MRM} & \begin{tabular}[c]{@{}c@{}}CNN (localization) \\ +AE (shape initialization) \\ + Deformable model\end{tabular} & SAX & LV; RV \\
 & \cite{Yang_2016_MICCAI} & CNN combined with Multi-atlas & SAX & LV \\
 & \cite{Ngo_2016_MedIA} & Level-set based segmentation with Deep belief networks & SAX & LV \\ 
 \midrule
\multirow{3}{*}{Atrial Segmentation} & \cite{Mortazi_2017_STACOM} & Multi-view CNN with adaptive fusion strategy & 3D scans & LA \\
 & \cite{Xiong_2019_TMI} & Patch-based dual-stream 2D FCN & LGE MRI & LA \\
 & \cite{Xia_2018_STACOM} & Two-stage pipeline; 3D U-net (localization) +3D U-net (segmentation) & LGE MRI & LA \\
 \midrule
\multirow{4}{*}{Scar Segmentation} & \cite{Yang_2018_MedPhy} & \begin{tabular}[c]{@{}c@{}}Fully automated;\\ Multi-atlas method for LA segmentation \\ followed by an AE to find the atrial scars\end{tabular} & LGE MRI & LA; atrial scars \\
 & \cite{Chen_2018_MICCAI} & \begin{tabular}[c]{@{}c@{}}Fully automated; \\ Multi-view Two-Task Recursive Attention Model\end{tabular} & LGE MRI & LA; atrial scars \\
\cmidrule(l){2-5}
 & \cite{Zabihollahy_2018_MI} & \begin{tabular}[c]{@{}c@{}}Semi-automated; \\ 2D CNN for scar tissue classification\end{tabular} & LGE MRI & Myocardial scars \\
 & \cite{Moccia_2019_Magma} & \begin{tabular}[c]{@{}c@{}}Semi-automated; \\ 2D FCN for scar segmentation\end{tabular} & LGE MRI & Myocardial scars \\
 & \cite{Xu_2018_MedIA} & \begin{tabular}[c]{@{}c@{}}Fully automated; \\  RNN for joint motion feature learning and scar segmentaion\end{tabular} & cine MRI & Myocardial scars \\
\midrule
Aorta Segmentation & \cite{Bai_2018_MICCAI} & \begin{tabular}[c]{@{}c@{}} RNN to learn temporal coherence; \\ Propagate labels from labeled frames to unlabeled adjacent frames\\ for semi-supervised learning;\end{tabular} & cine MRI & Aorta \\ 
\midrule
\multirow{3}{*}{Whole Heart Segmentation} & \cite{Yu_2017_MICCAI} & 3D U-net with deep supervision & 3D scans & Blood pool+Myocardium of the heart\\
  & \cite{Li_2017_RSAM} & 3D FCN with deep supervision & 3D scans & Blood pool+Myocardium of the heart\\
    & \cite{Wolterink_2017_RSAM} & dilated CNN with deep supervision & 3D scans & Blood pool+Myocardium of the heart\\
\bottomrule
\end{tabular}
}
\end{table*}
In this section, we provide a summary of deep learning-based applications for the three main imaging modalities: \gls{MRI}, \gls{CT}, and \gls{US} regarding specific applications for targeted structures. In general, these deep learning-based methods provide an efficient and effective way to segmenting particular organs or tissues (e.g. the \gls{LV}, coronary vessels, scars) in different modalities, facilitating follow-up quantitative analysis of cardiovascular structure and function. Among these works, a large portion of these methods are designed for ventricle segmentation, especially in MR and US domains. The objective of ventricle segmentation is to delineate the endocardium and epicardium of the \gls{LV} and/or \gls{RV}. These segmentation maps are important for deriving clinical indices, such as \gls{LVEDV}, \gls{LVESV}, \gls{RVEDV}, \gls{RVESV}, and \gls{EF}. In addition, these segmentation maps are essential for 3D shape analysis~\citep{Xue_2018_MedIA,Biffi_2018_MICCAI}, 3D+time motion analysis~\citep{Zheng_2019_MedIA} and survival prediction~\citep{Bello_2019_NMI}.
\medskip
\subsection{Cardiac MR Image Segmentation}
\label{SEC:MR image segmentation}

Cardiac \gls{MRI} is a non-invasive imaging technique that can visualize the structures within and around the heart. Compared to \gls{CT}, it does not require ionising radiation. Instead, it relies on the magnetic field in conjunction with radio-frequency waves to excite hydrogen nuclei in the heart, and then generates an image by measuring their response. By utilizing different imaging sequences, cardiac \gls{MRI} allows accurate quantification of both cardiac anatomy and function (e.g. cine imaging) and pathological tissues such as scars (\gls{LGE} imaging). Accordingly, cardiac MRI is currently regarded as the gold standard for quantitative cardiac analysis~\citep{Van_der_Geest_1999_JMRI}.

A group of representative deep learning based cardiac MR segmentation methods are shown in Table \ref{tab:MR}. From the table, one can see that a majority of works have focused on segmenting cardiac chambers (e.g. \gls{LV}, \gls{RV}, \gls{LA}). In contrast, there are relatively fewer works on segmenting abnormal cardiac tissue regions such as myocardial scars and atrial fibrosis. This is likely due to the limited relevant public datasets as well as the difficulty of the task. In addition, to the best of our knowledge, there are very few works that apply deep learning techniques to atrial wall segmentation, as also suggested by a recent survey paper~\citep{Karim_2018_MedIA}. In the following sections, we will describe and discuss these methods regarding different applications in detail.

\smallskip\subsubsection{Ventricle Segmentation}
\textbf{Vanilla FCN-based Segmentation:}
\cite{Tran_2016_Arxiv} was among the first ones to apply a \gls{FCN}~\citep{Shelhamer_2017_TPAMI} to segment the left ventricle, myocardium and right ventricle directly on short-axis cardiac \gls{MR} images. Their end-to-end approach based on \gls{FCN} achieved competitive segmentation performance, significantly outperforming traditional methods in terms of both speed and accuracy. In the following years, a number of works based on \glspl{FCN} have been proposed, aiming at achieving further improvements in segmentation performance. In this regard, one stream of work focuses on optimizing the network structure to enhance the feature learning capacity for segmentation~\citep{Khened_2019_MedIA,Li_2019_ITBE, Zhou_2018_RAL,Zhang_2019_Access, Cong_2018_JE, Jang_2017_STACOM,Fahmy_2019_JCMR}. For example,~\cite{Khened_2019_MedIA} developed a dense U-net with inception modules to combine multi-scale features for robust segmentation across images with large anatomical variability.~\cite{Jang_2017_STACOM,Yang_2017_STACOM, Sander_2019_MIP,Chen_2019_ISBI} investigated different loss functions such as weighted cross-entropy, weighted Dice loss, deep supervision loss and focal loss to improve the segmentation performance. Among these \gls{FCN}-based methods, the majority of approaches use 2D networks rather than 3D networks for segmentation. This is mainly due to the typical low through-plane resolution and motion artifacts of most cardiac \gls{MR} scans, which limits the applicability of 3D networks~\citep{Baumgartner_2017_STACOM}. 

\textbf{Introducing spatial or temporal context}:
One drawback of using 2D networks for cardiac segmentation is that these networks work slice by slice, and thus they do not leverage any inter-slice dependencies. As a result, 2D networks can fail to locate and segment the heart on challenging slices such as apical and basal slices where the contours of the ventricles are not well defined. To address this problem, a number of works have attempted to introduce additional contextual information to guide 2D \gls{FCN}. This contextual information can include shape priors learned from labels or multi-view images~\citep{Zotti_2017_STACOM, Zotti_2019_JBHI,Chen_2019_MICCAI}. Others extract spatial information from adjacent slices to assist the segmentation, using recurrent units (\glspl{RNN}) or multi-slice networks (2.5D networks)~\citep{Poudel_2016_HVSCMR,Patravali_2017_STACOM,Du_2019_JTEHM,Zheng_2018_TMI}. These networks can also be applied to leveraging information across different temporal frames in the cardiac cycle to improve spatial and temporal consistency of segmentation results~\citep{Yan_2018_MICCAI,Savioli_2018_SNAMS,Du_2019_JTEHM, Qin_2018_MICCAI,Wolterink_2017_STACOM}. 

\textbf{Applying anatomical constraints:} Another problem that may limit the segmentation performance of both 2D and 3D \glspl{FCN} is that they are typically trained with pixel-wise loss functions only (e.g. cross-entropy or soft-Dice losses). These pixel-wise loss functions may not be sufficient to learn features that represent the underlying anatomical structures. Several approaches therefore focus on designing and applying anatomical constraints to train the network to improve its prediction accuracy and robustness. These constraints are represented as regularization terms which take into account the topology~\citep{Clough_2019_IPMI}, contour and region information~\citep{Chen_2019_CVPR} or shape information~\citep{Oktay_2018_TMI, Yue_2019_MICCAI}, encouraging the network to generate more anatomically plausible segmentations. In addition to regularizing networks at training time, ~\cite{Painchaud_2019_MICCAI} proposed a variational \gls{AE} to correct inaccurate segmentations, in the post-processing stage.

\textbf{Multi-task learning}: Multi-task learning has also been explored to regularize \gls{FCN}-based cardiac ventricle segmentation during training by performing auxiliary tasks that are relevant to the main segmentation task, such as motion estimation~\citep{Qin_2018_MLMLR}, estimation of cardiac function~\citep{Dangi_2018_STACOM}, ventricle size classification~\citep{Zhang_2018_ICASSP} and image reconstruction~\citep{ChartsiasA_2018,Huang_2019_FIMH}. Training a network for multiple tasks simultaneously encourages the network to extract features which are useful across these tasks, resulting in improved learning efficiency and prediction accuracy.

\textbf{Multi-stage networks}: Recently, there is a growing interest in applying neural networks in a multi-stage pipeline which breaks down the segmentation problem into subtasks ~\citep{Vigneault_2018_MedIA,Zheng_2018_TMI,Li_2019_ISBI,Tan_2017_MedIA,Liao_2019_TCyber}. For example,~\cite{Zheng_2018_TMI,Li_2019_ISBI} proposed a \gls{ROI} localization network followed by a segmentation network. Likewise,~\cite{Vigneault_2018_MedIA} proposed a network called Omega-Net which consists of a U-net for cardiac chamber localization, a learnable transformation module to normalize image orientation and a series of U-nets for fine-grained segmentation. By explicitly localizing the \gls{ROI} and by rotating the input image into a canonical orientation, the proposed method better generalizes to images with varying sizes and orientations.

\textbf{Hybrid segmentation methods}: Another stream of work aims at combining neural networks with classical segmentation approaches, e.g. level-sets~\citep{Ngo_2016_MedIA,Duan_2018_MICCAI}, deformable models~\citep{Avendi_2016_MedIA, Avendi_2017_MRM,Medley_2019_ISBI}, atlas-based methods~\citep{ Yang_2016_MICCAI,Rohe_2017_STACOM} and graph-cut based methods~\citep{Lu_2019_ICCSP}. Here, neural networks are applied in the feature extraction and model initialization stages, reducing the dependency on manual interactions and improving the segmentation accuracy of the conventional segmentation methods deployed afterwards. For example, \cite{Avendi_2016_MedIA} proposed one of the first \gls{DL}-based methods for \gls{LV} segmentation in cardiac short-axis MR images. The authors first applied a \gls{CNN} to automatically detect the \gls{LV} and then used an \gls{AE} to estimate the shape of the LV. The estimated shape was then used to initialize follow-up deformable models for shape refinement. As a result, the proposed integrated deformable model converges faster than conventional deformable models and the segmentation achieves higher accuracy. In their later work, the authors extended this approach to segment \gls{RV}~\citep{Avendi_2017_MRM}. While these hybrid methods demonstrated better segmentation accuracy than previous non-deep learning methods, most of them still require an iterative optimization for shape refinement. Furthermore, these methods are often designed for one particular anatomical structure. As noted in the recent benchmark study~\citep{Bernard_2018_TMI}, most state-of-the-art segmentation algorithms for bi-ventricle segmentation are based on end-to-end \glspl{FCN}, which allows the simultaneous segmentation of the \gls{LV} and \gls{RV}.

To better illustrate these developments for cardiac ventricle segmentation from cardiac MR images, we collate a list of bi-ventricle segmentation methods that have been trained and tested on the \gls{ACDC} dataset, reported in Table~\ref{tab:ACDC_Online_Evaluation}. For ease of comparison, we only consider those methods which have been evaluated on the same online test set (50 subjects). As the \gls{ACDC} challenge organizers keep the online evaluation platform open to the public, our comparison not only includes the methods from the original challenge participants (summarized in the benchmark study paper from~\cite{Bernard_2018_TMI}) but also three segmentation algorithms that have been proposed after the challenge (i.e. \cite{Zotti_2019_JBHI,Li_2019_ISBI,Painchaud_2019_MICCAI}). From this comparison, one can see that top algorithms are the ensemble method proposed by \cite{Isensee_2017_STACOM} and the two-stage method proposed by \cite{Li_2019_ISBI}, both of which are based on \glspl{FCN}. In particular, compared to the traditional level-set method~\citep{Tziritas_2017_STACOM}, both methods achieved considerably higher accuracy even for the more challenging segmentation of the left ventricular myocardium (Myo), indicating the power of deep learning based approaches. 

\begin{table*}[!ht]
\resizebox{\textwidth}{!}{
\begin{threeparttable}
\caption{\textbf{Segmentation accuracy of state-of-the-art segmentation methods verified on the cardiac bi-ventricular segmentation challenge (ACDC) dataset~\citep{Bernard_2018_TMI}} All the methods were evaluated on the same test set (50 subjects). Bold numbers are the highest overall Dice values for the corresponding structure. LV: left ventricle, RV: right ventricle, Myo: left ventricular myocardium; ED: end-diastolic; ES: end-systolic. Last update: 2019.8.1.}
\label{tab:ACDC_Online_Evaluation}
\begin{tabular}{@{}lllll@{}}
\toprule
Methods & Description & LV & Myo & RV \\ \midrule
\cite{Isensee_2017_STACOM} & 2D U-net +3D U-net  (ensemble) & \textbf{0.950} & \textbf{0.911} & 0.923 \\
\cite{Li_2019_ISBI} & Two 2D FCNs for ROI detection and segmentation respectively; & 0.944 & \textbf{0.911} & \textbf{0.926} \\
\cite{Zotti_2019_JBHI} & 2D GridNet-MD with registered shape prior & 0.938 & 0.894 & 0.910 \\
\cite{Khened_2019_MedIA} & 2D Dense U-net with inception module & 0.941 & 0.894 & 0.907 \\
\cite{Baumgartner_2017_STACOM} & 2D U-net with cross entropy loss & 0.937 & 0.897 & 0.908 \\
\cite{Zotti_2017_STACOM} & 2D GridNet with registered shape prior & 0.931 & 0.890 & 0.912 \\
\cite{Jang_2017_STACOM} & 2D M-Net with weighted cross entropy loss & 0.940 & 0.885 & 0.907 \\
\cite{Painchaud_2019_MICCAI} & FCN followed by an AE for shape correction & 0.936 & 0.889 & 0.909 \\
\cite{Wolterink_2017_STACOM} & Multi-input 2D dilated FCN, segmenting paired ED and ES frames simultaneously & 0.940 & 0.885 & 0.900 \\
\cite{Patravali_2017_STACOM} & 2D U-net with a Dice loss & 0.920 & 0.890 & 0.865 \\
\cite{Rohe_2017_STACOM} & Multi-atlas based method combined with 3D CNN for registration & 0.929 & 0.868 & 0.881 \\
\cite{Tziritas_2017_STACOM} &  Level-set +\gls{MRF}; \emph{Non-deep learning method} & 0.907 & 0.798 & 0.803 \\
\cite{Yang_2017_STACOM} & 3D FCN with deep supervision & 0.820 & N/A & 0.780 \\ \bottomrule
\end{tabular}

\begin{tablenotes}
\small
\item Note that for simplicity, we report the average Dice scores for each structure over ED and ES phases. More detailed comparison for different phases can be found on the public leaderboard in the post testing part (\url{https://acdc.creatis.insa-lyon.fr}) as well as corresponding published works in this table.
\end{tablenotes}
\end{threeparttable}
}

\end{table*} 

\smallskip
\subsubsection{Atrial Segmentation}
\Glsfirst{AF} is one of the most common cardiac electrical disorders, affecting around 1 million people in the UK~\footnote{\url{https://www.nhs.uk/conditions/atrial-fibrillation/}}. Accordingly, atrial segmentation is of prime importance in the clinic, improving the assessment of the atrial anatomy in both pre-operative~\gls{AF} ablation planning and post-operative follow-up evaluations. In addition, the segmentation of atrium can be used as a basis for scar segmentation and atrial fibrosis quantification from \gls{LGE} images. Traditional methods such as region growing~\citep{Karim_2008_Online} and methods that employ strong priors (i.e. atlas-based label fusion~\citep{Tao_2016_JMRI} and non-rigid registration~\citep{Zhuang_2010_TMI}) have been applied in the past for automated left atrium segmentation. However, the accuracy of these methods highly relies on good initialization and ad-hoc pre-processing methods, which limits the widespread adoption in the clinic.

Recently, ~\cite{Bai_2018_JCMR} and~\cite{Vigneault_2018_MedIA} applied 2D \glspl{FCN} to directly segment the \gls{LA} and \gls{RA} from standard 2D long-axis images, i.e. \gls{2CH}, \gls{4CH} views. Notably, their networks can also be trained to segment ventricles from 2D short-axis stacks without any modifications to the network architecture. Likewise, \cite{Xiong_2019_TMI, Preetha_2018_STACOM, Bian_2018_STACOM,Chen_2018_STACOM} applied 2D FCNs to segment the atrium from 3D \gls{LGE} images in a slice-by-slice fashion, where they optimized the network structure for enhanced feature learning. 3D networks~\citep{Xia_2018_STACOM,Savioli_2018_STACOM,Jia_2018_STACOM,Vesal_2018_STACOM,Li_2018_STACOM} and multi-view \gls{FCN}~\citep{Mortazi_2017_STACOM,Yang_2018_EMBC} have also been explored to capture 3D global information from 3D LGE images for accurate atrium segmentation. 

In particular, \cite{Xia_2018_STACOM} proposed a fully automatic two-stage segmentation framework which contains a first 3D U-net to roughly locate the atrial center from down-sampled images followed by a second 3D U-net to accurately segment the atrium in the cropped portions of the original images at full resolution. Their multi-stage approach is both memory-efficient and accurate, ranking first in the \gls{LASC 2018} with a mean Dice score of 0.93 evaluated on a test set of 54 cases.

\smallskip
\subsubsection{Scar Segmentation}
Scar characterization is usually performed using \gls{LGE} MR imaging, a contrast-enhanced MR imaging technique. \gls{LGE} MR imaging enables the identification of myocardial scars and atrial fibrosis, allowing improved management of myocardial infarction and atrial fibrillation~\citep{Kim_1999_Circulation}. Prior to the advent of deep learning, scar segmentation was often performed using intensity thresholding-based or clustering methods which are sensitive to the local intensity changes~\citep{Zabihollahy_2018_MI}. The main limitation of these methods is that they usually require the manual segmentation of the region of interest to reduce the search space and the computational costs~\citep{Carminati_2016_JTI}. As a result, these semi-automated methods are not suitable for large-scale studies or clinical deployment.

Deep learning approaches have been combined with traditional segmentation methods for the purpose of scar segmentation: \cite{Yang_2017_MIUA,Yang_2018_MedPhy} applied an atlas-based method to identify the left atrium and then applied deep neural networks to detect fibrotic tissue in that region. Relatively to end-to-end approaches, \cite{Chen_2018_MICCAI} applied deep neural networks to segment both the left atrium and the atrial scars. In particular, the authors employed a multi-view CNN with a recursive attention module to fuse features from complementary views for better segmentation accuracy. Their approach achieved a mean Dice score of 0.90 for the \gls{LA} region and a mean Dice score of 0.78 for atrial scars. 

In the work of \cite{Fahmy_2018_JACC}, the authors applied a U-net based network to segment the myocardium and the scars at the same time from \gls{LGE} images acquired from patients with \gls{HCM}, achieving a fast segmentation speed. However, the reported segmentation accuracy for the scar regions was relatively low (mean Dice: 0.58). \cite{Zabihollahy_2018_MI,Moccia_2019_Magma} instead adopted a semi-automated method which requires a manual segmentation of the myocardium followed by the application of a 2D network to differentiate scars from normal myocardium. They reported higher segmentation accuracy on their test sets (mean Dice $>$0.68). At the moment, fully-automated scar segmentation is still a challenging task since the infarcted regions in patients can lead to kinematic variabilities and abnormalities in those contrast-enhanced images. Interestingly, \cite{Xu_2018_MedIA} developed an \gls{RNN} which leverages motion patterns to automatically delineate myocardial infarction area from cine \gls{MR} image sequences without contrast agents. Their method achieved a high overall Dice score of 0.90 when compared to the manual annotations on LGE MR images, providing a novel approach for infarction assessment.

\smallskip
\subsubsection{Aorta Segmentation}
The segmentation of the aortic lumen from cine MR images is essential for accurate mechanical and hemodynamic characterization of the aorta. One common challenge for this task is the typical sparsity of the annotations in aortic cine image sequences, where only a few frames have been annotated. To address the problem, \cite{Bai_2018_MICCAI} applied a non-rigid image registration method~\citep{Rueckert_1999_TMI} to propagate the labels from the annotated frames to the unlabeled neighboring ones in the cardiac cycle, effectively generating pseudo annotated frames that could be utilized for further training. This semi-supervised method achieved an average Dice metric of 0.96 for the ascending aorta and 0.95 for the descending aorta over a test set of 100 subjects. In addition, compared to a previous approach based on deformable models~\citep{Herment_2010_JMRI}, their approach based on \gls{FCN} and \gls{RNN} can directly perform the segmentation task on a whole image sequence without requiring the explicit estimation of the \gls{ROI}.

\smallskip
\subsubsection{Whole Heart Segmentation}
Apart from the above mentioned segmentation applications which target one particular structure,  deep learning can also be used to segment the main substructures of the heart in 3D MR images \citep{Yu_2017_MICCAI,Wolterink_2017_RSAM,Li_2017_RSAM,Shi_2018_MICCAI}. An early work from \cite{Yu_2017_MICCAI} adopted a 3D dense FCN to segment the myocardium and blood pool in the heart from 3D MR scans. Recently, more and more methods began to apply deep learning pipelines to segment more specific substructures (incl. four chambers, \gls{MYO}, aorta, \gls{PV}) in both 3D CT and MR images. This has been facilitated by the availability of public datasets for whole heart segmentation (\gls{MM-WHS}). In general, the segmentation task on MR images is harder than the one of CT images mainly because of the large variations in terms of image intensity distribution among different scanners. As mentioned in a recent benchmark study paper by \cite{Zhuang_2019_MedIA}, deep learning methods in general achieve better segmentation accuracy on CT images compared to that of MR images. We will discuss these segmentation methods in the next CT section in further detail (see section~\ref{sec: CT_whole_heart}).

\medskip\subsection{Cardiac CT Image Segmentation}
\gls{CT} is a non-invasive imaging technique that is performed routinely for disease diagnosis and treatment planning. In particular, cardiac CT scans are used for assessment of cardiac anatomy and specifically the coronary arteries. There are two main imaging modalities: non-contrast CT imaging and contrast-enhanced coronary CT angiography (CTA). Typically, non-contrast CT imaging exploits density of tissues to generate an image, such that different densities using various attenuation values such as soft tissues, calcium, fat, and air can be easily distinguished, and thus allows to estimate the amount of calcium present in the coronary arteries \citep{Kang_2012_JoEI}. In comparison, contrast-enhanced coronary CTA, which is acquired after the injection of a contrast agent, can provide excellent visualization of cardiac chambers, vessels and coronaries, and has been shown to be effective in detecting non-calcified coronary plaques. In the following sections, we will review some of the most commonly used deep learning-based cardiac CT segmentation methods. A summary of these approaches is presented in Table \ref{tab:CT}.
\begin{table*}[!ht]
    \centering
        \caption{\textbf{A summary of selected deep learning methods on cardiac CT segmentation.}}
\resizebox{\textwidth}{!}{
\begin{tabular}{cllcc}
\toprule
Application & Selected works & Description & Imaging Modality & Structure(s) \\ \midrule
\multirow{15}{*}{\makecell{ Cardiac Substructure \\ Segmentation }} & \textbf{Two-step segmentation} & & & \\ &  \cite{Zreik_2016_ISBI_CT} & patch based CNN & CTA & LV \\
&  \cite{Payer_2018_MMWHS} & a pipeline of two FCNs & MR/CT & WHS \\
&  \cite{Tong_2017_MMWHS} & deeply supervised 3D U-net & MR/CT & WHS \\
&  \cite{Wang_2018_arXiv} & {two-stage 3D U-net with dynamic ROI extraction} & MR/CT & WHS \\
&  \cite{Xu_2018_arXiv} & faster RCNN and U-net & CT & WHS \\ \cline{2-5}
& \textbf{Multi-view CNNs } & & & \\ & \cite{Wang_2017_MMWHS} & {orthogonal 2D U-nets with shape context} & MR/CT & WHS \\
&  \cite{Mortazi_2017_MMWHS} & {multi-planar FCNs with an adaptive fusion strategy} & MR/CT & WHS \\ \cline{2-5}
& \textbf{Hybrid loss } & & & \\ & \cite{Yang_2017c_MMWHS} & 3D U-net with deep supervision & MR/CT & WHS \\
&  \cite{Ye_2019_Access} & {3D deeply-supervised U-net with multi-depth fusion} & CT & WHS \\ \cline{2-5}
& \textbf{Others } & & & \\ & \cite{Zreik_2018_MedIA} & multi-scale FCN & CTA & Myo \\
& \cite{Joyce_2018_MIDL} & {unsupervised segmentation with GANs} & MR/CT & LV/RV/Myo \\ \midrule
\multirow{11}{*}{\makecell{ Coronary Artery \\ Segmentation }} & \textbf{End-to-end CNNs}& & & \\ & \cite{Moeskops_2016_MICCAI} & multi-task CNN & CTA & Vessel \\
&  \cite{Merkow_2016_MICCAI} & {3D U-net with deep multi-scale supervision} & CTA & Vessel \\
& \cite{Lee_2019_TMI} & template transformer network & CTA & Vessel \\ \cline{2-5}
& \textbf{{CNN as pre-/post-processing }} & & & \\ & \cite{Gulsun_2016_MICCAI} & CNN as path pruning & CTA & {coronary artery centerline} \\
&  \cite{Guo_2019_IPMI} & {multi-task FCN with a minimal patch extractor} & CTA & {coronary artery centerline} \\
& \cite{Shen_2019_IEEEAccess} & 3D FCN with level set & CTA & Vessel \\ \cline{2-5}
& \textbf{Others} & & & \\ & \cite{Wolterink_2019_MedIA} & {CNN to estimate direction classification
and radius regression} & CTA & {coronary artery centerline} \\
& \cite{Wolterink_2019_arXiv} & graph convolutional network & CTA & Vessel \\ \midrule
\multirow{11}{*}{\makecell{Coronary Artery \\ Calcium 
and Plaque \\ Segmentation }} & \textbf{Two-step segmentation } & & & \\ & \cite{Wolterink_2016_MedIA} & CNN pairs & CTA & CAC \\
& \cite{Lessmann_2016_MedicalImaging} & multi-view CNNs & CT & CAC \\
& \cite{Lessmann_2017_TMI} & two consecutive CNNs & CT & CAC \\
& \cite{Liu_2018_STACOM} & 3D vessel-focused ConvNets & CTA & {CAC/NCP/MCP} \\ \cline{2-5}
& \textbf{Direct segmentation } & & & \\ & \cite{Santini_2017_IFMBE} & patch based CNN & CT & CAC \\
& \cite{Shadmi_2018_ISBI} & U-net and FC DenseNet & CT & CAC \\
& \cite{Zhang_2019_Computing} & U-DenseNet & CT & CAC \\
& \cite{Ma_2019_arXiv} & DenseRAU-net & CT & CAC \\
\bottomrule
\end{tabular}
}
\label{tab:CT}
\end{table*}

\smallskip\subsubsection{Cardiac Substructure Segmentation}
\label{sec: CT_whole_heart}
Accurate delineation of cardiac substructures plays a crucial role in cardiac function analysis, providing important clinical variables such as \gls{EF}, myocardial mass, wall thickness etc. Typically, the cardiac substructures that are segmented include the \gls{LV}, \gls{RV}, \gls{LA}, \gls{RA}, \gls{MYO}, \gls{AO} and \gls{PA}. 
\newline
\textbf{Two-step segmentation}: One group of deep learning methods relies on a two-step segmentation procedure, where a \gls{ROI} is first extracted and then fed into a CNN for subsequent classification \citep{Zreik_2016_ISBI_CT,Dormer_2018_SPIE_CT}. For instance, \cite{Zreik_2016_ISBI_CT} proposed a two-step \gls{LV} segmentation process where a bounding box for the \gls{LV} is first detected using the method described in \citep{De_2017_TMI}, followed by a voxel classification within the defined bounding box using a patch-based CNN. More recently, \gls{FCN}, especially U-net \citep{Ronneberger_2015_MICCAI}, has become the method of choice for cardiac CT segmentation. \cite{Zhuang_2019_MedIA} provides a comparison of a group of methods \citep{Payer_2018_MMWHS,Wang_2017_MMWHS,Yang_2017b_MMWHS,Yang_2017c_MMWHS,Tong_2017_MMWHS, Mortazi_2017_MMWHS} for \gls{WHS} that have been evaluated on the \gls{MM-WHS} challenge. Several of these methods \citep{Payer_2018_MMWHS,Tong_2017_MMWHS,Xu_2018_arXiv,Wang_2018_arXiv} combine a localization network, which produces a coarse detection of the heart, with 3D FCNs applied to the detected \gls{ROI} for segmentation. This allows the segmentation network to focus on the anatomically relevant regions, and has shown to be effective for whole heart segmentation. In the \gls{MM-WHS} challenge the method of \cite{Payer_2018_MMWHS} ranked 1st. A summary of the comparison between the segmentation accuracy of the methods evaluated on \gls{MM-WHS} dataset is presented in Table \ref{tab:WHS}. For more details, please refer to \cite{Zhuang_2019_MedIA}.

\begin{table*}[!ht]
    \centering
    \caption{\textbf{Segmentation accuracy of methods validated on \gls{MM-WHS} dataset.} The training set contains 20 CT and 20 MRI whereas the test set contains 40 CT and 40 MRI. Reported numbers are Dice scores (CT/MRI) for different substructures on both CT and MRI scans. For more detailed comparisons, please refer to \cite{Zhuang_2019_MedIA}.}
    \scalebox{0.85}{
    \begin{tabular}{clllllllll}
    \toprule
    Methods  & LV & RV & LA & RA & MYO & AO & PA & WHS  \\ \midrule
    \cite{Payer_2018_MMWHS} & 91.8/{91.6} & \textbf{90.9}/{86.8} & 92.9/85.5 & \textbf{88.8}/88.1 & 88.1/77.8 & 93.3/88.8 & 84.0/73.1 & \textbf{90.8}/86.3 \\ 
    \cite{Yang_2017b_MMWHS} & 92.3/75.0 & 85.7/75.0 &\textbf{93.0}/82.6 & 87.1/85.9 & 85.6/65.8 & 89.4/80.9 & 83.5/72.6 & 89.0/78.3\\
    \cite{Mortazi_2017_MMWHS} & 90.4/87.1 & 88.3/83.0 & 91.6/81.1 & 83.6/75.9 & 85.1/74.7 & 90.7/83.9 & 78.4/71.5 & 87.9/81.8\\
    \cite{Tong_2017_MMWHS} & 89.3/70.2 & 81.0/68.0 & 88.9/67.6 & 81.2/65.4 & 83.7/62.3 & 86.8/59.9 & 69.8/47.0 & 84.9/67.4 \\
    \cite{Wang_2018_arXiv} & 80.0/86.3 & 78.6/84.9 & 90.4/85.2 & 79.4/84.0 & 72.9/74.4 & 87.4/82.4 & 64.8/78.8 & 80.6/83.2 \\
    \cite{Ye_2019_Access} & \textbf{94.4}/ - & 89.5/ - & 91.6/ - & 87.8/ - & \textbf{88.9}/ - & \textbf{96.7}/ - & \textbf{86.2}/ - & 90.7/ - \\
    \cite{Xu_2018_arXiv} &87.9/ - & 90.2/ - & 83.2/ - & 84.4/ - & 82.2/ - &91.3/ - & 82.1/ -& 85.9/ - \\
    \bottomrule
    \end{tabular}
    }
    \label{tab:WHS}
\end{table*}

\textbf{Multi-view CNNs}: Another line of research utilizes the volumetric information of the heart by training multi-planar CNNs (axial, sagittal, and coronal views) in a 2D fashion. Examples include \cite{Wang_2017_MMWHS} and \cite{Mortazi_2017_MMWHS} where three independent orthogonal CNNs were trained to segment different views. Specifically, \cite{Wang_2017_MMWHS} additionally incorporated shape context in the framework for the segmentation refinement, while \cite{Mortazi_2017_MMWHS} adopted an adaptive fusion strategy to combine multiple outputs utilising complementary information from different planes. 

\textbf{Hybrid loss}: Several methods employ a hybrid loss, where different loss functions (such as focal loss, Dice loss, and weighted categorical cross-entropy) are combined to address the class imbalance issue, e.g. the volume size imbalance among different ventricular structures, and to improve the segmentation performance \citep{Yang_2017c_MMWHS,Ye_2019_Access}.

In addition, the work of \cite{Zreik_2018_MedIA} has proposed a method for the automatic identification of patients with significant coronary artery stenoses through the segmentation and analysis of the \gls{LV} myocardium. In this work, a multi-scale FCN is first employed for myocardium segmentation, and then a convolutional autoencoder is used to characterize the \gls{LV} myocardium, followed by a \gls{SVM} to classify patients based on the extracted features.

\smallskip\subsubsection{Coronary Artery Segmentation}
Quantitative analysis of coronary arteries is an important step for the diagnosis of cardiovascular diseases, stenosis grading, blood flow simulation and surgical planning \citep{Zhang_2010_Thesis}. Though this topic has been studied for years \citep{Lesage_2009_MedIA}, only a small number of works investigate the use of deep learning in this context. Methods relating to coronary artery segmentation can be mainly divided into two categories: centerline extraction and lumen (i.e. vessel wall) segmentation. 

\textbf{CNNs as a post-/pre-processing step}: Coronary centerline extraction is a challenging task due to the presence of nearby cardiac structures and coronary veins as well as motion artifacts in cardiac CT. Several deep learning approaches employ CNNs as either a post-processing or pre-processing step for traditional methods. For instance, \cite{Gulsun_2016_MICCAI} formulated centerline extraction as finding the maximum flow paths in a steady state porous media flow, with a learning-based classifier estimating anisotropic vessel orientation tensors for flow computation. A CNN classifier was then employed to distinguish true coronary centerlines from leaks into non-coronary structures. \cite{Guo_2019_IPMI} proposed a multi-task FCN centerline extraction method that can generate a single-pixel-wide centerline, where the FCN simultaneously predicted centerline distance maps and endpoint confidence maps from coronary arteries and ascending aorta segmentation masks, which were then used as input to the subsequent minimal path extractor to obtain the final centerline extraction results. In contrast, unlike the aforementioned methods that used CNNs either as a pre-processing or post-processing step, \cite{Wolterink_2019_MedIA} proposed to address centerline extraction via a 3D dilated CNN, where the CNN was trained on patches to directly determine a posterior probability distribution over a discrete set of possible directions as well as to estimate the radius of an artery at the given point. 

\textbf{End-to-end CNNs}: With respect to the lumen or vessel wall segmentation, most deep learning based approaches use an end-to-end CNN segmentation scheme to predict dense segmentation probability maps \citep{Moeskops_2016_MICCAI,Merkow_2016_MICCAI,Huang_2018_EMBC,Shen_2019_IEEEAccess}. In particular, \cite{Moeskops_2016_MICCAI} proposed a multi-task segmentation framework where a single CNN can be trained to perform three different tasks including coronary artery segmentation in cardiac CTA and tissue segmentation in brain MR images. They showed that such a multi-task segmentation network in multiple modalities can achieve equivalent performance as a single task network. \cite{Merkow_2016_MICCAI} introduced deep multi-scale supervision into a 3D U-net architecture, enabling efficient multi-scale feature learning and precise voxel-level predictions. Besides, shape priors can also be incorporated into the network \citep{Lee_2019_TMI,Chen_2019_MIDL,Duan_2018_STACOM}. For instance, \cite{Lee_2019_TMI} explicitly enforced a roughly tubular shape prior for the vessel segments by introducing a template transformer network, through which a shape template can be deformed via network-based registration to produce an accurate segmentation of the input image, as well as to guarantee topological constraints. More recently, graph convolutional networks have also been investigated by \cite{Wolterink_2019_arXiv} for coronary artery segmentation in CTA, where vertices on the coronary lumen surface mesh were considered as graph nodes and the locations of these tubular surface mesh vertices were directly optimized. They showed that such method significantly outperformed a baseline network that used only fully-connected layers on healthy subjects (mean Dice score: 0.75 vs 0.67 ). Besides, the graph convolutional network used in their work is able to directly generate smooth surface meshes without post-processing steps.

\smallskip\subsubsection{Coronary Artery Calcium and Plaque Segmentation}
\Gls{CAC} is a direct risk factor for cardiovascular disease. Clinically, \gls{CAC} is quantified using the Agatston score \citep{Agatston_1990_JACC} which considers the lesion area and the weighted maximum density of the lesion \citep{deVos_2019_TMI}. Precise detection and segmentation of \gls{CAC} are thus important for the accurate prediction of the Agatston score and disease diagnosis. 

\textbf{Two-step segmentation}: One group of deep learning approaches to segmentation and automatic calcium scoring proposed to use a two-step segmentation scheme. For example, \cite{Wolterink_2016_MedIA} attempted to classify \gls{CAC} in cardiac CTA using a pair of CNNs, where the first CNN coarsely identified voxels likely to be \gls{CAC} within a ROI detected using \citep{De_2017_TMI} and then the second CNN further distinguished between \gls{CAC} and \gls{CAC}-like negatives more accurately. Similar to such a two-stage scheme, \cite{Lessmann_2016_MedicalImaging,Lessmann_2017_TMI} proposed to identify \gls{CAC} in low-dose chest CT, in which a ROI of the heart or potential calcifications were first localized followed by a \gls{CAC} classification process. 

\textbf{Direct segmentation}: More recently, several approaches \citep{Shadmi_2018_ISBI,Santini_2017_IFMBE,Ma_2019_arXiv,Zhang_2019_Computing} have been proposed for the direct segmentation of \gls{CAC} from non-contrast cardiac CT or chest CT: the majority of them employed  combinations of U-net \citep{Ronneberger_2015_MICCAI} and DenseNet \citep{Huang_2017_CVPR} for precise quantification of CAC which showed that a sensitivity over 90\% can be achieved \cite{Santini_2017_IFMBE}. These aforementioned approaches all follow the same workflow where the \gls{CAC} is first identified and then quantified. An alternative approach is to circumvent the intermediate segmentation and to perform direct quantification, such as in \citep{Cano_2018_MedicalImaging,deVos_2019_TMI}, which have proven that this approach is effective and promising.

Finally, for \gls{NCP} and \gls{MCP} in coronary arteries, only a limited number of works have been reported that investigate deep learning methods for segmentation and quantification  \citep{Zreik_2018_TMI,Liu_2018_STACOM}. Yet, this is a very important task from a clinical point of view, since these plaques can potentially rupture and obstruct an artery, causing ischemic events and severe cardiac damage. In contrast to CAC segmentation, \gls{NCP} and \gls{MCP} segmentation are more challenging due to their similar appearances and intensities as adjacent tissues. Therefore, robust and and accurate analysis often requires the generation of \gls{MPR} images that have been straightened along the centreline of the vesssel. Recently, \cite{Liu_2018_STACOM} proposed a vessel-focused 3D convolutional network with attention layers to segment three types of plaques on the extracted and reformatted coronary \gls{MPR} volumes. \cite{Zreik_2018_TMI} presented an automatic method for detection and characterization of coronary artery plaques as well as  determination of coronary artery stenosis significance, in which a multi-task convolutional \gls{RNN} was used to perform both plaque and stenosis classification by analyzing the features extracted along the coronary artery in an MPR image.

\subsection{Cardiac Ultrasound Image Segmentation}
\begin{table*}[!ht]
\centering
  \caption{\textbf{A summary of reviewed deep learning methods for US image segmentation.} A[X]C is short for Apical [X]-chamber view. PLAX/PSAX: parasternal long-axis/short-axis. CETUS: using the dataset from Challenge on Endocardial Three-dimensional Ultrasound Segmentation.}
\label{tab:echo-table}
\resizebox{\textwidth}{!}{

\begin{tabular}{@{}lllll@{}}
\toprule
Application & Selected works & Method & Structure & Imaging modality \\ \midrule
\multirow{17}{*}{\makecell{2D LV}} &

\textbf{Combined with deformable models} &  &  &  \\
 & \cite{Carneiro_2010_ISBI,  Carneiro_2012_TIP} & \begin{tabular}[c]{@{}l@{}}DBN with two-step approach: localization and fine segmentation\\
 \end{tabular} & LV & 2D A2C, A4C \\
 & \cite{Nascimento_2017_TIP} & \begin{tabular}[c]{@{}l@{}}\gls{DBN} and sparse manifold learning for the localization step\end{tabular} & LV & 2D A2C, A4C \\
 & \cite{Nascimento_2014_CVPR, Nascimento_2019_TPAMI} & \begin{tabular}[c]{@{}l@{}}\gls{DBN} and sparse manifold learning for one-step segmentation\end{tabular} & LV & 2D A2C, A4C \\
 & \cite{Veni_2018_ISBI} & \begin{tabular}[c]{@{}l@{}}\gls{FCN} (U-net) followed by level-set based deformable model\end{tabular} & LV & 2D A4C \\ \cmidrule(l){2-5}

 & \textbf{Utilizing temporal coherence} &  &  &  \\
 & \cite{Carneiro_2010_CVPR, Carneiro_2013_TPAMI} & \begin{tabular}[c]{@{}l@{}}\gls{DBN} and particle filtering for dynamic modeling\end{tabular} & LV & 2D A2C, A4C \\
 & \cite{Jafari_2018_DLMIA} & \begin{tabular}[c]{@{}l@{}}U-net and LSTM with additional optical flow input 
 \end{tabular} & LV & 2D A4C \\ \cmidrule(l){2-5} 
 
 & \textbf{Utilizing unlabeled data} &  &  &  \\
  & \cite{Carneiro_2011_ICCV,  Carneiro_2012_CVPR} & \begin{tabular}[c]{@{}l@{}}\gls{DBN} on-line retrain using external classifier as additional supervision \end{tabular} & LV & 2D A2C, A4C \\
 
  & \cite{Smistad_2017_IUS} & \begin{tabular}[c]{@{}l@{}}U-Net trained using labels generated by a Kalman filter based method 
 \end{tabular} & LV and LA & 2D A2C, A4C \\
 
 & \cite{Yu_2017_BiomedEng} & \begin{tabular}[c]{@{}l@{}}Dynamic CNN fine-tuning with mitral valve tracking to separate LV from LA 
 \end{tabular} & Fetal LV & 2D \\
 
 & \cite{Jafari_2019_ISBI} & \begin{tabular}[c]{@{}l@{}}U-net with TL-net~\citep{Girdhar_2016_ECCV} based shape constraint on unannotated frames\end{tabular} & LV & 2D A4C \\ \cmidrule(l){2-5} 
 
 & \textbf{Utilizing data from multiple domains} &  &  &  \\
 & \cite{Chen_2016_MICCAI} & FCN trained using annotated data of multiple anatomical structures & Fetal head and LV & 2D head, A2-5C \\\cmidrule(l){2-5} 
 
 & \textbf{Trained directly on large datasets} &  &  & \\
 & \cite{Smistad_2018_IUS} & \begin{tabular}[c]{@{}l@{}}Real time CNN view-classification and segmentation\end{tabular} & LV & 2D A2C, A4C \\
 & \cite{Leclerc_2018_IUS} & \begin{tabular}[c]{@{}l@{}}U-net trained on a large heterogeneous dataset\end{tabular} & LV & 2D A4C \\ \midrule
 
\multirow{4}{*}{\makecell{3D LV}} & 
\cite{Dong_2018_HindawiBiomed} & \begin{tabular}[c]{@{}l@{}}CNN for 2D coarse segmentation refined by 3D snake model\end{tabular} & LV & 3D (CETUS) \\
 & \cite{Oktay_2018_TMI} & U-net with TL-net based shape constraint & LV & 3D (CETUS) \\
 & \cite{Dong_2018_MICCAI_echo} & \begin{tabular}[c]{@{}l@{}}Atlas-based segmentation using \gls{DL} registration and adversarial training \end{tabular} & LV & 3D \\ \midrule

\multirow{2}{*}{Others}

&\cite{Ghesu_2016_TMI} & \begin{tabular}[c]{@{}l@{}}Marginal space learning and adaptive sparse neural network\end{tabular} & Aortic valves & 3D \\ 
 
 & \cite{Degel_2018_MICCAI} & \begin{tabular}[c]{@{}l@{}}V-net with TL-net based shape constraint and GAN-based domain adaptation \end{tabular} & LA & 3D \\
 
&\cite{Zhang_2018_Circulation} & \begin{tabular}[c]{@{}l@{}}CNN for view-classification, segmentation and disease detection\end{tabular} & Multi-chamber & 2D PLAX, PSAX, A2-4C \\ \bottomrule
\end{tabular}

}
\end{table*}
Cardiac ultrasound (\gls{US}) imaging, also known as echocardiography, is an indispensable clinical tool for the assessment of cardiovascular function. It is often used clinically as the first imaging examination owing to its portability, low cost and real-time capability. While a number of traditional methods such as active contours, level-sets and active shape models have been employed to automate the segmentation of anatomical structures in ultrasound images~\citep{Noble_2006_TMI}, the achieved accuracy is limited by various problems of ultrasound imaging such as low signal-to-noise ratio, varying speckle noise, low image contrast (especially between the myocardium and the blood pool), edge dropout and shadows cast by structures such as dense muscle and ribs.

As in cardiac MR and CT, several \gls{DL}-based methods have been recently proposed to improve the performance of cardiac ultrasound image segmentation in terms of both accuracy and speed. The majority of these DL-based approaches focus on \gls{LV} segmentation, with only few addressing the problem of aortic valve and \gls{LA} segmentation. A summary of the reviewed works can be found in Table~\ref{tab:echo-table}. 

\smallskip\subsubsection{2D LV segmentation} 
\textbf{Deep learning combined with deformable models}: The imaging quality of echocardiography makes voxel-wise tissue classification highly challenging. To address this challenge, deep learning has been combined with deformable model for LV segmentation in 2D images \citep{Carneiro_2010_ISBI, Carneiro_2012_TIP, Carneiro_2010_CVPR, Carneiro_2013_TPAMI, Nascimento_2014_CVPR, Nascimento_2019_TPAMI, Veni_2018_ISBI}. Features extracted by trained deep neural networks were used instead of handcrafted features to improve accuracy and robustness.

Several works applied deep learning in a two-stage pipeline which first localizes the target \gls{ROI} via rigid transformation of a bounding box, then segments the target structure within the \gls{ROI}. This two-stage pipeline reduces the search region of the segmentation and increases robustness of the overall segmentation framework. \cite{Carneiro_2010_ISBI, Carneiro_2012_TIP} first adopted this \gls{DL} framework to segment the \gls{LV} in apical long-axis echocardiograms. The method uses \gls{DBN} \citep{Hinton_2006_Science} to predict the rigid transformation parameters for localization and the deformable model parameters for segmentation. The results demonstrated the robustness of \gls{DBN}-based feature extraction to image appearance variations. \cite{Nascimento_2017_TIP} further reduced the training and inference complexity of the \gls{DBN}-based framework by using sparse manifold learning in the rigid detection step. 
 
To further reduce the computational complexity, some works perform segmentation in one step without resorting to the two-stage approach. \cite{Nascimento_2014_CVPR, Nascimento_2019_TPAMI} applied sparse manifold learning in segmentation, showing a reduced training and search complexity compared to their previous version of the method, while maintaining the same level of segmentation accuracy. \cite{Veni_2018_ISBI} applied a \gls{FCN} to produce coarse segmentation masks, which is then further refined by a level-set based method.

\textbf{Utilizing temporal coherence}: Cardiac ultrasound data is often recorded as a temporal sequence of images. Several approaches aim to leverage the coherence between temporally close frames to improve the accuracy and robustness of the LV segmentation. \cite{Carneiro_2010_CVPR, Carneiro_2013_TPAMI} proposed a dynamic modeling method based on a \gls{SMC} (or particle filtering) framework with a transition model, in which the segmentation of the current cardiac phase depends on previous phases. The results show that this approach performs better than the previous method \citep{Carneiro_2010_ISBI} which does not take temporal information into account. In a more recent work, \cite{Jafari_2018_DLMIA} combined U-net, \gls{LSTM} and inter-frame optical flow to utilize multiple frames for segmenting one target frame, demonstrating improvement in overall segmentation accuracy. The method was also shown to be more robust to image quality variations in a sequence than single-frame U-net.

\textbf{Utilizing unlabeled data}: Several works proposed to use non-DL based segmentation algorithms to help generating labels on unlabeled images, effectively increasing the amount of training data. To achieve this, \cite{Carneiro_2011_ICCV, Carneiro_2012_CVPR} proposed on-line retraining strategies where segmentation network (\gls{DBN}) is firstly initialized using a small set of labeled data and then applied to non-labeled data to propose annotations. The proposed annotations are then checked by external classifiers before being used to re-train the network. \cite{Smistad_2017_IUS} trained a U-net using images annotated by a Kalman filtering based method \citep{Smistad_2014_MIDAS} and illustrated the potential of using this strategy for pre-training. Alternatively, 
some works proposed to exploit unlabeled data without using additional segmentation algorithm. \cite{Yu_2017_BiomedEng} proposed to train a \gls{CNN} on a partially labeled dataset of multiple sequences, then fine-tuned the network for each individual sequence using manual segmentation of the first frame as well as CNN-produced label of other frames. \cite{Jafari_2019_ISBI} proposed a semi-supervised framework which enables training on both the labeled and unlabeled images. The framework uses an additional generative network, which is trained to generate ultrasound images from segmentation masks, as additional supervision for the unlabeled frames in the sequences. The generative network forces the segmentation network to predict segmentation that can be used to successfully generate the input ultrasound image.

\textbf{Utilizing data from multiple domains}: Apart from exploiting unlabeled data in the same domain, leveraging manually annotated data from multiple domains (e.g. different 2D ultrasound views with various anatomical structures) can also help to improve the segmentation in one particular domain. \cite{Chen_2016_MICCAI} proposed a novel \gls{FCN}-based network to utilize multi-domain data to learn generic feature representations. Combined with an iterative refinement scheme, the method has shown superior performance in detection and segmentation over traditional database-guided method \citep{Georgescu_2005_CVPR}, \gls{FCN} trained on single-domain and other multi-domain training strategies. 

\textbf{DL networks trained directly on large datasets}: The potential of \gls{CNN} in segmentation has motivated the collection and labeling of large-scale datasets. Several methods have since shown that deep learning methods, most notably \gls{CNN}-based methods, are capable of performing accurate segmentation directly without complex post-processing. \cite{Leclerc_2018_IUS} performed a study to investigate the effect of the size of annotated data for the segmentation of the LV in 2D ultrasound images using a simple U-net. The authors demonstrated that the U-net approach significantly benefits from larger amounts of training data. Furthermore, \cite{Smistad_2018_IUS} demonstrated the efficiency of \gls{CNN}-based methods by successfully performing real-time view-classification and segmentation.

\smallskip\subsubsection{3D LV segmentation}
Segmenting cardiac structures in 3D ultrasound is even more challenging than 2D. While having the potential to derive more accurate volume-related clinical indices, 3D echocardiograms suffer from lower temporal resolution and lower image quality compared to 2D echocardiograms. Moreover, 3D images dramatically increase the dimension of parameter space of neural networks, which poses computational challenges for deep learning methods. 

One way to reduce the computational cost is to avoid direct processing of 3D data in deep learning networks. \cite{Dong_2018_HindawiBiomed} proposed a two-stage method by first applying a 2D \gls{CNN} to produce coarse segmentation maps on 2D slices from a 3D volume. The coarse 2D segmentation maps are used to initialize a 3D shape model which is then refined by 3D deformable model method~\citep{Kass_1988_IJCV}. In addition, the authors used transfer learning to side-step the limited training data problem by pre-training network on a large natural image segmentation dataset and then fine-tuning to the LV segmentation task. 

Anatomical shape priors have been utilized to increase the robustness of deep learning-based segmentation methods to challenging 3D ultrasound images. \cite{Oktay_2018_TMI} proposed an anatomically constrained network where a shape constraint-based loss is introduced to train a 3D segmentation network. The shape constraint is based on the shape prior learned from segmentation maps using auto-encoders~\citep{Girdhar_2016_ECCV}. \cite{Dong_2018_MICCAI_echo} utilized shape prior more explicitly by combining a neural network with a conventional atlas-based segmentation framework. Adversarial training was also applied to encourage the method to produce more anatomically plausible segmentation maps, which contributes to its superior segmentation performance comparing to a standard voxel-wise classification 3D segmentation network~\citep{Milletari_2016_3DV}.

\smallskip\subsubsection{Left-atrium segmentation}
\cite{Degel_2018_MICCAI} adopted the aforementioned anatomical constrain in 3D \gls{LA} segmentation to tackle the domain shift problem caused by variation of imaging device, protocol and patient condition. In addition to the anatomically constraining network, the authors applied an adversarial training scheme \citep{Kamnitsas_2017_IPMI} to improve the generalizability of the model to unseen domain.

\smallskip\subsubsection{Multi-chamber segmentation}
Apart from LV segmentation, a few works \citep{Zhang_2018_Circulation, Smistad_2017_IUS, Leclerc_2019_TMI} applied deep learning methods to perform multi-chamber (including \gls{LV} and \gls{LA}) segmentation. In particular, \cite{Zhang_2018_Circulation} demonstrated the applicability of CNNs on three tasks: view classification, multi-chamber segmentation and detection of cardiovascular diseases. Comprehensive validation on a large (non-public) clinical dataset showed that clinical metrics derived from automatic segmentation are comparable or superior than manual segmentation. To resemble real clinical situations and thus encourages the development and evaluation of robust and clinically effective segmentation methods, a large-scale dataset for 2D cardiac ultrasound has been recently made public\citep{Leclerc_2019_TMI}. The dataset and evaluation platform were released following the preliminary data requirement investigation of deep learning methods \citep{Leclerc_2018_IUS}. The dataset is composed of apical 4-chamber view images annotated for \gls{LV} and \gls{LA} segmentation, with uneven imaging quality from 500 patients with varying conditions. Notably, the initial benchmarking \citep{Leclerc_2019_TMI} on this dataset has shown that modern encoder-decoder CNNs resulted in lower error than inter-observer error between human cardiologists. 

\smallskip\subsubsection{Aortic valve segmentation}
\cite{Ghesu_2016_TMI} proposed a framework based on \gls{MSL}, \glspl{DNN} and \gls{ASM} to segment the aortic valve in 3D cardiac ultrasound volumes. An adaptive sparsely-connected neural network with reduced number of parameters is used to predict a bounding box to locate the target structure, where the learning of the bounding box parameters is marginalized into sub-spaces to reduce computational complexity. This framework showed significant improvement over the previous non-DL \gls{MSL} \citep{Zheng_2008_TMI} method while achieving competitive run-time.  \medskip\subsection{Discussion}
\label{sec: Dicussion}
So far, we have presented and discussed recent progress of deep learning-based segmentation methods in the three modalities (i.e. MR, CT, US) that are commonly used in the assessment of cardiovascular disease. To summarize, current state-of-the-art segmentation methods are mainly based on CNNs that employ the \gls{FCN} or U-net architecture. In addition, there are several commonalities in the FCN-based methods for cardiac segmentation which can be categorized into four groups: 1) enhancing network feature learning by employing advanced building blocks in networks (e.g. inception module, dilated convolutions), most of which have been mentioned earlier (Sec.~\ref{SEC: advanced blocks}); 2) alleviating the problem of class imbalance with advanced loss functions (e.g. weighted loss functions); 3) improving the networks' generalization ability and robustness through a multi-stage pipeline, multi-task learning, or multi-view feature fusion; 4) forcing the network to generate more anatomically-plausible segmentation results by incorporating shape priors, applying adversarial loss or anatomical constraints to regularize the network during training. It is also worthwhile to highlight that for cardiac image sequence segmentation (e.g. cine MR images, 2D US sequences), leveraging spatial and temporal coherence from these sequences with advanced neural networks (e.g. RNN~\citep{Bai_2018_MICCAI,Jafari_2018_DLMIA}, multi-slice FCN~\citep{Zheng_2018_TMI}) has been explored and shown to be beneficial for improving the segmentation accuracy and temporal consistency of the segmentation maps.

While the results reported in the literature show that neural networks have become more sophisticated and powerful, it is also clear that performance has improved with the increase of publicly available training subjects. A number of DL-based methods (especially in MRI) have been trained and tested on public challenge datasets, which not only provide large amounts of data to exploit the capabilities of deep learning in this domain, but also a platform for transparent evaluation and comparison. In addition, many of the participants in these challenges have shared their code with other researchers via open-source community websites (e.g. Github). Transparent and fair benchmarking and sharing of code are both essential for continued progress in this domain. We summarize the existing public datasets in Table~\ref{tab:public datasets} and public code repositories in Table~\ref{tab:public code} for reference.

\begin{table*}[t]
\centering
\caption{\textbf{Summary of public datasets on cardiac segmentation for the three modalities.} Mostly are from the \gls{MICCAI} society.}
\label{tab:public datasets}
\resizebox{\textwidth}{!}{
\begin{tabular}{cccccc}
\toprule
Dataset Name/Reference & Year & Main modalities & \# & Target(s) & Main Pathology \\
\midrule
York~\citep{Andreopoulos_2008_York} & 2008 & cine MRI  & 33 & LV & cardiomyopathy, aortic regurgitation, enlarged
ventricles and ischemia \\
Sunnybrook~\citep{Radau_2009_Sunnybrook}& 2009 & cine MRI  & 45 & LV & hypertrophy, heart failure w./w.o infarction \\
LVSC~\citep{Suinesiaputra_LVSC_2011} & 2011 & cine MRI & 200 & LV & coronary artery disease, myocardial infarction. \\
RVSC~\citep{Petitjean_2015_MedIA} & 2012 & cine MRI  & 48 & RV &
\begin{tabular}[c]{@{}c@{}} myocarditis, ischaemic cardiomyopathy, \\ suspicion of arrhythmogenic, right ventricular dysplasia, \\ dilated cardiomyopathy, hypertrophic cardiomyopathy, aortic stenosis \end{tabular} \\
cDEMRIS~\citep{Karim_2013_LA_SCAR} & 2012 & LGE MRI & 60 & LA fibrosis and scar & atrial fibrillation \\
LVIC~\citep{Karim_2016_LVIS_Dataset} & 2012 & LGE MRI & 30 & Myocardial scars & ischaemic cardiomyopathy \\
LASC'13~\citep{Tobon-Gomez_2015_LASC}& 2013 & 3D MRI & 30 & LA & N/A \\
HVSMR~\citep{Pace_2015_HVSMR} & 2016 & 3D MRI & 4 & Blood pool, MYO & congenital heart defects \\
ACDC~\citep{Bernard_2018_TMI} & 2017 & MRI & 150 & LV; RV & mycardial infarction, dilated/ hypertrophic
cardiomyopathy, abnormal RV \\
LASC'18~\citep{LASC_2018} & 2018 & LGE MRI & 150 & LA & atrial fibrillation \\
\midrule
MM-WHS~\citep{Zhuang_2019_MedIA} & 2017 & CT/MRI & 60/60 & WHS &
\begin{tabular}[c]{@{}c@{}c@{}} myocardium infarction, atrial fibrillation, 
tricuspid regurgitation, \\ aortic valve stenosis, 
Alagille syndrome, \\ Williams syndrome,
dilated cardiomyopathy, aortic coarctation, \\ Tetralogy of
Fallot \end{tabular}\\
CAT08~\citep{Schaap_2009_centerline_dataset} & 2008 & CTA & 32 & Coronary artery centerline  &
Patients with presence of calcium scored as absent, modest or severe.\\
CLS12~\citep{Kirisli2013_Lumen_Stenosis_dataset} & 2012 & CTA & 48 & Coronary lumen 
and stenosis  &
Patients with different levels of coronary artery stenoses.\\
\midrule
CETUS~\citep{CETUS_2016}& 2014 & 3D US & 45 & LV & myocardial infarction, dilated cardiomyopathy \\
CAMUS~\citep{Leclerc_2019_TMI} & 2019 & 2D US & 500 & LV, LA & Patients with \gls{EF}$<45\%$ \\
\bottomrule
\end{tabular}
}
\end{table*}
\begin{table*}[!t]
    \centering
    \caption{\textbf{Public code for DL-based cardiac image segmentation.} SAX: short-axis view; WHS: whole heart segmentation.}
    \label{tab:public code}
\resizebox{\textwidth}{!}{
\begin{tabular}{ccccc}
\toprule
Modality & Application(s) & Authors & Basic Network &  Code Repo (If not specified, the repository is located under \url{github.com}) \\
\midrule
MR (SAX) & Bi-ventricular Segmentation & \cite{Tran_2016_Arxiv} & 2D FCN & \url{vuptran/cardiac-segmentation} \\
MR (SAX) & Bi-ventricular Segmentation & \cite{Baumgartner_2017_STACOM} & 2D/3D U-net & \url{baumgach/acdc_segmenter} \\
MR (SAX) & Bi-ventricular Segmentation; 1st rank in ACDC challenge & \cite{Isensee_2017_STACOM} & 2D+3D U-net (ensemble) & \url{MIC-DKFZ/ACDC2017} \\
MR (SAX) & Bi-ventricular Segmentation & \cite{Zheng_2018_TMI} & cascaded 2D U-net & \url{julien-zheng/CardiacSegmentationPropagation} \\
MR (SAX) & Bi-ventricular segmentation and Motion Estimation & \cite{Qin_2018_MICCAI} & 2D FCN, RNN & \url{cq615} \\
MR (SAX) & Biventricular Segmentation & \cite{Khened_2019_MedIA} & 2D U-net & \url{mahendrakhened}
 \\
MR (3D scans) & Blood pool+MYO Segmentation & \cite{Yu_2017_MICCAI} & 3D CNN & \url{yulequan/HeartSeg} \\
MR (Multi-view) & Four-chamber Segmentation and Aorta Segmentation & \cite{Bai_2018_JCMR, Bai_2018_MICCAI} & 2D FCN, RNN & \url{baiwenjia/ukbb_cardiac} \\
MR & Cardiac Segmentation and Motion Tracking & \cite{Duan_2019_TMI} & 2.5D FCN +Atlas-based & \url{j-duan/4Dsegment} \\
\midrule
LGE MRI & Left Atrial Segmentation & \cite{Chen_2018_STACOM} & 2D U-net & \url{cherise215/atria_segmentation_2018} \\
LGE MRI & Left Atrial Segmentation & \cite{Yu_2019_MICCAI} & 3D V-net & \url{yulequan/UA-MT} \\
\midrule
CT & WHS & \cite{Yang_2017c_MMWHS} & 3D U-net & \url{xy0806/miccai17-mmwhs-hybrid} \\
CT & WHS & \cite{Xu_2018_arXiv} & Faster RCNN, 3D U-net & \url{Wuziyi616/CFUN} \\
CT, MRI & Coronary arteries & \cite{Merkow_2016_MICCAI} & 3D U-net & \url{jmerkow/I2I} \\
CT, MRI & WHS & \cite{Dou_2018_IJCAI,Dou_2019_Access} & 2D CNN & \url{carrenD/Medical-Cross-Modality-Domain-Adaptation} \\
CT, MRI & WHS& \cite{Chen_2019_AAAI} & 2D CNN & \url{cchen-cc/SIFA} \\
\midrule
US & View Classification and Four-chamber Segmentation & \cite{Zhang_2018_Circulation} & 2D U-net & \url{bitbucket.org/rahuldeo/echocv} \\
\bottomrule
\end{tabular}
}
\end{table*}

\textbf{2D Networks vs 3D Networks:}
An interesting conclusion supported by Table~\ref{tab:public code} is that the target image type can affect the choice of network structures (i.e. 2D networks, 3D networks). For 3D imaging acquisitions such as LGE-MRI and CT images, 3D networks are preferred whereas 2D networks are more popular approaches for segmenting cardiac cine short-axis or long-axis image stacks. One reason for using 2D networks for the segmentation of short-axis or long-axis images is their typically large slice thickness (usually around 7–8 mm) which can further exacerbated by inter-slice gaps. In addition, breath-hold related motion artifacts between different slices may negatively affect 3D networks. A study conducted by \cite{Baumgartner_2017_STACOM} has shown that a 3D U-net performs worse than a 2D U-net when evaluated on the ACDC challenge dataset. By contrast, in the LASC'18 challenge mentioned in Table~\ref{tab:public datasets}, which uses high-resolution 3D images, most participants applied 3D networks and the best performance was achieved by a cascaded network based on the 3D U-net~\citep{Xia_2018_STACOM}. 

It is well known that training 3D networks is more difficult than training 2D networks. In general, 3D networks have significantly more parameters than 2D networks. Therefore, 3D networks are more difficult and computationally expensive to optimize as well as prone to over-fitting, especially if the training data is limited. As a result, several researchers have tried to carefully design the structure of network to reduce the number of parameters for a particular application and have also applied advanced techniques (e.g. deep supervision) to alleviate the over-fitting problem~\citep{Yu_2017_MICCAI,Xia_2018_STACOM}. For this reason, 2D-based networks (e.g. 2D U-net) are still the most popular segmentation approaches for all three modalities. 

In addition to 2D and 3D networks, several authors have proposed `2D+' networks that have been shown to be effective in segmenting structures from cardiac volumetric data. These `2D+' networks are mainly based on 2D networks, but are adapted with increased capacity to utilize 3D context. These networks include multi-view networks which leverage multi-planar information (i.e. coronal, sagittal, axial views)~\citep{Mortazi_2017_STACOM,Wang_2017_MMWHS}, multi-slice networks, and 2D FCNs combined with RNNs which incorporate context across multiple slices ~\citep{Duan_2019_TMI,Patravali_2017_STACOM,Poudel_2016_HVSCMR,Du_2019_JTEHM}. These `2D+'networks inherit the advantages of 2D networks while still being capable of leveraging through-plane spatial context for more robust segmentation with strong 3D consistency.

 \section{Challenges and Future Work}
\label{SEC:Challenges}
It is evident from the literature that deep learning  methods have matched or surpassed the previous state of the art in a various cardiac segmentation applications, mainly benefiting from the increased size of public datasets and the emergence of advanced network architectures as well as powerful hardware for computing.  Given this rapid process, one may wonder if deep learning methods can be directly deployed to real-world applications to reduce the workload of clinicians. The current literature suggests that there is still a long way to go. In the following paragraphs, we summarize several major challenges in the field of cardiac segmentation and some recently proposed approaches that attempt to address them. These challenges and related works also provide potential research directions for future work in this field.

\medskip\subsection{Scarcity of Labels}
One of the biggest challenges for deep learning approaches is the scarcity of annotated data. In this review, we found that the majority of studies uses a fully supervised approach to train their networks, which requires a large number of annotated images. In fact, annotating cardiac images is time consuming and often requires significant amounts of expertise. While data augmentation techniques such as cropping, padding, and geometric transformations (e.g. affine transformations) can be used to increase the size of training samples, their diversity may still be limited, failing to reflect the spectrum of real-world data distributions. Several methods have been proposed to overcome this challenge. These methods can be categorized into four classes: transfer learning with fine-tuning, weakly and semi-supervised learning, self-supervised learning, and unsupervised learning. 
\begin{itemize}
    \item {\textbf{Transfer learning with fine-tuning.} Transfer learning aims at reusing a model pre-trained on one task as a starting point to train for a second task. The key of transfer learning is to learn features in the first task that are related to the second task such that the network can quickly converge even with limited data. Several researchers have successfully demonstrated the use of transfer learning to improve the model generalization ability for cardiac ventricle segmentation across different scanners, where they first trained a model on a large dataset and then fine-tuned it on a small dataset~\citep{Bai_2018_JCMR, Khened_2019_MedIA, Cong_2018_JE,Fahmy_2019_JCMR,chen_2019_med3d}.
    }
    \item {\textbf{Weakly and semi-supervised learning.}} Weakly and semi-supervised learning methods aim at improving the learning accuracy by making use of both labeled and unlabeled or weakly-labeled data (e.g annotations in forms of scribbles or bounding boxes). In this context, several works have been proposed for cardiac ventricle segmentation in MR images. One approach is to estimate full labels on unlabeled or weakly labeled images for further training. For example,~\cite{Bai_2018_MICCAI,Qin_2018_MICCAI} utilized motion information to propagate labels from labeled frames to unlabeled frames in a cardiac cycle whereas \cite{Bai_2017_MICCAI,Can_2018_DLMIAML} applied the \gls{EM} algorithm to predict and refine the estimated labels recursively. Others have explored different approaches to regularize the network when training on unlabeled images, applying multi-task learning~\citep{ChartsiasA_2018}, or global constraints~\citep{Kervadec_2018_MeDIA}.
    \item {\textbf{Self-supervised learning.}} Another approach is self-supervised learning which aims at utilizing labels that are \emph{generated automatically} without human intervention. These labels, designed to encode some properties or semantics of the object, can provide strong supervisory signals to pre-train a network before fine-tuning for a given task. A very recent work from \cite{Bai_2019_MICCAI} has shown the effectiveness of self-supervised learning for cardiac MR image segmentation where the authors used auto-generated anatomical position labels to pre-train a segmentation network. Compared to a network trained from scratch, networks pre-trained on the self-supervised task performed better,  especially when the training data was extremely limited.
    \item {\textbf{Unsupervised learning.}} Unsupervised learning aims at learning without paired labeled data. Compared to the former three classes, there is limited literature about unsupervised learning methods for cardiac image segmentation, perhaps because of the difficulty of the task. An early attempt has been made which applied adversarial training to train a network segmenting \gls{LV} and \gls{RV} from CT and MR images without requiring a training set of paired images and labels~\citep{Joyce_2018_MIDL}.
\end{itemize}
Apart from utilizing unlabeled images for training neural networks, another interesting direction is active learning~\citep{Mahapatra_2018_MICCAI}, which tries to select the most representative images from a large-scale dataset, reducing redundant labeling workload and training cost. This technique is also related to incremental learning, which aims to improve the model performance with new classes added incrementally while avoiding a dramatic decrease in overall performance~\citep{Castro_2018_ECCV}. Given the increasing size of the available medical datasets, and the practical challenges of labeling and storing large amounts of images from various sources, it is of great interest to develop algorithms capable of distilling a large-scale cardiac dataset into a small one containing the most representative cases for labeling and training.

\medskip\subsection{Model Generalization Across Various Imaging Modalities, Scanners and Pathologies.}
Another common limitation in  DL-based methods is that they still lack generalization capabilities when presented with previously unseen samples (e.g. data from a new scanner, abnormal and pathological cases that have not been included in the training set). In other words, deep learning models tend to be biased by their respective training datasets. This limitation prevents models to be deployed in the real world and therefore diminishes their impact for improving clinical workflows. To improve the model performance across MR images acquired from multiple vendors and multiple scanners,~\cite{Tao_2019_Radiology} collected a large multi-vendor, multi-center, heterogeneous labeled training set from patients with cardiovascular diseases. However, this approach may not scale to the real world, as it implies the collection of a vastly large dataset covering all possible cases. Moreover, it still faces the aforementioned collecting and labeling challenge. 

\textbf{Unsupervised domain adaptation.}
Several researchers have recently started to investigate the use of unsupervised domain adaptation techniques that aim at optimizing the model performance on unseen datasets without additional labeling costs. Several works have successfully applied adversarial training for cross-modality segmentation tasks, adapting a cardiac segmentation model learned from MR images to CT images and vice versa~\citep{Dou_2018_IJCAI,Dou_2019_Access, Ouyang_2019_MICCAI, Chen_2019_AAAI}. These type of approaches can also be adopted for semi-supervised learning, where the target domain is a new set of unlabeled data of the same modality~\citep{Chen_2019_MICCAI_UDA}.

\textbf{Data augmentation.}
An alternative yet simple and effective approach is data augmentation. The main idea is to increase the variety of training images so that the
training set distribution is more close to the one of a test set in the real world. In general, this type of augmentation is achieved by applying a stack of geometric or photometric transformations to existing image-label pairs. Recently,  \cite{Chen_2019_Arxiv} have proposed a data normalization and augmentation pipeline which enables a neural network for cardiac MR image segmentation trained from a single-scanner dataset to generalize well across multi-scanner and multi-site datasets.~\cite{Zhang_2019_Arxiv} applied a similar data augmentation approach to improve the model generalization ability on unseen datasets. Their method has been verified on three tasks including left atrial segmentation from 3D MRI and left ventricle segmentation from 3D ultrasound images. However, effectively designing such a pipeline requires expertise, which may not be easy to be extended to other applications. Most recently, several researchers have began to investigate the use of generative models (e.g. \glspl{GAN}, variational AE~\citep{Kingma_2013_ICLR}), reinforcement learning~\citep{Ekin_2019_CVPR} and adversarial example generation~\citep{Volpi_2018_NIPS} that aim at directly learning data augmentation strategies from existing data. In particular, the generative model-based approach has been proven to be effective for one-shot brain segmentation~\citep{Zhao_2019_CVPR} and few-shot cardiac MR image segmentation~\citep{Chaitanya_2019_IPMI} and is thus worth to be explored for more applications in the future.

\medskip\subsection{Lack of Model Interpretability}
Unlike symbolic artificial intelligence systems, deep learning systems are difficult to interpret and not transparent. Once a network has been trained, it behaves like a `black box', providing predictions which are not directly interpretable. This issue makes the model unpredictable, intractable for model verification, and ultimately untrustworthy. Recent studies have shown that deep learning-based vision recognition systems can be attacked by images modified with nearly imperceptible perturbations~\citep{Szegedy_2013_Arxiv,Kurakin_2016_Arxiv,Goodfellow_2015_ICLR}. These attacks can also happen in medical scenarios, e.g. a \gls{DL}-based system may make a wrong diagnosis given an image with adversarial noise or even just small rotation, as demonstrated in a very recent paper \citep{Finlayson_2019_Science}. Although there is no denying that deep learning has become a very powerful tool for image analysis, building resilient algorithms robust to potential attacks remains an unsolved problem. One potential solution, instead of building the resilience into the model, is raising failure awareness of the deployed networks. This can be achieved by providing users with segmentation quality scores~\citep{Robinson_2019_JCMR} or confidence maps such as uncertainty maps~\citep{Sander_2019_MIP} and attention maps~\citep{Heo_2018_NIPS}. These scores or maps can be used as evidence to alert users when failure happens. For example, \cite{Sander_2019_MIP} built a network that is able to simultaneously predict the segmentation mask over cardiac structures and its associated spatial uncertainty map, where the latter one could be used to highlight potential incorrect regions. Such uncertainty information could alert human experts for further justification and refinement in a human-in-the-loop setting.       
\medskip\subsection{Future work}
\textbf{Smart imaging.} We have shown that deep learning-based methods are able to segment images in real-time with good accuracy. However, these algorithms can still fail on those image acquisitions with low image quality or significant artifacts. Although there have been several algorithms developed to avoid this problem by either checking the image quality before follow-up studies~\citep{Ruijsink_2019_JACC,Tarroni_2019_TMI}, or predicting the segmentation quality to detect failures~\citep{Peng_2012_ECCV,Robinson_2019_JCMR,Zhou_2019_Arxiv}, the development of algorithms that can give instant feedback to correct and optimize the image acquisition process is also important despite less explored. Improving the imaging quality can greatly improve the effectiveness of medical imaging as well as the accuracy of imaging-based diagnosis. For radiologists, however, finding the optimal imaging and reconstruction parameters to scan each patient can take a great amount of time. Therefore, a \gls{DL}-based system that has the potential of efficiently and effectively improving the image quality with less noise is of great need. Some researchers have utilized learning-based methods (mostly are deep learning-based) for better image resolution~\citep{Oktay_2016_MICCAI}, view planning~\citep{Alansary_2018_MICCAI}, motion correction~\citep{Dangi_2018_SPIE,Tarroni_2018_MICCAI}, artifacts reduction~\citep{Ilkay_2019_MIDL}, shadow detection~\citep{Meng_2019_TMI} and noise reduction~\citep{Wolterink_2017_TMI} after image acquisition. However, combining these algorithms with segmentation algorithms and seamlessly integrating them into an efficient, patient-specific imaging system for high-quality image analysis and diagnosis is still an open challenge. An alternative approach is to directly predict cardiac segmentation maps from undersampled k-space data to accelerate the whole procedure, which bypasses the image reconstruction stage~\citep{Schlemper_2018_MICCAI}.

\textbf{Data harmonization.}
A number of works have reported the existence of missing labels and inconsistent labeling protocols among different cardiac image datasets~\citep{Zheng_2018_TMI, Chen_2019_Arxiv}. Variations have been found in defining the end of basal slices as well as the endocardial wall of myocardium (some include papillary muscles as part of the endocardial contours whereas others do not). These inconsistencies can be a major obstacle for transferring, evaluating and deploying deep learning models trained from one domain (e.g. hospital) to another. Therefore, building a standard benchmark dataset like CheXpert~\citep{Irvin_2019_AAAI} that 1) is \emph{large} enough to have substantial data diversity that reflects the spectrum of real-world diversity; 2) has a \emph{standard} labeling protocol approved by experts, is indeed a need. However, directly building such a dataset from scratch is time-consuming and expensive. A more promising way might be developing an automated tool to combine existing datasets from multiple sources and then to harmonize them to a unified, high-quality dataset. This tool can not only open the door for crowd-sourcing but also enable the rapid deployment of those \gls{DL}-based segmentation models.

\textbf{Data privacy.}
As deep learning is a data-driven approach, an unavoidable and rife concern is about the data privacy. Regulations such as \gls{GDPR} now play an important role to protect users' privacy and have forced organizations to treat data ownership seriously. On the other hand, from a technical point of view, how to store, query, and process data such that there is no privacy concerns for building deep learning systems has now become an even more difficult but interesting challenge. Building a privacy-preserving algorithm requires to combine cryptography and deep learning together and to mix techniques from a wide range of subjects such as data analysis, distributed computing, federated learning, differential privacy, in order to achieve models with strong security, fast run time, and great generalizability~\citep{Dwork_2014_FTTCS,Abadi_2016_CCS,Bonawitz_2017_CCS,Ryffel_2018_PPML}. In this respect,~\cite{Papernot_2018_Arxiv} published a report for guidance, which summarized a set of best practices for improving the privacy and security of machine learning systems. Yet, this field is still in its infancy. \section{Conclusion}
\label{SEC:Conclusion}
In this review paper, we provided a comprehensive overview of these deep learning techniques used in three common imaging modalities (MRI, CT, US), covering a wide range of existing deep learning approaches (mostly are CNN-based) that are designed for segmenting different cardiac anatomical structures (e.g. cardiac ventricle, atria, vessel). In particular, we presented and discussed recent progress of deep learning-based segmentation methods in the three modalities, outlined future potential and the remaining limitations of these deep learning-based cardiac segmentation methods that may hinder widespread clinical deployment. We hope that this review can provide an intuitive understanding of those deep learning-based techniques that have made a significant contribution to cardiac image segmentation and also increase the awareness of common challenges in this field that call for future contribution. 

\section*{Abbreviations}
\textbf{Imaging-related terminology:} 
CT: computed tomography;
CTA: computed tomography angiography;
LAX: long-axis;
MPR: multi-planar reformatted;
MR: magnetic resonance;
MRI: magnetic resonance imaging;
LGE: late gadolinium enhancement;
RFCA: radio-frequency catheter ablation;
SAX: short-axis;
US: ultrasound;
2CH: 2-chamber;
3CH: 3-chamber;
4CH: 4-chamber.
\\
\textbf{Cardiac structures and indexes:} 
AF: atrial fibrillation;
AS: aortic stenosis;
AO: aorta;
CVD: cardiovascular diseases;
CAC: coronary artery calcium;  
DCM: dilated cardiomyopathy; 
ED: end-diastole;
ES: end-systole;
EF: ejection fraction;
HCM: hypertrophic cardiomyopathy; 
LA: left atrium;
LV: left ventricle;
LVEDV: left ventricular end-diastolic volume; 
LVESV: left ventricular end-systolic volume;
MCP: mixed-calcified plaque;
MI: myocardial infarction; 
MYO: myocardium;
NCP: non-calcified plaque;
PA: pulmonary artery;
PV: pulmonary vein;
RA: right atrium;
RV: right ventricle;
RVEDV: right ventricular end-diastolic volume;
RVESV: right ventricular end-systolic volume;
RVEF: right ventricular ejection fraction;
WHS: whole heart segmentation.
\\
\textbf{Machine learning terminology:}
AE: autoencoder;
ASM: active shape model;
BN: batch normalization;
CNN: convolutional neural network;
CRF: conditional random field;
DBN: deep belief network;
DL: deep learning;
DNN: deep neural network;
EM: expectation maximization;
FCN: fully convolutional neural network;
GAN: generative adversarial network;
GRU: gated recurrent units;
MSE: mean squared error;
MSL: marginal space learning;
MRF: markov random field;
LSTM: Long-short term memory;
ReLU: rectified linear unit;
RNN: recurrent neural network;
ROI: region-of-interest;
SMC: sequential monte carlo;
SRF: structured random forest;
SVM: support vector machine.
\\
\textbf{Cardiac image segmentation datasets:} 
ACDC: Automated Cardiac Diagnosis Challenge; 
CETUS: Challenge on Endocardial Three-dimensional Ultrasound Segmentation;
MM-WHS: Multi-Modality Whole Heart Segmentation;
LASC: Left Atrium Segmentation Challenge;
LVSC: Left Ventricle Segmentation Challenge;
RVSC: Right Ventricle Segmentation Challenge.
\\
\textbf{Others:}
EMBC: The International Engineering in Medicine and Biology Conference;
GDPR: The General Data Protection Regulation;
GPU: graphic processing unit;
FDA: United States Food and Drug Administration;
ISBI: The IEEE International Symposium on Biomedical Imaging;
MICCAI: International Conference on Medical Image Computing and Computer-assisted Intervention;
TPU: tensor processing unit;
WHO: World Health Organization.

\section*{Conflict of Interest Statement}
The authors declare that the research was conducted in the absence of any commercial or financial relationships that could be construed as a potential conflict of interest.

\section*{Author Contributions}
CC, WB, DR conceived and designed the work; CC, CQ, HQ searched and read the MR, CT, US literature, respectively; CC, CQ, HQ drafted the manuscript together; WB, DR, GT, JD provided critical revision with insightful and constructive comments to improve the manuscript; All authors read and approved the manuscript.

\section*{Funding}
This work is supported by the SmartHeart EPSRC Programme Grant
(EP/P001009/1). Huaqi Qiu is supported by the EPSRC Programme Grant (EP/R005982/1).
\onecolumn{
\section*{Data Availability Statement}
The datasets summarized in Table~\ref{tab:public datasets} can be found in their corresponding websites listed below:
\begin{enumerate}
	\item York: \tiny{\url{http://www.cse.yorku.ca/~mridataset/}}
	\item Sunnybrook:\tiny\url{http://www.cardiacatlas.org/studies/sunnybrook-cardiac-data/}
	\item LVSC:\tiny\url{http://www.cardiacatlas.org/challenges/lv-segmentation-challenge/}
	\item RVSC:\tiny\url{http://www.litislab.fr/?projet=1rvsc}
	\item cDEMRIS:\tiny\url{https://www.doc.ic.ac.uk/~rkarim/la_lv_framework/fibrosis}
	\item LVIC:\tiny\url{https://www.doc.ic.ac.uk/~rkarim/la\_lv\_framework/lv\_infarct}
	\item LASC'13:\tiny\url{www.cardiacatlas.org/challenges/left-atrium-segmentation-challenge/}
	\item HVSMR:\tiny\url{http://segchd.csail.mit.edu/}
	\item ACDC:\tiny\url{https://acdc.creatis.insa-lyon.fr/}
	\item LASC'18:\tiny\url{http://atriaseg2018.cardiacatlas.org/data/}
	\item MM-WHS:\tiny\url{http://www.sdspeople.fudan.edu.cn/zhuangxiahai/0/mmwhs17/}
	\item CAT08:\tiny\url{http://coronary.bigr.nl/centerlines/}
	\item CLS12:\tiny\url{http://coronary.bigr.nl/stenoses}
	\item CETUS:\tiny\url{https://www.creatis.insa-lyon.fr/Challenge/CETUS}
	\item CAMUS:\tiny\url{https://www.creatis.insa-lyon.fr/Challenge/camus}
\end{enumerate}
}

\section*{Acknowledgments}
We would like to thank our colleagues: Karl Hahn, Qingjie Meng, James Batten, and Jonathan Passerat-Palmbach who provided insight and expertise that greatly assisted the work, and also constructive and thoughtful comments from Turkay Kart that greatly improved the manuscript.

\bibliographystyle{frontiersinSCNS_ENG_HUMS}

\bibliography{test}

\begin{thebibliography}{249}
\providecommand{\natexlab}[1]{#1}
\expandafter\ifx\csname urlstyle\endcsname\relax
  \providecommand{\doi}[1]{doi:\discretionary{}{}{}#1}\else
  \providecommand{\doi}{doi:\discretionary{}{}{}\begingroup
  \urlstyle{rm}\Url}\fi
\providecommand{\selectlanguage}[1]{\relax}
\providecommand{\bibAnnoteFile}[1]{%
  \IfFileExists{#1}{\begin{quotation}\noindent\textsc{Key:} #1\\
  \textsc{Annotation:}\ \input{#1}\end{quotation}}{}}
\providecommand{\bibAnnote}[2]{%
  \begin{quotation}\noindent\textsc{Key:} #1\\
  \textsc{Annotation:}\ #2\end{quotation}}

\bibitem[{Abadi et~al.(2016)Abadi, Chu, Goodfellow, McMahan, Mironov, Talwar
  et~al.}]{Abadi_2016_CCS}
Abadi, M., Chu, A., Goodfellow, I.~J., McMahan, H.~B., Mironov, I., Talwar, K.,
  et~al. (2016).
\newblock Deep learning with differential privacy.
\newblock In \emph{the 2016 {ACM} {SIGSAC} Conference on Computer and
  Communications Security, Vienna, Austria, October 24-28, 2016}. 308--318
\bibAnnoteFile{Abadi_2016_CCS}

\bibitem[{Agatston et~al.(1990)Agatston, Janowitz, Hildner, Zusmer, Viamonte,
  and Detrano}]{Agatston_1990_JACC}
Agatston, A.~S., Janowitz, W.~R., Hildner, F.~J., Zusmer, N.~R., Viamonte, M.,
  and Detrano, R. (1990).
\newblock Quantification of coronary artery calcium using ultrafast computed
  tomography.
\newblock \emph{Journal of the American College of Cardiology} 15, 827--832
\bibAnnoteFile{Agatston_1990_JACC}

\bibitem[{Alansary et~al.(2018)Alansary, Folgoc, Vaillant, Oktay, Li, Bai
  et~al.}]{Alansary_2018_MICCAI}
Alansary, A., Folgoc, L.~L., Vaillant, G., Oktay, O., Li, Y., Bai, W., et~al.
  (2018).
\newblock Automatic view planning with multi-scale deep reinforcement learning
  agents.
\newblock In \emph{Medical Image Computing and Computer Assisted Intervention}.
  277--285.
\newblock \doi{10.1007/978-3-030-00928-1\_32}
\bibAnnoteFile{Alansary_2018_MICCAI}

\bibitem[{Andreopoulos and Tsotsos(2008)}]{Andreopoulos_2008_York}
Andreopoulos, A. and Tsotsos, J.~K. (2008).
\newblock Efficient and generalizable statistical models of shape and
  appearance for analysis of cardiac {MRI}.
\newblock \emph{Medical image analysis} 12, 335--357.
\newblock \doi{10.1016/j.media.2007.12.003}
\bibAnnoteFile{Andreopoulos_2008_York}

\bibitem[{Avendi et~al.(2016)Avendi, Kheradvar, and
  Jafarkhani}]{Avendi_2016_MedIA}
Avendi, M.~R., Kheradvar, A., and Jafarkhani, H. (2016).
\newblock A combined deep-learning and deformable-model approach to fully
  automatic segmentation of the left ventricle in cardiac mri.
\newblock \emph{Medical Image Analysis} 30, 108--119
\bibAnnoteFile{Avendi_2016_MedIA}

\bibitem[{Avendi et~al.(2017)Avendi, Kheradvar, and
  Jafarkhani}]{Avendi_2017_MRM}
Avendi, M.~R., Kheradvar, A., and Jafarkhani, H. (2017).
\newblock Automatic segmentation of the right ventricle from cardiac {MRI}
  using a learning-based approach.
\newblock \emph{Magnetic resonance in medicine: official journal of the Society
  of Magnetic Resonance in Medicine / Society of Magnetic Resonance in
  Medicine} 78, 2439--2448.
\newblock \doi{10.1002/mrm.26631}
\bibAnnoteFile{Avendi_2017_MRM}

\bibitem[{Bai et~al.(2019)Bai, Chen, Tarroni, Duan, Guitton, Petersen
  et~al.}]{Bai_2019_MICCAI}
Bai, W., Chen, C., Tarroni, G., Duan, J., Guitton, F., Petersen, S.~E., et~al.
  (2019).
\newblock {Self-Supervised} learning for cardiac {MR} image segmentation by
  anatomical position prediction.
\newblock In \emph{Medical Image Computing and Computer Assisted Intervention}.
  541--549.
\newblock \doi{10.1007/978-3-030-32245-8\_60}
\bibAnnoteFile{Bai_2019_MICCAI}

\bibitem[{Bai et~al.(2017)Bai, Oktay, Sinclair, Suzuki, Rajchl, Tarroni
  et~al.}]{Bai_2017_MICCAI}
Bai, W., Oktay, O., Sinclair, M., Suzuki, H., Rajchl, M., Tarroni, G., et~al.
  (2017).
\newblock Semi-supervised learning for {Network-Based} cardiac {MR} image
  segmentation.
\newblock In \emph{Medical Image Computing and Computer Assisted Intervention},
  eds. M.~Descoteaux, L.~Maier-Hein, A.~Franz, P.~Jannin, D.~L. Collins, and
  S.~Duchesne (Springer International Publishing), vol. 10434, 253--260
\bibAnnoteFile{Bai_2017_MICCAI}

\bibitem[{Bai et~al.(2018{\natexlab{a}})Bai, Sinclair, Tarroni, Oktay, Rajchl,
  Vaillant et~al.}]{Bai_2018_JCMR}
Bai, W., Sinclair, M., Tarroni, G., Oktay, O., Rajchl, M., Vaillant, G., et~al.
  (2018{\natexlab{a}}).
\newblock Automated cardiovascular magnetic resonance image analysis with fully
  convolutional networks.
\newblock \emph{Journal of Cardiovascular Magnetic Resonance} 20, 65
\bibAnnoteFile{Bai_2018_JCMR}

\bibitem[{Bai et~al.(2018{\natexlab{b}})Bai, Suzuki, Qin, Tarroni, Oktay,
  Matthews et~al.}]{Bai_2018_MICCAI}
Bai, W., Suzuki, H., Qin, C., Tarroni, G., Oktay, O., Matthews, P.~M., et~al.
  (2018{\natexlab{b}}).
\newblock Recurrent neural networks for aortic image sequence segmentation with
  sparse annotations.
\newblock In \emph{Medical Image Computing and Computer Assisted Intervention}.
  586--594.
\newblock \doi{10.1007/978-3-030-00937-3\_67}
\bibAnnoteFile{Bai_2018_MICCAI}

\bibitem[{Baumgartner et~al.(2017)Baumgartner, Koch, Pollefeys, and
  Konukoglu}]{Baumgartner_2017_STACOM}
Baumgartner, C.~F., Koch, L.~M., Pollefeys, M., and Konukoglu, E. (2017).
\newblock An exploration of {2D} and {3D} deep learning techniques for cardiac
  {MR} image segmentation.
\newblock In \emph{International Workshop on Statistical Atlases and
  Computational Models of the Heart}. 1--8
\bibAnnoteFile{Baumgartner_2017_STACOM}

\bibitem[{Bello et~al.(2019)Bello, Dawes, Duan, Biffi, de~Marvao, Howard
  et~al.}]{Bello_2019_NMI}
Bello, G.~A., Dawes, T. J.~W., Duan, J., Biffi, C., de~Marvao, A., Howard, L.
  S. G.~E., et~al. (2019).
\newblock Deep learning cardiac motion analysis for human survival prediction.
\newblock \emph{Nature machine intelligence} 1, 95--104.
\newblock \doi{10.1038/s42256-019-0019-2}
\bibAnnoteFile{Bello_2019_NMI}

\bibitem[{Bernard et~al.(2016)Bernard, Bosch, Heyde, Alessandrini, Barbosa,
  Camarasu-Pop et~al.}]{CETUS_2016}
Bernard, O., Bosch, J.~G., Heyde, B., Alessandrini, M., Barbosa, D.,
  Camarasu-Pop, S., et~al. (2016).
\newblock Standardized evaluation system for left ventricular segmentation
  algorithms in {3D} echocardiography.
\newblock \emph{IEEE transactions on medical imaging} 35, 967--977.
\newblock \doi{10.1109/TMI.2015.2503890}
\bibAnnoteFile{CETUS_2016}

\bibitem[{Bernard et~al.(2018)Bernard, Lalande, Zotti, Cervenansky, Yang, Heng
  et~al.}]{Bernard_2018_TMI}
Bernard, O., Lalande, A., Zotti, C., Cervenansky, F., Yang, X., Heng, P.-A.,
  et~al. (2018).
\newblock Deep learning techniques for automatic {MRI} cardiac
  {Multi-Structures} segmentation and diagnosis: Is the problem solved?
\newblock \emph{IEEE Transactions on Medical Imaging} 37, 2514--2525.
\newblock \doi{10.1109/TMI.2018.2837502}.
\newblock \url{https://acdc.creatis.insa-lyon.fr/}
\bibAnnoteFile{Bernard_2018_TMI}

\bibitem[{Bian et~al.(2018)Bian, Yang, Ma, Zheng, Liu, Nezafat
  et~al.}]{Bian_2018_STACOM}
Bian, C., Yang, X., Ma, J., Zheng, S., Liu, Y.-A., Nezafat, R., et~al. (2018).
\newblock Pyramid network with online hard example mining for accurate left
  atrium segmentation.
\newblock In \emph{Statistical Atlases and Computational Models of the Heart.
  Atrial Segmentation and {LV} Quantification Challenges 2018} (Springer
  International Publishing), 237--245.
\newblock \doi{10.1007/978-3-030-12029-0\_26}
\bibAnnoteFile{Bian_2018_STACOM}

\bibitem[{Biffi et~al.(2018)Biffi, Oktay, Tarroni, Bai, De~Marvao, Doumou
  et~al.}]{Biffi_2018_MICCAI}
Biffi, C., Oktay, O., Tarroni, G., Bai, W., De~Marvao, A., Doumou, G., et~al.
  (2018).
\newblock Learning interpretable anatomical features through deep generative
  models: Application to cardiac remodeling.
\newblock In \emph{Medical Image Computing and Computer Assisted Intervention}.
  vol. 11071 LNCS, 464--471.
\newblock \doi{10.1007/978-3-030-00934-2\_52}
\bibAnnoteFile{Biffi_2018_MICCAI}

\bibitem[{Bonawitz et~al.(2017)Bonawitz, Ivanov, Kreuter, Marcedone, McMahan,
  Patel et~al.}]{Bonawitz_2017_CCS}
Bonawitz, K., Ivanov, V., Kreuter, B., Marcedone, A., McMahan, H.~B., Patel,
  S., et~al. (2017).
\newblock Practical secure aggregation for privacy-preserving machine learning.
\newblock In \emph{the 2017 {ACM} {SIGSAC} Conference on Computer and
  Communications Security, {CCS} 2017, Dallas, TX, USA, October 30 - November
  03, 2017}. 1175--1191.
\newblock \doi{10.1145/3133956.3133982}
\bibAnnoteFile{Bonawitz_2017_CCS}

\bibitem[{Can et~al.(2018)Can, Chaitanya, Mustafa, Koch, Konukoglu, and
  Baumgartner}]{Can_2018_DLMIAML}
Can, Y.~B., Chaitanya, K., Mustafa, B., Koch, L.~M., Konukoglu, E., and
  Baumgartner, C.~F. (2018).
\newblock Learning to segment medical images with {Scribble-Supervision} alone.
\newblock In \emph{Deep Learning in Medical Image Analysis and Multimodal
  Learning for Clinical Decision Support} (Springer International Publishing),
  236--244
\bibAnnoteFile{Can_2018_DLMIAML}

\bibitem[{Cano-Espinosa et~al.(2018)Cano-Espinosa, Gonz{\'a}lez, Washko,
  Cazorla, and Est{\'e}par}]{Cano_2018_MedicalImaging}
Cano-Espinosa, C., Gonz{\'a}lez, G., Washko, G.~R., Cazorla, M., and
  Est{\'e}par, R. S.~J. (2018).
\newblock Automated {Agatston} score computation in {non-ECG} gated {CT} scans
  using deep learning.
\newblock In \emph{Medical Imaging 2018: Image Processing} (International
  Society for Optics and Photonics), vol. 10574, 105742K
\bibAnnoteFile{Cano_2018_MedicalImaging}

\bibitem[{Carminati et~al.(2016)Carminati, Boniotti, Fusini, Andreini, Pontone,
  Pepi et~al.}]{Carminati_2016_JTI}
Carminati, M.~C., Boniotti, C., Fusini, L., Andreini, D., Pontone, G., Pepi,
  M., et~al. (2016).
\newblock Comparison of image processing techniques for nonviable tissue
  quantification in late gadolinium enhancement cardiac magnetic resonance
  images.
\newblock \emph{Journal of thoracic imaging} 31, 168--176.
\newblock \doi{10.1097/RTI.0000000000000206}
\bibAnnoteFile{Carminati_2016_JTI}

\bibitem[{Carneiro et~al.(2010)Carneiro, Nascimento, and
  Freitas}]{Carneiro_2010_ISBI}
Carneiro, G., Nascimento, J., and Freitas, A. (2010).
\newblock Robust left ventricle segmentation from ultrasound data using deep
  neural networks and efficient search methods.
\newblock In \emph{2010 {IEEE} International Symposium on Biomedical Imaging:
  From Nano to Macro}. 1085--1088
\bibAnnoteFile{Carneiro_2010_ISBI}

\bibitem[{Carneiro and Nascimento(2010)}]{Carneiro_2010_CVPR}
Carneiro, G. and Nascimento, J.~C. (2010).
\newblock Multiple dynamic models for tracking the left ventricle of the heart
  from ultrasound data using particle filters and deep learning architectures.
\newblock In \emph{Conference on Computer Vision and Pattern Recognition}
  (IEEE), 2815--2822
\bibAnnoteFile{Carneiro_2010_CVPR}

\bibitem[{Carneiro and Nascimento(2011)}]{Carneiro_2011_ICCV}
Carneiro, G. and Nascimento, J.~C. (2011).
\newblock Incremental on-line semi-supervised learning for segmenting the left
  ventricle of the heart from ultrasound data.
\newblock In \emph{2011 International Conference on Computer Vision} (IEEE),
  1700--1707
\bibAnnoteFile{Carneiro_2011_ICCV}

\bibitem[{Carneiro and Nascimento(2012)}]{Carneiro_2012_CVPR}
Carneiro, G. and Nascimento, J.~C. (2012).
\newblock The use of on-line co-training to reduce the training set size in
  pattern recognition methods: Application to left ventricle segmentation in
  ultrasound.
\newblock In \emph{Conference on Computer Vision and Pattern Recognition}
  (IEEE), 948--955
\bibAnnoteFile{Carneiro_2012_CVPR}

\bibitem[{Carneiro and Nascimento(2013)}]{Carneiro_2013_TPAMI}
Carneiro, G. and Nascimento, J.~C. (2013).
\newblock Combining multiple dynamic models and deep learning architectures for
  tracking the left ventricle endocardium in ultrasound data.
\newblock \emph{IEEE transactions on pattern analysis and machine intelligence}
  35, 2592--2607
\bibAnnoteFile{Carneiro_2013_TPAMI}

\bibitem[{Carneiro et~al.(2012)Carneiro, Nascimento, and
  Freitas}]{Carneiro_2012_TIP}
Carneiro, G., Nascimento, J.~C., and Freitas, A. (2012).
\newblock The segmentation of the left ventricle of the heart from ultrasound
  data using deep learning architectures and derivative-based search methods.
\newblock \emph{IEEE Transactions on Image Processing} 21, 968--982.
\newblock \doi{10.1109/TIP.2011.2169273}
\bibAnnoteFile{Carneiro_2012_TIP}

\bibitem[{Castro et~al.(2018)Castro, Mar{\'\i}n-Jim{\'e}nez, Guil, Schmid, and
  Alahari}]{Castro_2018_ECCV}
Castro, F.~M., Mar{\'\i}n-Jim{\'e}nez, M.~J., Guil, N., Schmid, C., and
  Alahari, K. (2018).
\newblock End-to-end incremental learning.
\newblock In \emph{European Conference on Computer Vision}. 241--257.
\newblock \doi{10.1007/978-3-030-01258-8\_15}
\bibAnnoteFile{Castro_2018_ECCV}

\bibitem[{Chaitanya et~al.(2019)Chaitanya, Karani, Baumgartner, Donati, Becker,
  and Konukoglu}]{Chaitanya_2019_IPMI}
Chaitanya, K., Karani, N., Baumgartner, C., Donati, O., Becker, A., and
  Konukoglu, E. (2019).
\newblock {Semi-Supervised} and {Task-Driven} data augmentation.
\newblock In \emph{International Conference on Information Processing in
  Medical Imaging}. 29--41.
\newblock \doi{10.1007/978-3-030-20351-1\_3}
\bibAnnoteFile{Chaitanya_2019_IPMI}

\bibitem[{Chartsias et~al.(2018)Chartsias, Joyce, Papanastasiou, Semple,
  Williams, Newby et~al.}]{ChartsiasA_2018}
Chartsias, A., Joyce, T., Papanastasiou, G., Semple, S., Williams, M., Newby,
  D., et~al. (2018).
\newblock Factorised spatial representation learning: Application in
  semi-supervised myocardial segmentation.
\newblock In \emph{Medical Image Computing and Computer Assisted Intervention}.
  vol. 11071 LNCS, 490--498
\bibAnnoteFile{ChartsiasA_2018}

\bibitem[{Chen et~al.(2019{\natexlab{a}})Chen, Bai, Davies, Bhuva, Manisty,
  Moon et~al.}]{Chen_2019_Arxiv}
Chen, C., Bai, W., Davies, R.~H., Bhuva, A.~N., Manisty, C., Moon, J.~C.,
  et~al. (2019{\natexlab{a}}).
\newblock Improving the generalizability of convolutional neural network-based
  segmentation on {CMR} images.
\newblock \emph{Arxiv Preprint} abs/1907.01268.
\newblock Available at \url{http://arxiv.org/abs/1907.01268} (Accessed
  September 1, 2019)
\bibAnnoteFile{Chen_2019_Arxiv}

\bibitem[{Chen et~al.(2018{\natexlab{a}})Chen, Bai, and
  Rueckert}]{Chen_2018_STACOM}
Chen, C., Bai, W., and Rueckert, D. (2018{\natexlab{a}}).
\newblock Multi-task learning for left atrial segmentation on {GE-MRI}.
\newblock In \emph{9th International Workshop, {STACOM} 2018, Held in
  Conjunction with {MICCAI} 2018, Granada, Spain, September 16, 2018,}
  (Springer International Publishing), 292--301.
\newblock \doi{10.1007/978-3-030-12029-0\_32}
\bibAnnoteFile{Chen_2018_STACOM}

\bibitem[{Chen et~al.(2019{\natexlab{b}})Chen, Biffi, Tarroni, Petersen, Bai,
  and Rueckert}]{Chen_2019_MICCAI}
Chen, C., Biffi, C., Tarroni, G., Petersen, S., Bai, W., and Rueckert, D.
  (2019{\natexlab{b}}).
\newblock Learning shape priors for robust cardiac {MR} segmentation from
  multi-view images.
\newblock In \emph{Medical Image Computing and Computer Assisted Intervention}.
  523--531
\bibAnnoteFile{Chen_2019_MICCAI}

\bibitem[{Chen et~al.(2019{\natexlab{c}})Chen, Dou, Zhou, Qin, and
  Heng}]{Chen_2019_AAAI}
Chen, C., Dou, Q., Zhou, J., Qin, J., and Heng, P.~A. (2019{\natexlab{c}}).
\newblock Synergistic image and feature adaptation: Towards {Cross-Modality}
  domain adaptation for medical image segmentation.
\newblock \emph{Conference on Artificial Intelligence} ,
  865--872\doi{10.1609/aaai.v33i01.3301865}
\bibAnnoteFile{Chen_2019_AAAI}

\bibitem[{Chen et~al.(2016)Chen, Zheng, Park, Heng, and
  Zhou}]{Chen_2016_MICCAI}
Chen, H., Zheng, Y., Park, J.-H., Heng, P.-A., and Zhou, S.~K. (2016).
\newblock Iterative multi-domain regularized deep learning for anatomical
  structure detection and segmentation from ultrasound images.
\newblock In \emph{Medical Image Computing and {Computer-Assisted} Intervention
  -- {MICCAI} 2016} (Springer International Publishing), 487--495
\bibAnnoteFile{Chen_2016_MICCAI}

\bibitem[{Chen et~al.(2018{\natexlab{b}})Chen, Yang, Gao, Ni, Firmin, and
  {others}}]{Chen_2018_MICCAI}
Chen, J., Yang, G., Gao, Z., Ni, H., Firmin, D., and {others}
  (2018{\natexlab{b}}).
\newblock Multiview two-task recursive attention model for left atrium and
  atrial scars segmentation.
\newblock In \emph{Medical Image Computing and Computer Assisted Intervention}.
  455--463.
\newblock \doi{10.1007/978-3-030-00934-2\_51}
\bibAnnoteFile{Chen_2018_MICCAI}

\bibitem[{Chen et~al.(2019{\natexlab{d}})Chen, Zhang, Zhang, Zhao, Mohiaddin,
  Wong et~al.}]{Chen_2019_MICCAI_UDA}
Chen, J., Zhang, H., Zhang, Y., Zhao, S., Mohiaddin, R., Wong, T., et~al.
  (2019{\natexlab{d}}).
\newblock Discriminative consistent domain generation for semi-supervised
  learning.
\newblock In \emph{Medical Image Computing and Computer Assisted Intervention}.
  595--604.
\newblock \doi{10.1007/978-3-030-32245-8\_66}
\bibAnnoteFile{Chen_2019_MICCAI_UDA}

\bibitem[{Chen et~al.(2017)Chen, Papandreou, Schroff, and
  Adam}]{Chen_2017_Arxiv}
Chen, L.-C., Papandreou, G., Schroff, F., and Adam, H. (2017).
\newblock Rethinking atrous convolution for semantic image segmentation.
\newblock \emph{Arxiv Preprint} abs/1706.05587.
\newblock Available at \url{http://arxiv.org/abs/1706.05587}(Accessed September
  1, 2019)
\bibAnnoteFile{Chen_2017_Arxiv}

\bibitem[{Chen et~al.(2019{\natexlab{e}})Chen, Fang, and Liu}]{Chen_2019_ISBI}
Chen, M., Fang, L., and Liu, H. (2019{\natexlab{e}}).
\newblock {FR-NET}: Focal loss constrained deep residual networks for
  segmentation of cardiac {MRI}.
\newblock In \emph{International Symposium on Biomedical Imaging}. 764--767.
\newblock \doi{10.1109/ISBI.2019.8759556}
\bibAnnoteFile{Chen_2019_ISBI}

\bibitem[{Chen et~al.(2019{\natexlab{f}})Chen, Ma, and Zheng}]{chen_2019_med3d}
Chen, S., Ma, K., and Zheng, Y. (2019{\natexlab{f}}).
\newblock Med3d: Transfer learning for 3d medical image analysis.
\newblock \emph{Arxiv Preprint} abs/1904.00625.
\newblock Available at \url{http://arxiv.org/abs/1904.00625} (Accessed
  September 1, 2019)
\bibAnnoteFile{chen_2019_med3d}

\bibitem[{Chen et~al.(2019{\natexlab{g}})Chen, Williams, Vallabhaneni, Czanner,
  Williams, and Zheng}]{Chen_2019_CVPR}
Chen, X., Williams, B.~M., Vallabhaneni, S.~R., Czanner, G., Williams, R., and
  Zheng, Y. (2019{\natexlab{g}}).
\newblock Learning active contour models for medical image segmentation.
\newblock In \emph{Conference on Computer Vision and Pattern Recognition}.
  11632--11640
\bibAnnoteFile{Chen_2019_CVPR}

\bibitem[{Chen et~al.(2019{\natexlab{h}})Chen, Lin, Wang, Lee, Lee, Wang
  et~al.}]{Chen_2019_MIDL}
Chen, Y.-C., Lin, Y.-C., Wang, C.-P., Lee, C.-Y., Lee, W.-J., Wang, T.-D.,
  et~al. (2019{\natexlab{h}}).
\newblock Coronary artery segmentation in cardiac {CT} angiography using {3D}
  multi-channel {U-net}.
\newblock In \emph{Medical Imaging with Deep Learning}. 1907.12246
\bibAnnoteFile{Chen_2019_MIDL}

\bibitem[{Cho et~al.(2014)Cho, van Merrienboer, Gulcehre, Bahdanau, Bougares,
  Schwenk et~al.}]{Cho_2014_EMNLP}
Cho, K., van Merrienboer, B., Gulcehre, C., Bahdanau, D., Bougares, F.,
  Schwenk, H., et~al. (2014).
\newblock Learning phrase representations using {RNN} {Encoder-Decoder} for
  statistical machine translation.
\newblock In \emph{Conference on Empirical Methods in Natural Language
  Processing} ({ACL}), 1724--1734
\bibAnnoteFile{Cho_2014_EMNLP}

\bibitem[{{\c C}i{\c c}ek et~al.(2016){\c C}i{\c c}ek, Abdulkadir, Lienkamp,
  Brox, and Ronneberger}]{Cicek_2016_MICCAI}
{\c C}i{\c c}ek, {\"O}., Abdulkadir, A., Lienkamp, S.~S., Brox, T., and
  Ronneberger, O. (2016).
\newblock {3D} {U-Net}: Learning dense volumetric segmentation from sparse
  annotation.
\newblock In \emph{Medical Image Computing and Computer Assisted Intervention}.
  424--432.
\newblock \doi{10.1007/978-3-319-46723-8\_49}
\bibAnnoteFile{Cicek_2016_MICCAI}

\bibitem[{Ciresan and Giusti(2012)}]{Ciresan_2012_NIPS}
Ciresan, D.~C. and Giusti, A. (2012).
\newblock Deep neural networks segment neuronal membranes in electron
  microscopy images.
\newblock In \emph{Conference on Neural Information Processing Systems}.
  2852--2860
\bibAnnoteFile{Ciresan_2012_NIPS}

\bibitem[{Clough et~al.(2019)Clough, Oksuz, Byrne, Schnabel, and
  King}]{Clough_2019_IPMI}
Clough, J.~R., Oksuz, I., Byrne, N., Schnabel, J.~A., and King, A.~P. (2019).
\newblock Explicit topological priors for deep-learning based image
  segmentation using persistent homology.
\newblock In \emph{Information Processing in Medical Imaging}. vol. 11492 LNCS,
  16--28.
\newblock \doi{10.1007/978-3-030-20351-1\_2}
\bibAnnoteFile{Clough_2019_IPMI}

\bibitem[{Cong and Zhang(2018)}]{Cong_2018_JE}
Cong, C. and Zhang, H. (2018).
\newblock {Invert-U-Net} {DNN} segmentation model for {MRI} cardiac left
  ventricle segmentation.
\newblock \emph{The Journal of Engineering} 2018, 1463--1467.
\newblock \doi{10.1049/joe.2018.8302}
\bibAnnoteFile{Cong_2018_JE}

\bibitem[{Cubuk et~al.(2019)Cubuk, Zoph, Mane, Vasudevan, and
  Le}]{Ekin_2019_CVPR}
Cubuk, E.~D., Zoph, B., Mane, D., Vasudevan, V., and Le, Q.~V. (2019).
\newblock Autoaugment: Learning augmentation policies from data.
\newblock In \emph{Conference on Computer Vision and Pattern Recognition}.
  1--14
\bibAnnoteFile{Ekin_2019_CVPR}

\bibitem[{Dangi et~al.(2018{\natexlab{a}})Dangi, Linte, and
  Yaniv}]{Dangi_2018_SPIE}
Dangi, S., Linte, C.~A., and Yaniv, Z. (2018{\natexlab{a}}).
\newblock Cine cardiac {MRI} slice misalignment correction towards full {3D}
  left ventricle segmentation.
\newblock In \emph{Proceedings of SPIE The International Society for Optical
  Engineering}. 1057607.
\newblock \doi{10.1117/12.2294936}
\bibAnnoteFile{Dangi_2018_SPIE}

\bibitem[{Dangi et~al.(2018{\natexlab{b}})Dangi, Yaniv, and
  Linte}]{Dangi_2018_STACOM}
Dangi, S., Yaniv, Z., and Linte, C.~A. (2018{\natexlab{b}}).
\newblock Left ventricle segmentation and quantification from cardiac cine {MR}
  images via multi-task learning.
\newblock In \emph{Statistical Atlases and Computational Models of the Heart.
  Atrial Segmentation and {LV} Quantification Challenges - 9th International
  Workshop, {STACOM} 2018, Held in Conjunction with {MICCAI} 2018, Granada,
  Spain, September 16, 2018, Revised Selected Papers}. 21--31.
\newblock \doi{10.1007/978-3-030-12029-0\_3}
\bibAnnoteFile{Dangi_2018_STACOM}

\bibitem[{de~Vos et~al.(2017)de~Vos, Wolterink, De~Jong, Leiner, Viergever, and
  Isgum}]{De_2017_TMI}
de~Vos, B.~D., Wolterink, J.~M., De~Jong, P.~A., Leiner, T., Viergever, M.~A.,
  and Isgum, I. (2017).
\newblock Convnet-based localization of anatomical structures in {3-D} medical
  images.
\newblock \emph{IEEE Transactions on Medical Imaging} 36, 1470--1481
\bibAnnoteFile{De_2017_TMI}

\bibitem[{de~Vos et~al.(2019)de~Vos, Wolterink, Leiner, de~Jong, Lessmann, and
  Isgum}]{deVos_2019_TMI}
de~Vos, B.~D., Wolterink, J.~M., Leiner, T., de~Jong, P.~A., Lessmann, N., and
  Isgum, I. (2019).
\newblock Direct automatic coronary calcium scoring in cardiac and chest {CT}.
\newblock \emph{IEEE Transactions on Medical Imaging} 38, 2127--2138
\bibAnnoteFile{deVos_2019_TMI}

\bibitem[{Degel et~al.(2018)Degel, Navab, and Albarqouni}]{Degel_2018_MICCAI}
Degel, M.~A., Navab, N., and Albarqouni, S. (2018).
\newblock Domain and geometry agnostic {CNNs} for left atrium segmentation in
  {3D} ultrasound.
\newblock In \emph{Medical Image Computing and Computer Assisted Intervention
  -- {MICCAI} 2018} (Springer International Publishing), 630--637
\bibAnnoteFile{Degel_2018_MICCAI}

\bibitem[{Dong et~al.(2018{\natexlab{a}})Dong, Luo, Wang, Cao, Li, and
  Zhang}]{Dong_2018_HindawiBiomed}
Dong, S., Luo, G., Wang, K., Cao, S., Li, Q., and Zhang, H.
  (2018{\natexlab{a}}).
\newblock A combined fully convolutional networks and deformable model for
  automatic left ventricle segmentation based on {3D} echocardiography.
\newblock \emph{Biomed Res. Int.} 2018, 5682365.
\newblock \doi{10.1155/2018/5682365}
\bibAnnoteFile{Dong_2018_HindawiBiomed}

\bibitem[{Dong et~al.(2018{\natexlab{b}})Dong, Luo, Wang, Cao, Mercado,
  Shmuilovich et~al.}]{Dong_2018_MICCAI_echo}
Dong, S., Luo, G., Wang, K., Cao, S., Mercado, A., Shmuilovich, O., et~al.
  (2018{\natexlab{b}}).
\newblock {VoxelAtlasGAN}: {3D} left ventricle segmentation on echocardiography
  with atlas guided generation and {Voxel-to-Voxel} discrimination.
\newblock In \emph{Medical Image Computing and Computer Assisted Intervention
  -- {MICCAI} 2018} (Springer International Publishing), 622--629
\bibAnnoteFile{Dong_2018_MICCAI_echo}

\bibitem[{Dormer et~al.(2018)Dormer, Ma, Halicek, Reilly, Schreibmann, and
  Fei}]{Dormer_2018_SPIE_CT}
Dormer, J.~D., Ma, L., Halicek, M., Reilly, C.~M., Schreibmann, E., and Fei, B.
  (2018).
\newblock Heart chamber segmentation from {CT} using convolutional neural
  networks.
\newblock In \emph{Medical Imaging 2018: Biomedical Applications in Molecular,
  Structural, and Functional Imaging, Houston, Texas, United States, 10-15
  February 2018}. 105782S
\bibAnnoteFile{Dormer_2018_SPIE_CT}

\bibitem[{Dou et~al.(2019)Dou, Ouyang, Chen, Chen, Glocker, Zhuang
  et~al.}]{Dou_2019_Access}
Dou, Q., Ouyang, C., Chen, C., Chen, H., Glocker, B., Zhuang, X., et~al.
  (2019).
\newblock Pnp-adanet: Plug-and-play adversarial domain adaptation network at
  unpaired cross-modality cardiac segmentation.
\newblock \emph{{IEEE} Access} 7, 99065--99076.
\newblock \doi{10.1109/ACCESS.2019.2929258}
\bibAnnoteFile{Dou_2019_Access}

\bibitem[{Dou et~al.(2018)Dou, Ouyang, Chen, Chen, and Heng}]{Dou_2018_IJCAI}
Dou, Q., Ouyang, C., Chen, C., Chen, H., and Heng, P.-A. (2018).
\newblock Unsupervised {Cross-Modality} domain adaptation of {ConvNets} for
  biomedical image segmentations with adversarial loss.
\newblock In \emph{International Joint Conferences on Artificial Intelligence}.
  691--697
\bibAnnoteFile{Dou_2018_IJCAI}

\bibitem[{Du et~al.(2019)Du, Yin, Tang, Zhang, and Li}]{Du_2019_JTEHM}
Du, X., Yin, S., Tang, R., Zhang, Y., and Li, S. (2019).
\newblock {Cardiac-DeepIED}: Automatic pixel-level deep segmentation for
  cardiac bi-ventricle using improved end-to-end encoder-decoder network.
\newblock \emph{IEEE journal of translational engineering in health and
  medicine} 7, 1--10.
\newblock \doi{10.1109/JTEHM.2019.2900628}
\bibAnnoteFile{Du_2019_JTEHM}

\bibitem[{Duan et~al.(2019)Duan, Bello, Schlemper, Bai, Dawes, Biffi
  et~al.}]{Duan_2019_TMI}
Duan, J., Bello, G., Schlemper, J., Bai, W., Dawes, T. J.~W., Biffi, C., et~al.
  (2019).
\newblock Automatic {3D} bi-ventricular segmentation of cardiac images by a
  shape-constrained multi-task deep learning approach.
\newblock \emph{IEEE Transactions on Medical Imaging} PP, 1.
\newblock \doi{10.1109/TMI.2019.2894322}
\bibAnnoteFile{Duan_2019_TMI}

\bibitem[{Duan et~al.(2018{\natexlab{a}})Duan, Schlemper, Bai, Dawes, Bello,
  Doumou et~al.}]{Duan_2018_MICCAI}
Duan, J., Schlemper, J., Bai, W., Dawes, T. J.~W., Bello, G., Doumou, G.,
  et~al. (2018{\natexlab{a}}).
\newblock Deep nested level sets: Fully automated segmentation of cardiac {MR}
  images in patients with pulmonary hypertension.
\newblock In \emph{Medical Image Computing and Computer Assisted Intervention}.
  595--603
\bibAnnoteFile{Duan_2018_MICCAI}

\bibitem[{Duan et~al.(2018{\natexlab{b}})Duan, Feng, Lu, and
  Zhou}]{Duan_2018_STACOM}
Duan, Y., Feng, J., Lu, J., and Zhou, J. (2018{\natexlab{b}}).
\newblock Context aware {3D} fully convolutional networks for coronary artery
  segmentation.
\newblock In \emph{International Workshop on Statistical Atlases and
  Computational Models of the Heart} (Springer), 85--93
\bibAnnoteFile{Duan_2018_STACOM}

\bibitem[{Dwork and Roth(2014)}]{Dwork_2014_FTTCS}
Dwork, C. and Roth, A. (2014).
\newblock The algorithmic foundations of differential privacy.
\newblock \emph{Foundations and Trends in Theoretical Computer Science} 9,
  211--407
\bibAnnoteFile{Dwork_2014_FTTCS}

\bibitem[{Fahmy et~al.(2019)Fahmy, El-Rewaidy, Nezafat, Nakamori, and
  Nezafat}]{Fahmy_2019_JCMR}
Fahmy, A.~S., El-Rewaidy, H., Nezafat, M., Nakamori, S., and Nezafat, R.
  (2019).
\newblock Automated analysis of cardiovascular magnetic resonance myocardial
  native {T1} mapping images using fully convolutional neural networks.
\newblock \emph{Journal of Cardiovascular Magnetic Resonance} 21, 1--12.
\newblock \doi{10.1186/s12968-018-0516-1}
\bibAnnoteFile{Fahmy_2019_JCMR}

\bibitem[{Fahmy et~al.(2018)Fahmy, Rausch, Neisius, Chan, Maron, Appelbaum
  et~al.}]{Fahmy_2018_JACC}
Fahmy, A.~S., Rausch, J., Neisius, U., Chan, R.~H., Maron, M.~S., Appelbaum,
  E., et~al. (2018).
\newblock Automated cardiac {MR} scar quantification in hypertrophic
  cardiomyopathy using deep convolutional neural networks.
\newblock \emph{JACC. Cardiovascular imaging} 11, 1917--1918.
\newblock \doi{10.1016/j.jcmg.2018.04.030}
\bibAnnoteFile{Fahmy_2018_JACC}

\bibitem[{Finlayson et~al.(2019)Finlayson, Bowers, Ito, Zittrain, Beam, and
  Kohane}]{Finlayson_2019_Science}
Finlayson, S.~G., Bowers, J.~D., Ito, J., Zittrain, J.~L., Beam, A.~L., and
  Kohane, I.~S. (2019).
\newblock Adversarial attacks on medical machine learning.
\newblock \emph{Science} 363, 1287--1289.
\newblock \doi{10.1126/science.aaw4399}
\bibAnnoteFile{Finlayson_2019_Science}

\bibitem[{Gandhi et~al.(2018)Gandhi, Mosleh, Shen, and
  Chow}]{Gandhi_2018_Echocardiography}
Gandhi, S., Mosleh, W., Shen, J., and Chow, C.-M. (2018).
\newblock Automation, machine learning, and artificial intelligence in
  echocardiography: A brave new world.
\newblock \emph{Echocardiography} 35, 1402--1418.
\newblock \doi{10.1111/echo.14086}
\bibAnnoteFile{Gandhi_2018_Echocardiography}

\bibitem[{Georgescu et~al.(2005)Georgescu, Zhou, Comaniciu, and
  Gupta}]{Georgescu_2005_CVPR}
Georgescu, B., Zhou, X.~S., Comaniciu, D., and Gupta, A. (2005).
\newblock Database-guided segmentation of anatomical structures with complex
  appearance.
\newblock In \emph{2005 {IEEE} Computer Society Conference on Computer Vision
  and Pattern Recognition ({CVPR'05})} (IEEE), vol.~2, 429--436
\bibAnnoteFile{Georgescu_2005_CVPR}

\bibitem[{Ghesu et~al.(2016)Ghesu, Krubasik, Georgescu, Singh, {Yefeng Zheng},
  Hornegger et~al.}]{Ghesu_2016_TMI}
Ghesu, F.~C., Krubasik, E., Georgescu, B., Singh, V., {Yefeng Zheng},
  Hornegger, J., et~al. (2016).
\newblock Marginal space deep learning: Efficient architecture for volumetric
  image parsing.
\newblock \emph{IEEE Transactions on Medical Imaging} 35, 1217--1228
\bibAnnoteFile{Ghesu_2016_TMI}

\bibitem[{Girdhar et~al.(2016)Girdhar, Fouhey, Rodriguez, and
  Gupta}]{Girdhar_2016_ECCV}
Girdhar, R., Fouhey, D.~F., Rodriguez, M., and Gupta, A. (2016).
\newblock Learning a predictable and generative vector representation for
  objects.
\newblock In \emph{European Conference on Computer Vision} (Springer
  International Publishing), 484--499
\bibAnnoteFile{Girdhar_2016_ECCV}

\bibitem[{Goodfellow(2016)}]{Goodfellow_2016_MIT}
Goodfellow, I. (2016).
\newblock \emph{Deep learning}.
\newblock Adaptive computation and machine learning (Cambridge, Massachusetts ;
  London, England: The MIT Press)
\bibAnnoteFile{Goodfellow_2016_MIT}

\bibitem[{Goodfellow et~al.(2014)Goodfellow, Pouget-Abadie, Mirza, Xu,
  Warde-Farley, Ozair et~al.}]{Goodfellow_2014_NIPS}
Goodfellow, I.~J., Pouget-Abadie, J., Mirza, M., Xu, B., Warde-Farley, D.,
  Ozair, S., et~al. (2014).
\newblock Generative adversarial networks.
\newblock In \emph{Conference on Neural Information Processing Systems} (Curran
  Associates, Inc.), 2672--2680
\bibAnnoteFile{Goodfellow_2014_NIPS}

\bibitem[{Goodfellow et~al.(2015)Goodfellow, Shlens, and
  Szegedy}]{Goodfellow_2015_ICLR}
Goodfellow, I.~J., Shlens, J., and Szegedy, C. (2015).
\newblock Explaining and harnessing adversarial examples.
\newblock In \emph{International Conference on Learning Representations}. 43405
\bibAnnoteFile{Goodfellow_2015_ICLR}

\bibitem[{Greenspan et~al.(2016)Greenspan, Van~Ginneken, and
  Summers}]{Greenspan_2016_TMI}
Greenspan, H., Van~Ginneken, B., and Summers, R.~M. (2016).
\newblock Guest editorial deep learning in medical imaging: Overview and future
  promise of an exciting new technique.
\newblock \emph{IEEE Transactions on Medical Imaging} 35, 1153--1159
\bibAnnoteFile{Greenspan_2016_TMI}

\bibitem[{G{\"u}ls{\"u}n et~al.(2016)G{\"u}ls{\"u}n, Funka-Lea, Sharma, Rapaka,
  and Zheng}]{Gulsun_2016_MICCAI}
G{\"u}ls{\"u}n, M.~A., Funka-Lea, G., Sharma, P., Rapaka, S., and Zheng, Y.
  (2016).
\newblock Coronary centerline extraction via optimal flow paths and {CNN} path
  pruning.
\newblock In \emph{Medical Image Computing and Computer Assisted Intervention}
  (Springer), 317--325
\bibAnnoteFile{Gulsun_2016_MICCAI}

\bibitem[{Guo et~al.(2019)Guo, Bai, Lu, Wang, Cao, Song et~al.}]{Guo_2019_IPMI}
Guo, Z., Bai, J., Lu, Y., Wang, X., Cao, K., Song, Q., et~al. (2019).
\newblock Deepcenterline: A multi-task fully convolutional network for
  centerline extraction.
\newblock In \emph{International Conference on Information Processing in
  Medical Imaging} (Springer), 441--453
\bibAnnoteFile{Guo_2019_IPMI}

\bibitem[{He et~al.(2015)He, Zhang, Ren, and Sun}]{He_2015_ICCV}
He, K., Zhang, X., Ren, S., and Sun, J. (2015).
\newblock Delving deep into rectifiers: Surpassing {Human-Level} performance on
  {ImageNet} classification.
\newblock In \emph{International Conference on Computer Vision} ({IEEE}
  Computer Society), 1026--1034
\bibAnnoteFile{He_2015_ICCV}

\bibitem[{He et~al.(2016)He, Zhang, Ren, and Sun}]{He_2016_CVPR}
He, K., Zhang, X., Ren, S., and Sun, J. (2016).
\newblock Deep residual learning for image recognition.
\newblock In \emph{Conference on Computer Vision and Pattern Recognition}.
  770--778
\bibAnnoteFile{He_2016_CVPR}

\bibitem[{Heo et~al.(2018)Heo, Lee, Kim, Lee, Kim, Yang et~al.}]{Heo_2018_NIPS}
Heo, J., Lee, H.~B., Kim, S., Lee, J., Kim, K.~J., Yang, E., et~al. (2018).
\newblock {Uncertainty-Aware} attention for reliable interpretation and
  prediction.
\newblock In \emph{Advances in Neural Information Processing Systems 31}, eds.
  S.~Bengio, H.~Wallach, H.~Larochelle, K.~Grauman, N.~Cesa-Bianchi, and
  R.~Garnett (Curran Associates, Inc.). 909--918
\bibAnnoteFile{Heo_2018_NIPS}

\bibitem[{Herment et~al.(2010)Herment, Kachenoura, Lefort, Bensalah, Dogui,
  Frouin et~al.}]{Herment_2010_JMRI}
Herment, A., Kachenoura, N., Lefort, M., Bensalah, M., Dogui, A., Frouin, F.,
  et~al. (2010).
\newblock Automated segmentation of the aorta from phase contrast {MR} images:
  validation against expert tracing in healthy volunteers and in patients with
  a dilated aorta.
\newblock \emph{Journal of magnetic resonance imaging} 31, 881--888.
\newblock \doi{10.1002/jmri.22124}
\bibAnnoteFile{Herment_2010_JMRI}

\bibitem[{Hinton and Salakhutdinov(2006)}]{Hinton_2006_Science}
Hinton, G.~E. and Salakhutdinov, R.~R. (2006).
\newblock Reducing the dimensionality of data with neural networks.
\newblock \emph{Science} 313, 504--507
\bibAnnoteFile{Hinton_2006_Science}

\bibitem[{Hochreiter and Schmidhuber(1997)}]{Hochreiter_1997_NC}
Hochreiter, S. and Schmidhuber, J. (1997).
\newblock Long short-term memory.
\newblock \emph{Neural computation} 9, 1735--1780.
\newblock \doi{10.1162/neco.1997.9.8.1735}
\bibAnnoteFile{Hochreiter_1997_NC}

\bibitem[{Hu et~al.(2018)Hu, Shen, and Sun}]{Hu_2018_CVPR}
Hu, J., Shen, L., and Sun, G. (2018).
\newblock {Squeeze-and-Excitation} networks.
\newblock In \emph{Conference on Computer Vision and Pattern Recognition}.
  7132--7141.
\newblock \doi{10.1109/CVPR.2018.00745}
\bibAnnoteFile{Hu_2018_CVPR}

\bibitem[{Huang et~al.(2017)Huang, Liu, van~der Maaten, and
  Weinberger}]{Huang_2017_CVPR}
Huang, G., Liu, Z., van~der Maaten, L., and Weinberger, K.~Q. (2017).
\newblock Densely connected convolutional networks.
\newblock In \emph{Conference on Computer Vision and Pattern Recognition}.
  2261--2269
\bibAnnoteFile{Huang_2017_CVPR}

\bibitem[{Huang et~al.(2019)Huang, Yang, Yi, Axel, and
  Metaxas}]{Huang_2019_FIMH}
Huang, Q., Yang, D., Yi, J., Axel, L., and Metaxas, D. (2019).
\newblock {FR-Net}: Joint reconstruction and segmentation in compressed sensing
  cardiac {MRI}.
\newblock In \emph{Functional Imaging and Modelling of the Heart} (Springer
  International Publishing), 352--360.
\newblock \doi{10.1007/978-3-030-21949-9\_38}
\bibAnnoteFile{Huang_2019_FIMH}

\bibitem[{Huang et~al.(2018)Huang, Huang, Lin, Huang, Chi, Zhou
  et~al.}]{Huang_2018_EMBC}
Huang, W., Huang, L., Lin, Z., Huang, S., Chi, Y., Zhou, J., et~al. (2018).
\newblock Coronary artery segmentation by deep learning neural networks on
  computed tomographic coronary angiographic images.
\newblock In \emph{2018 40th Annual International Conference of the IEEE
  Engineering in Medicine and Biology Society (EMBC)} (IEEE), 608--611
\bibAnnoteFile{Huang_2018_EMBC}

\bibitem[{Ioffe and Szegedy(2015)}]{Ioffe_2015_ICML}
Ioffe, S. and Szegedy, C. (2015).
\newblock Batch normalization: accelerating deep network training by reducing
  internal covariate shift.
\newblock In \emph{ICML} (JMLR.org), 448--456
\bibAnnoteFile{Ioffe_2015_ICML}

\bibitem[{Irvin et~al.(2019)Irvin, Rajpurkar, Ko, Yu, Ciurea-Ilcus, Chute
  et~al.}]{Irvin_2019_AAAI}
Irvin, J., Rajpurkar, P., Ko, M., Yu, Y., Ciurea-Ilcus, S., Chute, C., et~al.
  (2019).
\newblock {CheXpert}: A large chest radiograph dataset with uncertainty labels
  and expert comparison.
\newblock In \emph{Conference on Artificial Intelligence}. 590--597
\bibAnnoteFile{Irvin_2019_AAAI}

\bibitem[{Isensee et~al.(2017)Isensee, Jaeger, Full, Wolf, Engelhardt, and
  Maier-Hein}]{Isensee_2017_STACOM}
Isensee, F., Jaeger, P.~F., Full, P.~M., Wolf, I., Engelhardt, S., and
  Maier-Hein, K.~H. (2017).
\newblock Automatic cardiac disease assessment on {cine-MRI} via {Time-Series}
  segmentation and domain specific features.
\newblock In \emph{Statistical Atlases and Computational Models of the Heart.
  {ACDC} and {MMWHS} Challenges} (Springer International Publishing), 120--129.
\newblock \doi{10.1007/978-3-319-75541-0\_13}
\bibAnnoteFile{Isensee_2017_STACOM}

\bibitem[{Jafari et~al.(2019)Jafari, Girgis, Abdi, Liao, Pesteie, Rohling
  et~al.}]{Jafari_2019_ISBI}
Jafari, M.~H., Girgis, H., Abdi, A.~H., Liao, Z., Pesteie, M., Rohling, R.,
  et~al. (2019).
\newblock {Semi-Supervised} learning for cardiac left ventricle segmentation
  using conditional deep generative models as prior.
\newblock In \emph{2019 {IEEE} 16th International Symposium on Biomedical
  Imaging ({ISBI} 2019)} (IEEE), 649--652
\bibAnnoteFile{Jafari_2019_ISBI}

\bibitem[{Jafari et~al.(2018)Jafari, Girgis, Liao, Behnami, Abdi, Vaseli
  et~al.}]{Jafari_2018_DLMIA}
Jafari, M.~H., Girgis, H., Liao, Z., Behnami, D., Abdi, A., Vaseli, H., et~al.
  (2018).
\newblock A unified framework integrating recurrent {Fully-Convolutional}
  networks and optical flow for segmentation of the left ventricle in
  echocardiography data.
\newblock In \emph{Deep Learning in Medical Image Analysis and Multimodal
  Learning for Clinical Decision Support} (Springer International Publishing),
  29--37
\bibAnnoteFile{Jafari_2018_DLMIA}

\bibitem[{Jang et~al.(2017)Jang, Hong, Ha, Kim, and Chang}]{Jang_2017_STACOM}
Jang, Y., Hong, Y., Ha, S., Kim, S., and Chang, H.-J. (2017).
\newblock Automatic segmentation of {LV} and {RV} in cardiac {MRI}.
\newblock In \emph{International Workshop on Statistical Atlases and
  Computational Models of the Heart} (Springer), 161--169
\bibAnnoteFile{Jang_2017_STACOM}

\bibitem[{Jia et~al.(2018)Jia, Despinasse, Wang, Delingette, Pennec, Ja{\"\i}s
  et~al.}]{Jia_2018_STACOM}
Jia, S., Despinasse, A., Wang, Z., Delingette, H., Pennec, X., Ja{\"\i}s, P.,
  et~al. (2018).
\newblock Automatically segmenting the left atrium from cardiac images using
  successive {3D} {U-Nets} and a contour loss.
\newblock In \emph{International Workshop on Statistical Atlases and
  Computational Models of the Heart}. 221--229
\bibAnnoteFile{Jia_2018_STACOM}

\bibitem[{Joyce et~al.(2018)Joyce, Chartsias, and Tsaftaris}]{Joyce_2018_MIDL}
Joyce, T., Chartsias, A., and Tsaftaris, S.~A. (2018).
\newblock Deep {Multi-Class} segmentation without {Ground-Truth} labels.
\newblock In \emph{Medical Imaging with Deep Learning}. 1--9
\bibAnnoteFile{Joyce_2018_MIDL}

\bibitem[{Kamnitsas et~al.(2017{\natexlab{a}})Kamnitsas, Bai, Ferrante,
  Mcdonagh, and Sinclair}]{Kamnitsas_2017_Arxiv}
Kamnitsas, K., Bai, W., Ferrante, E., Mcdonagh, S., and Sinclair, M.
  (2017{\natexlab{a}}).
\newblock Ensembles of multiple models and architectures for robust brain
  tumour segmentation.
\newblock In \emph{Brainlesion: Glioma, Multiple Sclerosis, Stroke and
  Traumatic Brain Injuries - Third International Workshop, BrainLes 2017, Held
  in Conjunction with {MICCAI} 2017, Quebec City, QC, Canada, September 14,
  2017, Revised Selected Papers}. 450--462
\bibAnnoteFile{Kamnitsas_2017_Arxiv}

\bibitem[{Kamnitsas et~al.(2017{\natexlab{b}})Kamnitsas, Baumgartner, Ledig,
  Newcombe, Simpson, Kane et~al.}]{Kamnitsas_2017_IPMI}
Kamnitsas, K., Baumgartner, C., Ledig, C., Newcombe, V., Simpson, J., Kane, A.,
  et~al. (2017{\natexlab{b}}).
\newblock Unsupervised domain adaptation in brain lesion segmentation with
  adversarial networks.
\newblock In \emph{International Conference on Information Processing in
  Medical Imaging} (Springer International Publishing), 597--609
\bibAnnoteFile{Kamnitsas_2017_IPMI}

\bibitem[{Kang et~al.(2012)Kang, Woo, Kuo, Slomka, Dey, and
  Germano}]{Kang_2012_JoEI}
Kang, D., Woo, J., Kuo, C.~J., Slomka, P.~J., Dey, D., and Germano, G. (2012).
\newblock Heart chambers and whole heart segmentation techniques.
\newblock \emph{Journal of Electronic Imaging} 21, 010901
\bibAnnoteFile{Kang_2012_JoEI}

\bibitem[{Karim et~al.(2016)Karim, Bhagirath, Claus, James~Housden, Chen,
  Karimaghaloo et~al.}]{Karim_2016_LVIS_Dataset}
Karim, R., Bhagirath, P., Claus, P., James~Housden, R., Chen, Z., Karimaghaloo,
  Z., et~al. (2016).
\newblock Evaluation of state-of-the-art segmentation algorithms for left
  ventricle infarct from late gadolinium enhancement {MR} images.
\newblock \emph{Medical image analysis} 30, 95--107.
\newblock \doi{10.1016/j.media.2016.01.004}
\bibAnnoteFile{Karim_2016_LVIS_Dataset}

\bibitem[{Karim et~al.(2018)Karim, Blake, Inoue, Tao, Jia, James~Housden
  et~al.}]{Karim_2018_MedIA}
Karim, R., Blake, L.-E., Inoue, J., Tao, Q., Jia, S., James~Housden, R., et~al.
  (2018).
\newblock Algorithms for left atrial wall segmentation and thickness --
  evaluation on an open-source {CT} and {MRI} image database.
\newblock \emph{Medical Image Analysis} 50, 36--53.
\newblock \doi{10.1016/j.media.2018.08.004}
\bibAnnoteFile{Karim_2018_MedIA}

\bibitem[{Karim et~al.(2013)Karim, Housden, Balasubramaniam, Chen, Perry, Uddin
  et~al.}]{Karim_2013_LA_SCAR}
Karim, R., Housden, R.~J., Balasubramaniam, M., Chen, Z., Perry, D., Uddin, A.,
  et~al. (2013).
\newblock Evaluation of current algorithms for segmentation of scar tissue from
  late gadolinium enhancement cardiovascular magnetic resonance of the left
  atrium: an open-access grand challenge.
\newblock \emph{Journal of cardiovascular magnetic resonance: official journal
  of the Society for Cardiovascular Magnetic Resonance} 15, 105.
\newblock \doi{10.1186/1532-429X-15-105}
\bibAnnoteFile{Karim_2013_LA_SCAR}

\bibitem[{Karim et~al.(2008)Karim, Mohiaddin, and Rueckert}]{Karim_2008_Online}
Karim, R., Mohiaddin, R., and Rueckert, D. (2008).
\newblock Left atrium segmentation for atrial fibrillation ablation.
\newblock In \emph{Medical Imaging 2008: Visualization, Image-Guided
  Procedures, and Modeling, San Diego, California, United States, 16-21
  February 2008} ({SPIE}), March 2008, 69182U.
\newblock \doi{10.1117/12.771023}
\bibAnnoteFile{Karim_2008_Online}

\bibitem[{Kass et~al.(1988)Kass, Witkin, and Terzopoulos}]{Kass_1988_IJCV}
Kass, M., Witkin, A., and Terzopoulos, D. (1988).
\newblock Snakes: Active contour models.
\newblock \emph{Int. J. Comput. Vis.} 1, 321--331
\bibAnnoteFile{Kass_1988_IJCV}

\bibitem[{Kervadec et~al.(2019)Kervadec, Dolz, Tang, Granger, Boykov, and
  Ben~Ayed}]{Kervadec_2018_MeDIA}
Kervadec, H., Dolz, J., Tang, M., Granger, E., Boykov, Y., and Ben~Ayed, I.
  (2019).
\newblock {Constrained-CNN} losses for weakly supervised segmentation.
\newblock In \emph{Medical Image Analysis}. vol.~54, 88--99.
\newblock \doi{10.1016/j.media.2019.02.009}
\bibAnnoteFile{Kervadec_2018_MeDIA}

\bibitem[{Khened et~al.(2019)Khened, Kollerathu, and
  Krishnamurthi}]{Khened_2019_MedIA}
Khened, M., Kollerathu, V.~A., and Krishnamurthi, G. (2019).
\newblock Fully convolutional multi-scale residual {DenseNets} for cardiac
  segmentation and automated cardiac diagnosis using ensemble of classifiers.
\newblock \emph{Medical Image Analysis} 51, 21--45.
\newblock \doi{10.1016/j.media.2018.10.004}
\bibAnnoteFile{Khened_2019_MedIA}

\bibitem[{Kim et~al.(1999)Kim, Fieno, Parrish, Harris, Chen, Simonetti
  et~al.}]{Kim_1999_Circulation}
Kim, R.~J., Fieno, D.~S., Parrish, T.~B., Harris, K., Chen, E.~L., Simonetti,
  O., et~al. (1999).
\newblock Relationship of {MRI} delayed contrast enhancement to irreversible
  injury, infarct age, and contractile function.
\newblock \emph{Circulation} 100, 1992--2002.
\newblock \doi{10.1161/01.cir.100.19.1992}
\bibAnnoteFile{Kim_1999_Circulation}

\bibitem[{Kingma and Welling(2013)}]{Kingma_2013_ICLR}
Kingma, D.~P. and Welling, M. (2013).
\newblock {Auto-Encoding} variational bayes.
\newblock In \emph{International Conference on Learning Representations}. 1--14
\bibAnnoteFile{Kingma_2013_ICLR}

\bibitem[{Kiri{\c s}li et~al.(2013)Kiri{\c s}li, Schaap, Metz, Dharampal,
  Meijboom, Papadopoulou et~al.}]{Kirisli2013_Lumen_Stenosis_dataset}
Kiri{\c s}li, H.~A., Schaap, M., Metz, C.~T., Dharampal, A.~S., Meijboom,
  W.~B., Papadopoulou, S.~L., et~al. (2013).
\newblock Standardized evaluation framework for evaluating coronary artery
  stenosis detection, stenosis quantification and lumen segmentation algorithms
  in computed tomography angiography.
\newblock \emph{Medical Image Analysis} 17, 859--876.
\newblock \doi{10.1016/j.media.2013.05.007}.
\newblock \url{http://coronary.bigr.nl/stenoses}
\bibAnnoteFile{Kirisli2013_Lumen_Stenosis_dataset}

\bibitem[{Kurakin et~al.(2017)Kurakin, Goodfellow, and
  Bengio}]{Kurakin_2016_Arxiv}
Kurakin, A., Goodfellow, I., and Bengio, S. (2017).
\newblock Adversarial examples in the physical world.
\newblock In \emph{5th International Conference on Learning Representations,
  {ICLR} 2017,Toulon, France, April 24-26, 2017, Workshop Track Proceedings}.
  1--14
\bibAnnoteFile{Kurakin_2016_Arxiv}

\bibitem[{Leclerc et~al.(2018)Leclerc, Smistad, Grenier, Lartizien, Ostvik,
  Espinosa et~al.}]{Leclerc_2018_IUS}
Leclerc, S., Smistad, E., Grenier, T., Lartizien, C., Ostvik, A., Espinosa, F.,
  et~al. (2018).
\newblock Deep learning applied to {Multi-Structure} segmentation in {2D}
  echocardiography: A preliminary investigation of the required database size.
\newblock In \emph{2018 {IEEE} International Ultrasonics Symposium ({IUS})}
  (IEEE), 1--4
\bibAnnoteFile{Leclerc_2018_IUS}

\bibitem[{Leclerc et~al.(2019)Leclerc, Smistad, Pedrosa, Ostvik, Cervenansky,
  Espinosa et~al.}]{Leclerc_2019_TMI}
Leclerc, S., Smistad, E., Pedrosa, J., Ostvik, A., Cervenansky, F., Espinosa,
  F., et~al. (2019).
\newblock Deep learning for segmentation using an open {Large-Scale} dataset in
  {2D} echocardiography.
\newblock \emph{IEEE Trans. Med. Imaging}
  \url{https://www.creatis.insa-lyon.fr/Challenge/camus}
\bibAnnoteFile{Leclerc_2019_TMI}

\bibitem[{Lee et~al.(2015)Lee, Xie, Gallagher, Zhang, and Tu}]{Lee_2015_AIS}
Lee, C.-Y., Xie, S., Gallagher, P., Zhang, Z., and Tu, Z. (2015).
\newblock {Deeply-Supervised} nets.
\newblock In \emph{Artificial Intelligence and Statistics}. 562--570
\bibAnnoteFile{Lee_2015_AIS}

\bibitem[{Lee et~al.(2019)Lee, Petersen, Pawlowski, Glocker, and
  Schaap}]{Lee_2019_TMI}
Lee, M. C.~H., Petersen, K., Pawlowski, N., Glocker, B., and Schaap, M. (2019).
\newblock {TETRIS}: Template transformer networks for image segmentation with
  shape priors.
\newblock \emph{IEEE transactions on medical imaging}
\bibAnnoteFile{Lee_2019_TMI}

\bibitem[{Lesage et~al.(2009)Lesage, Angelini, Bloch, and
  Funka-Lea}]{Lesage_2009_MedIA}
Lesage, D., Angelini, E.~D., Bloch, I., and Funka-Lea, G. (2009).
\newblock A review of {3D} vessel lumen segmentation techniques: Models,
  features and extraction schemes.
\newblock \emph{Medical Image Analysis} 13, 819--845
\bibAnnoteFile{Lesage_2009_MedIA}

\bibitem[{Lessmann et~al.(2016)Lessmann, I{\v{s}}gum, Setio, de~Vos, Ciompi,
  de~Jong et~al.}]{Lessmann_2016_MedicalImaging}
Lessmann, N., I{\v{s}}gum, I., Setio, A.~A., de~Vos, B.~D., Ciompi, F.,
  de~Jong, P.~A., et~al. (2016).
\newblock Deep convolutional neural networks for automatic coronary calcium
  scoring in a screening study with low-dose chest {CT}.
\newblock In \emph{Medical Imaging 2016: Computer-Aided Diagnosis}
  (International Society for Optics and Photonics), vol. 9785, 978511
\bibAnnoteFile{Lessmann_2016_MedicalImaging}

\bibitem[{Lessmann et~al.(2017)Lessmann, van Ginneken, Zreik, de~Jong, de~Vos,
  Viergever et~al.}]{Lessmann_2017_TMI}
Lessmann, N., van Ginneken, B., Zreik, M., de~Jong, P.~A., de~Vos, B.~D.,
  Viergever, M.~A., et~al. (2017).
\newblock Automatic calcium scoring in low-dose chest {CT} using deep neural
  networks with dilated convolutions.
\newblock \emph{IEEE Transactions on Medical Imaging} 37, 615--625
\bibAnnoteFile{Lessmann_2017_TMI}

\bibitem[{Li et~al.(2019{\natexlab{a}})Li, Tong, Liao, Si, Chen, Wang
  et~al.}]{Li_2019_ISBI}
Li, C., Tong, Q., Liao, X., Si, W., Chen, S., Wang, Q., et~al.
  (2019{\natexlab{a}}).
\newblock {APCP-NET}: Aggregated parallel {Cross-Scale} pyramid network for
  {CMR} segmentation.
\newblock In \emph{2019 {IEEE} 16th International Symposium on Biomedical
  Imaging ({ISBI} 2019)}. 784--788.
\newblock \doi{10.1109/ISBI.2019.8759147}
\bibAnnoteFile{Li_2019_ISBI}

\bibitem[{Li et~al.(2018)Li, Tong, Liao, Si, Sun, Wang et~al.}]{Li_2018_STACOM}
Li, C., Tong, Q., Liao, X., Si, W., Sun, Y., Wang, Q., et~al. (2018).
\newblock Attention based hierarchical aggregation network for {3D} left atrial
  segmentation: 9th international workshop, {STACOM} 2018, held in conjunction
  with {MICCAI} 2018, granada, spain, september 16, 2018, revised selected
  papers.
\newblock In \emph{Statistical Atlases and Computational Models of the Heart.
  Atrial Segmentation and {LV} Quantification Challenges}, eds. M.~Pop,
  M.~Sermesant, J.~Zhao, S.~Li, K.~McLeod, A.~Young, K.~Rhode, and T.~Mansi
  (Cham: Springer International Publishing), vol. 11395 of \emph{Lecture Notes
  in Computer Science}. 255--264.
\newblock \doi{10.1007/978-3-030-12029-0\_28}
\bibAnnoteFile{Li_2018_STACOM}

\bibitem[{Li et~al.(2019{\natexlab{b}})Li, Yu, Gu, Liu, and Li}]{Li_2019_ITBE}
Li, J., Yu, Z., Gu, Z., Liu, H., and Li, Y. (2019{\natexlab{b}}).
\newblock {Dilated-Inception} net: {Multi-Scale} feature aggregation for
  cardiac right ventricle segmentation.
\newblock \emph{IEEE transactions on bio-medical engineering} ,
  1--1\doi{10.1109/TBME.2019.2906667}
\bibAnnoteFile{Li_2019_ITBE}

\bibitem[{Li et~al.(2017)Li, Zhang, Shi, and Wang}]{Li_2017_RSAM}
Li, J., Zhang, R., Shi, L., and Wang, D. (2017).
\newblock Automatic {Whole-Heart} segmentation in congenital heart disease
  using {Deeply-Supervised} {3D} {FCN}.
\newblock In \emph{Reconstruction, Segmentation, and Analysis of Medical
  Images} (Springer International Publishing), 111--118.
\newblock \doi{10.1007/978-3-319-52280-7\_11}
\bibAnnoteFile{Li_2017_RSAM}

\bibitem[{Liao et~al.(2019)Liao, Chen, Hu, and Song}]{Liao_2019_TCyber}
Liao, F., Chen, X., Hu, X., and Song, S. (2019).
\newblock Estimation of the volume of the left ventricle from {MRI} images
  using deep neural networks.
\newblock \emph{IEEE transactions on cybernetics} 49, 495--504.
\newblock \doi{10.1109/TCYB.2017.2778799}
\bibAnnoteFile{Liao_2019_TCyber}

\bibitem[{Lieman-Sifry et~al.(2017)Lieman-Sifry, Le, Lau, Sall, and
  Golden}]{Lieman-Sifry_2017_FIMH}
Lieman-Sifry, J., Le, M., Lau, F., Sall, S., and Golden, D. (2017).
\newblock {FastVentricle}: Cardiac segmentation with {ENet}.
\newblock In \emph{Functional Imaging and Modelling of the Heart}, eds. M.~Pop
  and G.~A. Wright (Cham: Springer International Publishing), vol. 10263 LNCS
  of \emph{Lecture Notes in Computer Science}, 127--138.
\newblock \doi{10.1007/978-3-319-59448-4\_13}
\bibAnnoteFile{Lieman-Sifry_2017_FIMH}

\bibitem[{Litjens et~al.(2017)Litjens, Kooi, Bejnordi, Setio, Ciompi,
  Ghafoorian et~al.}]{LITJENS_2017_MedIA}
Litjens, G., Kooi, T., Bejnordi, B.~E., Setio, A. A.~A., Ciompi, F.,
  Ghafoorian, M., et~al. (2017).
\newblock A survey on deep learning in medical image analysis.
\newblock \emph{Medical Image Analysis} 42, 60 -- 88
\bibAnnoteFile{LITJENS_2017_MedIA}

\bibitem[{Liu et~al.(2018)Liu, Jin, Feng, Du, Lu, and Zhou}]{Liu_2018_STACOM}
Liu, J., Jin, C., Feng, J., Du, Y., Lu, J., and Zhou, J. (2018).
\newblock A vessel-focused {3D} convolutional network for automatic
  segmentation and classification of coronary artery plaques in cardiac {CTA}.
\newblock In \emph{International Workshop on Statistical Atlases and
  Computational Models of the Heart} (Springer), 131--141
\bibAnnoteFile{Liu_2018_STACOM}

\bibitem[{Long et~al.(2015)Long, Shelhamer, and Darrell}]{Long_2014_CVPR}
Long, J., Shelhamer, E., and Darrell, T. (2015).
\newblock Fully convolutional networks for semantic segmentation.
\newblock In \emph{Conference on Computer Vision and Pattern Recognition}.
  3431--3440
\bibAnnoteFile{Long_2014_CVPR}

\bibitem[{Lu et~al.(2019)Lu, Chen, Li, and Qiao}]{Lu_2019_ICCSP}
Lu, X., Chen, X., Li, W., and Qiao, Y. (2019).
\newblock Graph cut segmentation of the right ventricle in cardiac {MRI} using
  multi-scale feature learning.
\newblock In \emph{Proceedings of the 3rd International Conference on
  Cryptography, Security and Privacy} (ACM), 231--235.
\newblock \doi{10.1145/3309074.3309117}
\bibAnnoteFile{Lu_2019_ICCSP}

\bibitem[{Luc et~al.(2016)Luc, Couprie, Chintala, and
  Verbeek}]{Luc_2016_NIPS_workshop}
Luc, P., Couprie, C., Chintala, S., and Verbeek, J. (2016).
\newblock Semantic segmentation using adversarial networks.
\newblock In \emph{NIPS Workshop on Adversarial Training}. 1--12
\bibAnnoteFile{Luc_2016_NIPS_workshop}

\bibitem[{Ma and Zhang(2019)}]{Ma_2019_arXiv}
Ma, J. and Zhang, R. (2019).
\newblock Automatic calcium scoring in cardiac and chest {CT} using
  {DenseRAUnet}.
\newblock \emph{Arxiv Preprint} abs/1907.11392.
\newblock Available at \url{http://arxiv.org/abs/1907.11392} (Accessed
  September 1, 2019)
\bibAnnoteFile{Ma_2019_arXiv}

\bibitem[{Mahapatra et~al.(2018)Mahapatra, Bozorgtabar, Thiran, and
  Reyes}]{Mahapatra_2018_MICCAI}
Mahapatra, D., Bozorgtabar, B., Thiran, J.-P., and Reyes, M. (2018).
\newblock Efficient active learning for image classification and segmentation
  using a sample selection and conditional generative adversarial network.
\newblock In \emph{Medical Image Computing and Computer Assisted Intervention}
  (Springer International Publishing), 580--588.
\newblock \doi{10.1007/978-3-030-00934-2\_65}
\bibAnnoteFile{Mahapatra_2018_MICCAI}

\bibitem[{Mazurowski et~al.(2019)Mazurowski, Buda, Saha, and
  Bashir}]{Mazurowski_2019_JMRI}
Mazurowski, M.~A., Buda, M., Saha, A., and Bashir, M.~R. (2019).
\newblock Deep learning in radiology: An overview of the concepts and a survey
  of the state of the art with focus on {MRI}.
\newblock \emph{Journal of magnetic resonance imaging} 49, 939--954.
\newblock \doi{10.1002/jmri.26534}
\bibAnnoteFile{Mazurowski_2019_JMRI}

\bibitem[{Medley et~al.(2019)Medley, Santiago, and
  Nascimento}]{Medley_2019_ISBI}
Medley, D.~O., Santiago, C., and Nascimento, J.~C. (2019).
\newblock Segmenting the left ventricle in cardiac in cardiac {MRI}: From
  handcrafted to deep region based descriptors.
\newblock In \emph{2019 {IEEE} 16th International Symposium on Biomedical
  Imaging ({ISBI} 2019)}. 644--648.
\newblock \doi{10.1109/ISBI.2019.8759179}
\bibAnnoteFile{Medley_2019_ISBI}

\bibitem[{Meng et~al.(2019)Meng, Sinclair, Zimmer, Hou, Rajchl, Toussaint
  et~al.}]{Meng_2019_TMI}
Meng, Q., Sinclair, M., Zimmer, V., Hou, B., Rajchl, M., Toussaint, N., et~al.
  (2019).
\newblock Weakly supervised estimation of shadow confidence maps in fetal
  ultrasound imaging.
\newblock \emph{IEEE Transactions on Medical Imaging}
\bibAnnoteFile{Meng_2019_TMI}

\bibitem[{Merkow et~al.(2016)Merkow, Marsden, Kriegman, and
  Tu}]{Merkow_2016_MICCAI}
Merkow, J., Marsden, A., Kriegman, D., and Tu, Z. (2016).
\newblock Dense volume-to-volume vascular boundary detection.
\newblock In \emph{Medical Image Computing and Computer Assisted Intervention}
  (Springer), 371--379
\bibAnnoteFile{Merkow_2016_MICCAI}

\bibitem[{Milletari et~al.(2016)Milletari, Navab, and
  Ahmadi}]{Milletari_2016_3DV}
Milletari, F., Navab, N., and Ahmadi, S. (2016).
\newblock {V-Net}: Fully convolutional neural networks for volumetric medical
  image segmentation.
\newblock In \emph{2016 Fourth International Conference on {3D} Vision
  ({3DV})}. 565--571.
\newblock \doi{10.1109/3DV.2016.79}
\bibAnnoteFile{Milletari_2016_3DV}

\bibitem[{Moccia et~al.(2019)Moccia, Banali, Martini, Muscogiuri, Pontone, Pepi
  et~al.}]{Moccia_2019_Magma}
Moccia, S., Banali, R., Martini, C., Muscogiuri, G., Pontone, G., Pepi, M.,
  et~al. (2019).
\newblock Development and testing of a deep learning-based strategy for scar
  segmentation on {CMR-LGE} images.
\newblock \emph{Magnetic Resonance Materials in Physics, Biology and Medicine}
  32, 187--195.
\newblock \doi{10.1007/s10334-018-0718-4}
\bibAnnoteFile{Moccia_2019_Magma}

\bibitem[{Moeskops et~al.(2016)Moeskops, Wolterink, van~der Velden, Gilhuijs,
  Leiner, Viergever et~al.}]{Moeskops_2016_MICCAI}
Moeskops, P., Wolterink, J.~M., van~der Velden, B.~H., Gilhuijs, K.~G., Leiner,
  T., Viergever, M.~A., et~al. (2016).
\newblock Deep learning for multi-task medical image segmentation in multiple
  modalities.
\newblock In \emph{Medical Image Computing and Computer Assisted Intervention}
  (Springer), 478--486
\bibAnnoteFile{Moeskops_2016_MICCAI}

\bibitem[{Mortazi et~al.(2017{\natexlab{a}})Mortazi, Burt, and
  Bagci}]{Mortazi_2017_MMWHS}
Mortazi, A., Burt, J., and Bagci, U. (2017{\natexlab{a}}).
\newblock Multi-planar deep segmentation networks for cardiac substructures
  from {MRI} and {CT}.
\newblock In \emph{International Workshop on Statistical Atlases and
  Computational Models of the Heart} (Springer), 199--206
\bibAnnoteFile{Mortazi_2017_MMWHS}

\bibitem[{Mortazi et~al.(2017{\natexlab{b}})Mortazi, Karim, Rhode, Burt, and
  Bagci}]{Mortazi_2017_STACOM}
Mortazi, A., Karim, R., Rhode, K., Burt, J., and Bagci, U.
  (2017{\natexlab{b}}).
\newblock {CardiacNET}: Segmentation of left atrium and proximal pulmonary
  veins from {MRI} using multi-view {CNN}.
\newblock In \emph{Lecture Notes in Computer Science (including subseries
  Lecture Notes in Artificial Intelligence and Lecture Notes in
  Bioinformatics)}. 377--385.
\newblock \doi{10.1007/978-3-319-66185-8\_43}
\bibAnnoteFile{Mortazi_2017_STACOM}

\bibitem[{Nascimento and Carneiro(2014)}]{Nascimento_2014_CVPR}
Nascimento, J.~C. and Carneiro, G. (2014).
\newblock Non-rigid segmentation using sparse low dimensional manifolds and
  deep belief networks.
\newblock In \emph{Proceedings of the {IEEE} Conference on Computer Vision and
  Pattern Recognition} (cv-foundation.org), 288--295
\bibAnnoteFile{Nascimento_2014_CVPR}

\bibitem[{Nascimento and Carneiro(2017)}]{Nascimento_2017_TIP}
Nascimento, J.~C. and Carneiro, G. (2017).
\newblock Deep learning on sparse manifolds for faster object segmentation.
\newblock \emph{IEEE Transactions on Image Processing} 26, 4978--4990
\bibAnnoteFile{Nascimento_2017_TIP}

\bibitem[{Nascimento and Carneiro(2019)}]{Nascimento_2019_TPAMI}
Nascimento, J.~C. and Carneiro, G. (2019).
\newblock One shot segmentation: unifying rigid detection and non-rigid
  segmentation using elastic regularization.
\newblock \emph{IEEE Trans. Pattern Anal. Mach. Intell.}
  \doi{10.1109/TPAMI.2019.2922959}
\bibAnnoteFile{Nascimento_2019_TPAMI}

\bibitem[{Ngo et~al.(2017)Ngo, Lu, and Carneiro}]{Ngo_2016_MedIA}
Ngo, T.~A., Lu, Z., and Carneiro, G. (2017).
\newblock Combining deep learning and level set for the automated segmentation
  of the left ventricle of the heart from cardiac cine magnetic resonance.
\newblock \emph{Medical Image Analysis} 35, 159--171
\bibAnnoteFile{Ngo_2016_MedIA}

\bibitem[{Noble and Boukerroui(2006)}]{Noble_2006_TMI}
Noble, J.~A. and Boukerroui, D. (2006).
\newblock Ultrasound image segmentation: a survey.
\newblock \emph{IEEE Transactions on Medical Imaging} 25, 987--1010.
\newblock \doi{10.1109/TMI.2006.877092}
\bibAnnoteFile{Noble_2006_TMI}

\bibitem[{Oksuz et~al.(2019)Oksuz, Clough, Bai, Ruijsink, Puyol-Ant{\'o}n, Cruz
  et~al.}]{Ilkay_2019_MIDL}
Oksuz, I., Clough, J., Bai, W., Ruijsink, B., Puyol-Ant{\'o}n, E., Cruz, G.,
  et~al. (2019).
\newblock High-quality segmentation of low quality cardiac {MR} images using
  k-space artefact correction.
\newblock In \emph{Medical Imaging with Deep Learning}, eds. M.~J. Cardoso,
  A.~Feragen, B.~Glocker, E.~Konukoglu, I.~Oguz, G.~Unal, and T.~Vercauteren
  (London, United Kingdom: PMLR), vol. 102 of \emph{Proceedings of Machine
  Learning Research}, 380--389
\bibAnnoteFile{Ilkay_2019_MIDL}

\bibitem[{Oktay et~al.(2016)Oktay, Bai, Lee, Guerrero, Kamnitsas, Caballero
  et~al.}]{Oktay_2016_MICCAI}
Oktay, O., Bai, W., Lee, M., Guerrero, R., Kamnitsas, K., Caballero, J., et~al.
  (2016).
\newblock Multi-input cardiac image super-resolution using convolutional neural
  networks.
\newblock In \emph{Medical Image Computing and Computer Assisted Intervention}.
  vol. 9902 LNCS, 246--254.
\newblock \doi{10.1007/978-3-319-46726-9\_29}
\bibAnnoteFile{Oktay_2016_MICCAI}

\bibitem[{Oktay et~al.(2018{\natexlab{a}})Oktay, Ferrante, Kamnitsas, Heinrich,
  Bai, Caballero et~al.}]{Oktay_2018_TMI}
Oktay, O., Ferrante, E., Kamnitsas, K., Heinrich, M., Bai, W., Caballero, J.,
  et~al. (2018{\natexlab{a}}).
\newblock Anatomically constrained neural networks (acnns): Application to
  cardiac image enhancement and segmentation.
\newblock \emph{IEEE Transactions on Medical Imaging} 37, 384--395
\bibAnnoteFile{Oktay_2018_TMI}

\bibitem[{Oktay et~al.(2018{\natexlab{b}})Oktay, Schlemper, Folgoc, Lee,
  Heinrich, Misawa et~al.}]{Oktay_2018_MIDL}
Oktay, O., Schlemper, J., Folgoc, L.~L., Lee, M., Heinrich, M., Misawa, K.,
  et~al. (2018{\natexlab{b}}).
\newblock Attention {U-Net}: Learning where to look for the pancreas.
\newblock In \emph{Medical Imaging with Deep Learning}. 1804.03999
\bibAnnoteFile{Oktay_2018_MIDL}

\bibitem[{Ouyang et~al.(2019)Ouyang, Kamnitsas, Biffi, Duan, and
  Rueckert}]{Ouyang_2019_MICCAI}
Ouyang, C., Kamnitsas, K., Biffi, C., Duan, J., and Rueckert, D. (2019).
\newblock Data efficient unsupervised domain adaptation for {Cross-Modality}
  image segmentation.
\newblock In \emph{Medical Image Computing and Computer Assisted Intervention}.
  669--677
\bibAnnoteFile{Ouyang_2019_MICCAI}

\bibitem[{Pace et~al.(2015)Pace, Dalca, Geva, Powell, Moghari, and
  Golland}]{Pace_2015_HVSMR}
Pace, D.~F., Dalca, A.~V., Geva, T., Powell, A.~J., Moghari, M.~H., and
  Golland, P. (2015).
\newblock Interactive {Whole-Heart} segmentation in congenital heart disease.
\newblock \emph{Medical image computing and computer-assisted intervention:
  MICCAI ... International Conference on Medical Image Computing and
  Computer-Assisted Intervention} 9351, 80--88.
\newblock \doi{10.1007/978-3-319-24574-4\_10}.
\newblock \url{http://segchd.csail.mit.edu/}
\bibAnnoteFile{Pace_2015_HVSMR}

\bibitem[{Painchaud et~al.(2019)Painchaud, Skandarani, Judge, Bernard, Lalande,
  and Jodoin}]{Painchaud_2019_MICCAI}
Painchaud, N., Skandarani, Y., Judge, T., Bernard, O., Lalande, A., and Jodoin,
  P.-M. (2019).
\newblock Cardiac {MRI} segmentation with strong anatomical guarantees.
\newblock In \emph{Medical Image Computing and Computer Assisted Intervention}.
  632--640
\bibAnnoteFile{Painchaud_2019_MICCAI}

\bibitem[{Papernot(2018)}]{Papernot_2018_Arxiv}
Papernot, N. (2018).
\newblock A marauder's map of security and privacy in machine learning: An
  overview of current and future research directions for making machine
  learning secure and private.
\newblock In \emph{the 11th {ACM} Workshop on Artificial Intelligence and
  Security, {CCS} 2018, Toronto, ON, Canada, October 19, 2018}. 1.
\newblock \doi{10.1145/3270101.3270102}
\bibAnnoteFile{Papernot_2018_Arxiv}

\bibitem[{Patravali et~al.(2017)Patravali, Jain, and
  Chilamkurthy}]{Patravali_2017_STACOM}
Patravali, J., Jain, S., and Chilamkurthy, S. (2017).
\newblock 2d-3d fully convolutional neural networks for cardiac mr
  segmentation.
\newblock In \emph{Statistical Atlases and Computational Models of the Heart.
  {ACDC} and {MMWHS} Challenges - 8th International Workshop, {STACOM} 2017,
  Held in Conjunction with {MICCAI} 2017, Quebec City, Canada, September 10-14,
  2017, Revised Selected Papers} (Springer), 130--139
\bibAnnoteFile{Patravali_2017_STACOM}

\bibitem[{Payer et~al.(2018)Payer, {\v S}tern, Bischof, and
  Urschler}]{Payer_2018_MMWHS}
Payer, C., {\v S}tern, D., Bischof, H., and Urschler, M. (2018).
\newblock Multi-label whole heart segmentation using {CNNs} and anatomical
  label configurations.
\newblock In \emph{International Workshop on Statistical Atlases and
  Computational Models of the Heart} (Springer International Publishing),
  190--198
\bibAnnoteFile{Payer_2018_MMWHS}

\bibitem[{Peng and Zhang(2012)}]{Peng_2012_ECCV}
Peng, B. and Zhang, L. (2012).
\newblock Evaluation of image segmentation quality by adaptive ground truth
  composition.
\newblock In \emph{European Conference on Computer Vision} (Springer Berlin
  Heidelberg), 287--300.
\newblock \doi{10.1007/978-3-642-33712-3\_21}
\bibAnnoteFile{Peng_2012_ECCV}

\bibitem[{Peng et~al.(2016)Peng, Lekadir, Gooya, Shao, Petersen, and
  Frangi}]{Peng_2016_MAGMA}
Peng, P., Lekadir, K., Gooya, A., Shao, L., Petersen, S.~E., and Frangi, A.~F.
  (2016).
\newblock A review of heart chamber segmentation for structural and functional
  analysis using cardiac magnetic resonance imaging.
\newblock \emph{Magnetic Resonance Materials in Physics, Biology and Medicine}
  29, 155—195.
\newblock \doi{10.1007/s10334-015-0521-4}
\bibAnnoteFile{Peng_2016_MAGMA}

\bibitem[{Petitjean et~al.(2015)Petitjean, Zuluaga, Bai, Dacher, Grosgeorge,
  Caudron et~al.}]{Petitjean_2015_MedIA}
Petitjean, C., Zuluaga, M.~A., Bai, W., Dacher, J.-N., Grosgeorge, D., Caudron,
  J., et~al. (2015).
\newblock Right ventricle segmentation from cardiac {MRI}: a collation study.
\newblock \emph{Medical image analysis} 19, 187--202.
\newblock \doi{10.1016/j.media.2014.10.004}
\bibAnnoteFile{Petitjean_2015_MedIA}

\bibitem[{Poudel et~al.(2016)Poudel, Lamata, and Montana}]{Poudel_2016_HVSCMR}
Poudel, R. P.~K., Lamata, P., and Montana, G. (2016).
\newblock Recurrent fully convolutional neural networks for multi-slice {MRI}
  cardiac segmentation.
\newblock In \emph{1st International Workshops on Reconstruction and Analysis
  of Moving Body Organs, RAMBO 2016 and 1st International Workshops on
  Whole-Heart and Great Vessel Segmentation from 3D Cardiovascular MRI in
  Congenital Heart Disease, HVSMR 2016}. 83--94
\bibAnnoteFile{Poudel_2016_HVSCMR}

\bibitem[{Preetha et~al.(2018)Preetha, Haridasan, Abdi, and
  Engelhardt}]{Preetha_2018_STACOM}
Preetha, C.~J., Haridasan, S., Abdi, V., and Engelhardt, S. (2018).
\newblock Segmentation of the left atrium from {3D} {Gadolinium-Enhanced} {MR}
  images with convolutional neural networks.
\newblock In \emph{Statistical Atlases and Computational Models of the Heart.
  Atrial Segmentation and {LV} Quantification Challenges} (Springer
  International Publishing), 265--272.
\newblock \doi{10.1007/978-3-030-12029-0\_29}
\bibAnnoteFile{Preetha_2018_STACOM}

\bibitem[{Qin et~al.(2018{\natexlab{a}})Qin, Bai, Schlemper, Petersen,
  Piechnik, Neubauer et~al.}]{Qin_2018_MICCAI}
Qin, C., Bai, W., Schlemper, J., Petersen, S.~E., Piechnik, S.~K., Neubauer,
  S., et~al. (2018{\natexlab{a}}).
\newblock Joint learning of motion estimation and segmentation for cardiac mr
  image sequences.
\newblock In \emph{Medical Image Computing and Computer Assisted Intervention}.
  472--480
\bibAnnoteFile{Qin_2018_MICCAI}

\bibitem[{Qin et~al.(2018{\natexlab{b}})Qin, Bai, Schlemper, Petersen,
  Piechnik, Neubauer et~al.}]{Qin_2018_MLMLR}
Qin, C., Bai, W., Schlemper, J., Petersen, S.~E., Piechnik, S.~K., Neubauer,
  S., et~al. (2018{\natexlab{b}}).
\newblock Joint motion estimation and segmentation from undersampled cardiac
  {MR} image.
\newblock In \emph{Machine Learning for Medical Image Reconstruction} (Springer
  International Publishing), 55--63.
\newblock \doi{10.1007/978-3-030-00129-2\_7}
\bibAnnoteFile{Qin_2018_MLMLR}

\bibitem[{Radau and {Others}(2009)}]{Radau_2009_Sunnybrook}
Radau, P. and {Others} (2009).
\newblock Evaluation framework for algorithms segmenting short axis cardiac
  {MRI}.
\newblock \emph{The MIDAS Journal - Cardiac MR Left Ventricle Segmentation
  Challenge} \url{http://www.cardiacatlas.org/studies/sunnybrook-cardiac-data/}
\bibAnnoteFile{Radau_2009_Sunnybrook}

\bibitem[{Robinson et~al.(2019)Robinson, Valindria, Bai, Oktay, Kainz, Suzuki
  et~al.}]{Robinson_2019_JCMR}
Robinson, R., Valindria, V.~V., Bai, W., Oktay, O., Kainz, B., Suzuki, H.,
  et~al. (2019).
\newblock Automated quality control in image segmentation: application to the
  {UK} biobank cardiovascular magnetic resonance imaging study.
\newblock \emph{Journal of Cardiovascular Magnetic Resonance} 21, 18.
\newblock \doi{10.1186/s12968-019-0523-x}
\bibAnnoteFile{Robinson_2019_JCMR}

\bibitem[{Roh{\'e} et~al.(2017)Roh{\'e}, Sermesant, and
  Pennec}]{Rohe_2017_STACOM}
Roh{\'e}, M.-M., Sermesant, M., and Pennec, X. (2017).
\newblock Automatic {Multi-Atlas} segmentation of myocardium with {SVF-Net}.
\newblock In \emph{Statistical Atlases and Computational Models of the Heart.
  {ACDC} and {MMWHS} Challenges} (Springer International Publishing), 170--177.
\newblock \doi{10.1007/978-3-319-75541-0\_18}
\bibAnnoteFile{Rohe_2017_STACOM}

\bibitem[{Ronneberger and Brox(2015)}]{Ronneberger_2015_MICCAI}
Ronneberger, F.~P., Olaf and Brox, T. (2015).
\newblock {U-Net}: Convolutional networks for biomedical image segmentation.
\newblock In \emph{Medical Image Computing and Computer Assisted Intervention}
  (Springer), 234--241
\bibAnnoteFile{Ronneberger_2015_MICCAI}

\bibitem[{Rueckert et~al.(1999)Rueckert, Sonoda, Hayes, Hill, Leach, and
  Hawkes}]{Rueckert_1999_TMI}
Rueckert, D., Sonoda, L.~I., Hayes, C., Hill, D.~L., Leach, M.~O., and Hawkes,
  D.~J. (1999).
\newblock Nonrigid registration using free-form deformations: application to
  breast {MR} images.
\newblock \emph{IEEE Transactions on Medical Imaging} 18, 712--721.
\newblock \doi{10.1109/42.796284}
\bibAnnoteFile{Rueckert_1999_TMI}

\bibitem[{Ruijsink et~al.(2019)Ruijsink, Puyol-Ant{\'o}n, Oksuz, Sinclair, Bai,
  Schnabel et~al.}]{Ruijsink_2019_JACC}
Ruijsink, B., Puyol-Ant{\'o}n, E., Oksuz, I., Sinclair, M., Bai, W., Schnabel,
  J.~A., et~al. (2019).
\newblock Fully automated, {Quality-Controlled} cardiac analysis from {CMR}:
  Validation and {Large-Scale} application to characterize cardiac function.
\newblock \emph{Journal of the American College of Cardiology}
  \doi{10.1016/j.jcmg.2019.05.030}
\bibAnnoteFile{Ruijsink_2019_JACC}

\bibitem[{Ryffel et~al.(2018)Ryffel, Trask, Dahl, Wagner, Mancuso, Rueckert
  et~al.}]{Ryffel_2018_PPML}
Ryffel, T., Trask, A., Dahl, M., Wagner, B., Mancuso, J., Rueckert, D., et~al.
  (2018).
\newblock A generic framework for privacy preserving deep learning.
\newblock In \emph{Privacy preserving machine learning}. 1--8
\bibAnnoteFile{Ryffel_2018_PPML}

\bibitem[{Sander et~al.(2019)Sander, de~Vos, Wolterink, and I{\v
  s}gum}]{Sander_2019_MIP}
Sander, J., de~Vos, B.~D., Wolterink, J.~M., and I{\v s}gum, I. (2019).
\newblock Towards increased trustworthiness of deep learning segmentation
  methods on cardiac {MRI}.
\newblock In \emph{Medical Imaging 2019: Image Processing} (International
  Society for Optics and Photonics), vol. 10949, 1094919.
\newblock \doi{10.1117/12.2511699}
\bibAnnoteFile{Sander_2019_MIP}

\bibitem[{Santini et~al.(2017)Santini, Della~Latta, Martini, Valvano, Gori,
  Ripoli et~al.}]{Santini_2017_IFMBE}
Santini, G., Della~Latta, D., Martini, N., Valvano, G., Gori, A., Ripoli, A.,
  et~al. (2017).
\newblock An automatic deep learning approach for coronary artery calcium
  segmentation.
\newblock In \emph{International Federation for Medical and Biological
  Engineering}. vol.~65, 374--377
\bibAnnoteFile{Santini_2017_IFMBE}

\bibitem[{Savioli et~al.(2018)Savioli, Montana, and
  Lamata}]{Savioli_2018_STACOM}
Savioli, N., Montana, G., and Lamata, P. (2018).
\newblock {V-FCNN}: Volumetric fully convolution neural network for automatic
  atrial segmentation.
\newblock In \emph{Statistical Atlases and Computational Models of the Heart.
  Atrial Segmentation and {LV} Quantification Challenges} (Springer
  International Publishing), 273--281.
\newblock \doi{10.1007/978-3-030-12029-0\_30}
\bibAnnoteFile{Savioli_2018_STACOM}

\bibitem[{{Savioli} et~al.(2018){Savioli}, {Vieira}, {Lamata}, and
  {Montana}}]{Savioli_2018_SNAMS}
{Savioli}, N., {Vieira}, M.~S., {Lamata}, P., and {Montana}, G. (2018).
\newblock Automated segmentation on the entire cardiac cycle using a deep
  learning work - flow.
\newblock In \emph{2018 Fifth International Conference on Social Networks
  Analysis, Management and Security (SNAMS)}. 153--158.
\newblock \doi{10.1109/SNAMS.2018.8554962}
\bibAnnoteFile{Savioli_2018_SNAMS}

\bibitem[{Savioli et~al.(2018)Savioli, Vieira, Lamata, and
  Montana}]{Savioli_2018_Arxiv}
Savioli, N., Vieira, M.~S., Lamata, P., and Montana, G. (2018).
\newblock A generative adversarial model for right ventricle segmentation.
\newblock \emph{Arxiv Preprint} abs/1810.03969.
\newblock Available at \url{http://arxiv.org/abs/1810.03969} (Accessed
  September 1, 2019)
\bibAnnoteFile{Savioli_2018_Arxiv}

\bibitem[{Schaap et~al.(2009)Schaap, Metz, van Walsum, van~der Giessen,
  Weustink, Mollet et~al.}]{Schaap_2009_centerline_dataset}
Schaap, M., Metz, C.~T., van Walsum, T., van~der Giessen, A.~G., Weustink,
  A.~C., Mollet, N.~R., et~al. (2009).
\newblock Standardized evaluation methodology and reference database for
  evaluating coronary artery centerline extraction algorithms.
\newblock \emph{Medical image analysis} 13, 701--714.
\newblock \doi{10.1016/j.media.2009.06.003}.
\newblock \url{:http://coronary.bigr.nl/centerlines/}
\bibAnnoteFile{Schaap_2009_centerline_dataset}

\bibitem[{Schlemper et~al.(2018)Schlemper, Oktay, Bai, Castro, Duan, Qin
  et~al.}]{Schlemper_2018_MICCAI}
Schlemper, J., Oktay, O., Bai, W., Castro, D.~C., Duan, J., Qin, C., et~al.
  (2018).
\newblock Cardiac {MR} segmentation from undersampled k-space using deep latent
  representation learning.
\newblock In \emph{Medical Image Computing and Computer Assisted Intervention}
  (Springer International Publishing), 259--267.
\newblock \doi{10.1007/978-3-030-00928-1\_30}
\bibAnnoteFile{Schlemper_2018_MICCAI}

\bibitem[{Shadmi et~al.(2018)Shadmi, Mazo, Bregman-Amitai, and
  Elnekave}]{Shadmi_2018_ISBI}
Shadmi, R., Mazo, V., Bregman-Amitai, O., and Elnekave, E. (2018).
\newblock Fully-convolutional deep-learning based system for coronary calcium
  score prediction from non-contrast chest {CT}.
\newblock In \emph{International Symposium on Biomedical Imaging} (IEEE),
  24--28
\bibAnnoteFile{Shadmi_2018_ISBI}

\bibitem[{Shelhamer et~al.(2017)Shelhamer, Long, and
  Darrell}]{Shelhamer_2017_TPAMI}
Shelhamer, E., Long, J., and Darrell, T. (2017).
\newblock Fully convolutional networks for semantic segmentation.
\newblock \emph{IEEE transactions on pattern analysis and machine intelligence}
  39, 640--651.
\newblock \doi{10.1109/TPAMI.2016.2572683}
\bibAnnoteFile{Shelhamer_2017_TPAMI}

\bibitem[{Shen et~al.(2017)Shen, Wu, and Suk}]{Shen_2017_Review}
Shen, D., Wu, G., and Suk, H.-I. (2017).
\newblock Deep learning in medical image analysis.
\newblock \emph{Annual review of biomedical engineering} 19, 221--248
\bibAnnoteFile{Shen_2017_Review}

\bibitem[{Shen et~al.(2019)Shen, Fang, Gao, Xiong, Zhong, and
  Tang}]{Shen_2019_IEEEAccess}
Shen, Y., Fang, Z., Gao, Y., Xiong, N., Zhong, C., and Tang, X. (2019).
\newblock Coronary arteries segmentation based on {3D FCN} with attention gate
  and level set function.
\newblock \emph{IEEE Access} 7, 42826--42835
\bibAnnoteFile{Shen_2019_IEEEAccess}

\bibitem[{Shi et~al.(2018)Shi, Zeng, Zhang, Zhuang, Li, Yang
  et~al.}]{Shi_2018_MICCAI}
Shi, Z., Zeng, G., Zhang, L., Zhuang, X., Li, L., Yang, G., et~al. (2018).
\newblock Bayesian {VoxDRN}: A probabilistic deep voxelwise dilated residual
  network for whole heart segmentation from {3D} {MR} images.
\newblock In \emph{Medical Image Computing and Computer Assisted Intervention}
  (Springer International Publishing), 569--577.
\newblock \doi{10.1007/978-3-030-00937-3\_65}
\bibAnnoteFile{Shi_2018_MICCAI}

\bibitem[{Simonyan and Zisserman(2015)}]{Simonyan_2015_ICLR}
Simonyan, K. and Zisserman, A. (2015).
\newblock Very deep convolutional networks for large-scale image recognition.
\newblock In \emph{International Conference on Learning Representations}. 14.
\newblock \doi{10.1016/j.infsof.2008.09.005}
\bibAnnoteFile{Simonyan_2015_ICLR}

\bibitem[{Smistad and Lindseth(2014)}]{Smistad_2014_MIDAS}
Smistad, E. and Lindseth, F. (2014).
\newblock Real-time tracking of the left ventricle in {3D} ultrasound using
  kalman filter and mean value coordinates.
\newblock \emph{Medical Image Segmentation for Improved Surgical Navigation} ,
  189
\bibAnnoteFile{Smistad_2014_MIDAS}

\bibitem[{Smistad et~al.(2017)Smistad, Ostvik, Haugen, and
  Lovstakken}]{Smistad_2017_IUS}
Smistad, E., Ostvik, A., Haugen, B.~O., and Lovstakken, L. (2017).
\newblock {2D} left ventricle segmentation using deep learning.
\newblock In \emph{2017 {IEEE} International Ultrasonics Symposium ({IUS})}
  (IEEE), 1--4
\bibAnnoteFile{Smistad_2017_IUS}

\bibitem[{Smistad et~al.(2018)Smistad, {\O}stvik, Mjal~Salte, Leclerc, Bernard,
  and Lovstakken}]{Smistad_2018_IUS}
Smistad, E., {\O}stvik, A., Mjal~Salte, I., Leclerc, S., Bernard, O., and
  Lovstakken, L. (2018).
\newblock Fully automatic {Real-Time} ejection fraction and {MAPSE}
  measurements in {2D} echocardiography using deep neural networks.
\newblock In \emph{2018 {IEEE} International Ultrasonics Symposium ({IUS})}
  (IEEE), 1--4
\bibAnnoteFile{Smistad_2018_IUS}

\bibitem[{Srivastava et~al.(2014)Srivastava, Hinton, Krizhevsky, Sutskever, and
  Salakhutdinov}]{Srivastava_2014_JMLR}
Srivastava, N., Hinton, G., Krizhevsky, A., Sutskever, I., and Salakhutdinov,
  R. (2014).
\newblock Dropout: A simple way to prevent neural networks from overfitting.
\newblock \emph{Journal of machine learning research: JMLR} 15, 1929--1958
\bibAnnoteFile{Srivastava_2014_JMLR}

\bibitem[{Suinesiaputra et~al.(2014)Suinesiaputra, Cowan, Al-Agamy, Elattar,
  Ayache, Fahmy et~al.}]{Suinesiaputra_LVSC_2011}
Suinesiaputra, A., Cowan, B.~R., Al-Agamy, A.~O., Elattar, M.~A., Ayache, N.,
  Fahmy, A.~S., et~al. (2014).
\newblock A collaborative resource to build consensus for automated left
  ventricular segmentation of cardiac {MR} images.
\newblock \emph{Medical image analysis} 18, 50--62.
\newblock \doi{10.1016/j.media.2013.09.001}.
\newblock
  \url{http://www.cardiacatlas.org/challenges/lv-segmentation-challenge/}
\bibAnnoteFile{Suinesiaputra_LVSC_2011}

\bibitem[{Szegedy et~al.(2015)Szegedy, Liu, Jia, Sermanet, Reed, Anguelov
  et~al.}]{Szegedy_2015_CVPR}
Szegedy, C., Liu, W., Jia, Y., Sermanet, P., Reed, S., Anguelov, D., et~al.
  (2015).
\newblock Going deeper with convolutions.
\newblock In \emph{Conference on Computer Vision and Pattern Recognition}. 1--9
\bibAnnoteFile{Szegedy_2015_CVPR}

\bibitem[{Szegedy et~al.(2014)Szegedy, Zaremba, Sutskever, Bruna, Erhan,
  Goodfellow et~al.}]{Szegedy_2013_Arxiv}
Szegedy, C., Zaremba, W., Sutskever, I., Bruna, J., Erhan, D., Goodfellow, I.,
  et~al. (2014).
\newblock Intriguing properties of neural networks.
\newblock In \emph{2nd International Conference on Learning Representations,
  {ICLR} 2014, Banff, AB, Canada, April 14-16, 2014, Conference Track
  Proceedings}. 1--10
\bibAnnoteFile{Szegedy_2013_Arxiv}

\bibitem[{Tan et~al.(2017)Tan, Liew, Lim, and McLaughlin}]{Tan_2017_MedIA}
Tan, L.~K., Liew, Y.~M., Lim, E., and McLaughlin, R.~A. (2017).
\newblock Convolutional neural network regression for short-axis left ventricle
  segmentation in cardiac cine mr sequences.
\newblock \emph{Medical Image Analysis} 39, 78--86
\bibAnnoteFile{Tan_2017_MedIA}

\bibitem[{Tao et~al.(2016)Tao, Ipek, Shahzad, Berendsen, Nazarian, and van~der
  Geest}]{Tao_2016_JMRI}
Tao, Q., Ipek, E.~G., Shahzad, R., Berendsen, F.~F., Nazarian, S., and van~der
  Geest, R.~J. (2016).
\newblock Fully automatic segmentation of left atrium and pulmonary veins in
  late gadolinium-enhanced {MRI}: Towards objective atrial scar assessment.
\newblock \emph{Journal of magnetic resonance imaging} 44, 346--354.
\newblock \doi{10.1002/jmri.25148}
\bibAnnoteFile{Tao_2016_JMRI}

\bibitem[{Tao et~al.(2019)Tao, Yan, Wang, Paiman, Shamonin, Garg
  et~al.}]{Tao_2019_Radiology}
Tao, Q., Yan, W., Wang, Y., Paiman, E. H.~M., Shamonin, D.~P., Garg, P., et~al.
  (2019).
\newblock Deep learning--based method for fully automatic quantification of
  left ventricle function from cine {MR} images: A multivendor, multicenter
  study.
\newblock \emph{Radiology} 290, 180513.
\newblock \doi{10.1148/radiol.2018180513}
\bibAnnoteFile{Tao_2019_Radiology}

\bibitem[{Tarroni et~al.(2019)Tarroni, Oktay, Bai, Schuh, Suzuki,
  Passerat-Palmbach et~al.}]{Tarroni_2019_TMI}
Tarroni, G., Oktay, O., Bai, W., Schuh, A., Suzuki, H., Passerat-Palmbach, J.,
  et~al. (2019).
\newblock {Learning-Based} quality control for cardiac {MR} images.
\newblock \emph{IEEE Transactions on Medical Imaging} 38, 1127--1138.
\newblock \doi{10.1109/TMI.2018.2878509}
\bibAnnoteFile{Tarroni_2019_TMI}

\bibitem[{Tarroni et~al.(2018)Tarroni, Oktay, Sinclair, Bai, Schuh, Suzuki
  et~al.}]{Tarroni_2018_MICCAI}
Tarroni, G., Oktay, O., Sinclair, M., Bai, W., Schuh, A., Suzuki, H., et~al.
  (2018).
\newblock A comprehensive approach for learning-based fully-automated
  inter-slice motion correction for short-axis cine cardiac {MR} image stacks.
\newblock In \emph{Medical Image Computing and Computer Assisted Intervention}.
  268--276
\bibAnnoteFile{Tarroni_2018_MICCAI}

\bibitem[{Tavakoli and Amini(2013)}]{Tavakoli_2013_CVIU}
Tavakoli, V. and Amini, A.~A. (2013).
\newblock A survey of shaped-based registration and segmentation techniques for
  cardiac images.
\newblock \emph{Computer Vision and Image Understanding} 117, 966--989.
\newblock \doi{10.1016/j.cviu.2012.11.017}
\bibAnnoteFile{Tavakoli_2013_CVIU}

\bibitem[{Tobon-Gomez et~al.(2015)Tobon-Gomez, Geers, Peters, Weese, Pinto,
  Karim et~al.}]{Tobon-Gomez_2015_LASC}
Tobon-Gomez, C., Geers, A.~J., Peters, J., Weese, J., Pinto, K., Karim, R.,
  et~al. (2015).
\newblock Benchmark for algorithms segmenting the left atrium from {3D} {CT}
  and {MRI} datasets.
\newblock \emph{IEEE transactions on medical imaging} 34, 1460--1473.
\newblock \doi{10.1109/TMI.2015.2398818}.
\newblock
  \url{www.cardiacatlas.org/challenges/left-atrium-segmentation-challenge/}
\bibAnnoteFile{Tobon-Gomez_2015_LASC}

\bibitem[{Tong et~al.(2017)Tong, Ning, Si, Liao, and Qin}]{Tong_2017_MMWHS}
Tong, Q., Ning, M., Si, W., Liao, X., and Qin, J. (2017).
\newblock {3D deeply-supervised U-net based whole heart segmentation}.
\newblock In \emph{International Workshop on Statistical Atlases and
  Computational Models of the Heart} (Springer), 224--232
\bibAnnoteFile{Tong_2017_MMWHS}

\bibitem[{Tran(2016)}]{Tran_2016_Arxiv}
Tran, P.~V. (2016).
\newblock A fully convolutional neural network for cardiac segmentation in
  {Short-Axis} {MRI}.
\newblock \emph{Arxiv Preprint} abs/1604.00494.
\newblock Available at \url{http://arxiv.org/abs/1604.00494} (Accessed
  September 1, 2019)
\bibAnnoteFile{Tran_2016_Arxiv}

\bibitem[{Tziritas and Grinias(2017)}]{Tziritas_2017_STACOM}
Tziritas, G. and Grinias, E. (2017).
\newblock Fast fully-automatic localization of left ventricle and myocardium in
  mri using mrf model optimization, substructures tracking and b-spline
  smoothing.
\newblock In \emph{International Workshop on Statistical Atlases and
  Computational Models of the Heart}. 91--100
\bibAnnoteFile{Tziritas_2017_STACOM}

\bibitem[{Van Der~Geest and Reiber(1999)}]{Van_der_Geest_1999_JMRI}
Van Der~Geest, R.~J. and Reiber, J.~H. (1999).
\newblock Quantification in cardiac {MRI}.
\newblock \emph{Journal of magnetic resonance imaging} 10, 602--608
\bibAnnoteFile{Van_der_Geest_1999_JMRI}

\bibitem[{Vaswani et~al.(2017)Vaswani, Shazeer, Parmar, Uszkoreit, Jones, Gomez
  et~al.}]{Vaswani_2017_NIPS}
Vaswani, A., Shazeer, N., Parmar, N., Uszkoreit, J., Jones, L., Gomez, A.~N.,
  et~al. (2017).
\newblock Attention is all you need.
\newblock In \emph{Conference on Neural Information Processing Systems}, eds.
  I.~Guyon, U.~V. Luxburg, S.~Bengio, H.~Wallach, R.~Fergus, S.~Vishwanathan,
  and R.~Garnett (Curran Associates, Inc.), 5998--6008
\bibAnnoteFile{Vaswani_2017_NIPS}

\bibitem[{Veni et~al.(2018)Veni, Moradi, Bulu, Narayan, and
  Syeda-Mahmood}]{Veni_2018_ISBI}
Veni, G., Moradi, M., Bulu, H., Narayan, G., and Syeda-Mahmood, T. (2018).
\newblock Echocardiography segmentation based on a shape-guided deformable
  model driven by a fully convolutional network prior.
\newblock In \emph{2018 {IEEE} 15th International Symposium on Biomedical
  Imaging ({ISBI} 2018)} (IEEE), 898--902
\bibAnnoteFile{Veni_2018_ISBI}

\bibitem[{Vesal et~al.(2018)Vesal, Ravikumar, and Maier}]{Vesal_2018_STACOM}
Vesal, S., Ravikumar, N., and Maier, A. (2018).
\newblock Dilated convolutions in neural networks for left atrial segmentation
  in {3D} gadolinium {Enhanced-MRI}.
\newblock In \emph{Statistical Atlases and Computational Models of the Heart.
  Atrial Segmentation and {LV} Quantification Challenges - 9th International
  Workshop, {STACOM} 2018, Held in Conjunction with {MICCAI} 2018, Granada,
  Spain, September 16, 2018, Revised Selected Papers}. 319--328
\bibAnnoteFile{Vesal_2018_STACOM}

\bibitem[{Vigneault et~al.(2018)Vigneault, Xie, Ho, Bluemke, and
  Noble}]{Vigneault_2018_MedIA}
Vigneault, D.~M., Xie, W., Ho, C.~Y., Bluemke, D.~A., and Noble, J.~A. (2018).
\newblock {$\Omega$-Net} ({Omega-Net)}: Fully automatic, multi-view cardiac
  {MR} detection, orientation, and segmentation with deep neural networks.
\newblock \emph{Medical Image Analysis} 48, 95--106.
\newblock \doi{10.1016/j.media.2018.05.008}
\bibAnnoteFile{Vigneault_2018_MedIA}

\bibitem[{Volpi et~al.(2018)Volpi, Namkoong, Sener, Duchi, Murino, and
  Savarese}]{Volpi_2018_NIPS}
Volpi, R., Namkoong, H., Sener, O., Duchi, J.~C., Murino, V., and Savarese, S.
  (2018).
\newblock Generalizing to unseen domains via adversarial data augmentation.
\newblock In \emph{Conference on Neural Information Processing Systems}.
  5339--5349
\bibAnnoteFile{Volpi_2018_NIPS}

\bibitem[{Wang et~al.(2018)Wang, MacGillivray, Macnaught, Yang, and
  Newby}]{Wang_2018_arXiv}
Wang, C., MacGillivray, T., Macnaught, G., Yang, G., and Newby, D. (2018).
\newblock A two-stage {3D} unet framework for multi-class segmentation on full
  resolution image.
\newblock \emph{Arxiv Preprint} , 1--10Available at
  \url{http://arxiv.org/abs/1804.04341} (Accessed September 1, 2019)
\bibAnnoteFile{Wang_2018_arXiv}

\bibitem[{Wang and Smedby(2017)}]{Wang_2017_MMWHS}
Wang, C. and Smedby, {\"O}. (2017).
\newblock Automatic whole heart segmentation using deep learning and shape
  context.
\newblock In \emph{Statistical Atlases and Computational Models of the Heart.
  {ACDC} and {MMWHS} Challenges - 8th International Workshop, {STACOM} 2017,
  Held in Conjunction with {MICCAI} 2017, Quebec City, Canada, September 10-14,
  2017, Revised Selected Papers} (Springer), 242--249
\bibAnnoteFile{Wang_2017_MMWHS}

\bibitem[{Wolterink et~al.(2016)Wolterink, Leiner, de~Vos, van Hamersvelt,
  Viergever, and I{\v{s}}gum}]{Wolterink_2016_MedIA}
Wolterink, J.~M., Leiner, T., de~Vos, B.~D., van Hamersvelt, R.~W., Viergever,
  M.~A., and I{\v{s}}gum, I. (2016).
\newblock Automatic coronary artery calcium scoring in cardiac {CT} angiography
  using paired convolutional neural networks.
\newblock \emph{Medical Image Analysis} 34, 123--136
\bibAnnoteFile{Wolterink_2016_MedIA}

\bibitem[{Wolterink et~al.(2019{\natexlab{a}})Wolterink, Leiner, and
  I{\v{s}}gum}]{Wolterink_2019_arXiv}
Wolterink, J.~M., Leiner, T., and I{\v{s}}gum, I. (2019{\natexlab{a}}).
\newblock Graph convolutional networks for coronary artery segmentation in
  cardiac {CT} angiography.
\newblock \emph{Arxiv Preprint} abs/1908.05343.
\newblock Available at \url{http://arxiv.org/abs/1908.05343} (Accessed
  September 1, 2019)
\bibAnnoteFile{Wolterink_2019_arXiv}

\bibitem[{Wolterink et~al.(2017{\natexlab{a}})Wolterink, Leiner, Viergever, and
  I{\v s}gum}]{Wolterink_2017_RSAM}
Wolterink, J.~M., Leiner, T., Viergever, M.~A., and I{\v s}gum, I.
  (2017{\natexlab{a}}).
\newblock Dilated convolutional neural networks for cardiovascular {MR}
  segmentation in congenital heart disease.
\newblock In \emph{Reconstruction, Segmentation, and Analysis of Medical
  Images}. 95--102
\bibAnnoteFile{Wolterink_2017_RSAM}

\bibitem[{Wolterink et~al.(2017{\natexlab{b}})Wolterink, Leiner, Viergever, and
  Isgum}]{Wolterink_2017_TMI}
Wolterink, J.~M., Leiner, T., Viergever, M.~A., and Isgum, I.
  (2017{\natexlab{b}}).
\newblock Generative adversarial networks for noise reduction in {Low-Dose}
  {CT}.
\newblock \emph{IEEE Transactions on Medical Imaging} 36, 2536--2545.
\newblock \doi{10.1109/TMI.2017.2708987}
\bibAnnoteFile{Wolterink_2017_TMI}

\bibitem[{Wolterink et~al.(2017{\natexlab{c}})Wolterink, Leiner, Viergever, and
  Išgum}]{Wolterink_2017_STACOM}
Wolterink, J.~M., Leiner, T., Viergever, M.~A., and Išgum, I.
  (2017{\natexlab{c}}).
\newblock Automatic segmentation and disease classification using cardiac cine
  mr images.
\newblock In \emph{Statistical Atlases and Computational Models of the Heart.
  {ACDC} and {MMWHS} Challenges - 8th International Workshop, {STACOM} 2017,
  Held in Conjunction with {MICCAI} 2017, Quebec City, Canada, September 10-14,
  2017, Revised Selected Papers}. vol. 10663 LNCS of \emph{Lecture Notes in
  Computer Science (including subseries Lecture Notes in Artificial
  Intelligence and Lecture Notes in Bioinformatics)}, 101--110
\bibAnnoteFile{Wolterink_2017_STACOM}

\bibitem[{Wolterink et~al.(2019{\natexlab{b}})Wolterink, van Hamersvelt,
  Viergever, Leiner, and I{\v{s}}gum}]{Wolterink_2019_MedIA}
Wolterink, J.~M., van Hamersvelt, R.~W., Viergever, M.~A., Leiner, T., and
  I{\v{s}}gum, I. (2019{\natexlab{b}}).
\newblock Coronary artery centerline extraction in cardiac {CT} angiography
  using a {CNN}-based orientation classifier.
\newblock \emph{Medical Image Analysis} 51, 46--60
\bibAnnoteFile{Wolterink_2019_MedIA}

\bibitem[{Xia et~al.(2018)Xia, Yao, Hu, and Hao}]{Xia_2018_STACOM}
Xia, Q., Yao, Y., Hu, Z., and Hao, A. (2018).
\newblock Automatic {3D} atrial segmentation from {GE-MRIs} using volumetric
  fully convolutional networks.
\newblock In \emph{Statistical Atlases and Computational Models of the Heart.
  Atrial Segmentation and {LV} Quantification Challenges} (Springer
  International Publishing), 211--220.
\newblock \doi{10.1007/978-3-030-12029-0\_23}
\bibAnnoteFile{Xia_2018_STACOM}

\bibitem[{Xiong et~al.(2019)Xiong, Fedorov, Fu, Cheng, Macleod, and
  Zhao}]{Xiong_2019_TMI}
Xiong, Z., Fedorov, V.~V., Fu, X., Cheng, E., Macleod, R., and Zhao, J. (2019).
\newblock Fully automatic left atrium segmentation from late gadolinium
  enhanced magnetic resonance imaging using a dual fully convolutional neural
  network.
\newblock \emph{IEEE Transactions on Medical Imaging} 38, 515--524.
\newblock \doi{10.1109/TMI.2018.2866845}
\bibAnnoteFile{Xiong_2019_TMI}

\bibitem[{Xu et~al.(2018{\natexlab{a}})Xu, Xu, Gao, Zhao, Zhang, Zhang
  et~al.}]{Xu_2018_MedIA}
Xu, C., Xu, L., Gao, Z., Zhao, S., Zhang, H., Zhang, Y., et~al.
  (2018{\natexlab{a}}).
\newblock Direct delineation of myocardial infarction without contrast agents
  using a joint motion feature learning architecture.
\newblock \emph{Medical image analysis} 50, 82--94.
\newblock \doi{10.1016/j.media.2018.09.001}
\bibAnnoteFile{Xu_2018_MedIA}

\bibitem[{Xu et~al.(2018{\natexlab{b}})Xu, Wu, and Feng}]{Xu_2018_arXiv}
Xu, Z., Wu, Z., and Feng, J. (2018{\natexlab{b}}).
\newblock {CFUN}: Combining faster {R-CNN} and {U-net} network for efficient
  whole heart segmentation.
\newblock \emph{Arxiv Preprint} abs/1812.04914.
\newblock Available at \url{http://arxiv.org/abs/1812.04914} (Accessed
  September 1, 2019)
\bibAnnoteFile{Xu_2018_arXiv}

\bibitem[{Xue et~al.(2018)Xue, Brahm, Pandey, Leung, and Li}]{Xue_2018_MedIA}
Xue, W., Brahm, G., Pandey, S., Leung, S., and Li, S. (2018).
\newblock Full left ventricle quantification via deep multitask relationships
  learning.
\newblock \emph{Medical Image Analysis} 43, 54--65.
\newblock \doi{10.1016/j.media.2017.09.005}
\bibAnnoteFile{Xue_2018_MedIA}

\bibitem[{Yan et~al.(2018)Yan, Wang, Li, van~der Geest, and
  Tao}]{Yan_2018_MICCAI}
Yan, W., Wang, Y., Li, Z., van~der Geest, R.~J., and Tao, Q. (2018).
\newblock Left ventricle segmentation via {Optical-Flow-} net from {Short-Axis}
  cine {MRI} : Preserving the temporal coherence of cardiac motion.
\newblock In \emph{Medical Image Computing and Computer Assisted Intervention}
  (Springer International Publishing), vol. 11073 LNCS, 613--621.
\newblock \doi{10.1007/978-3-030-00937-3\_70}
\bibAnnoteFile{Yan_2018_MICCAI}

\bibitem[{Yang et~al.(2018{\natexlab{a}})Yang, Chen, Gao, Zhang, Firmin, and
  {others}}]{Yang_2018_EMBC}
Yang, G., Chen, J., Gao, Z., Zhang, H., Firmin, D., and {others}
  (2018{\natexlab{a}}).
\newblock Multiview sequential learning and dilated residual learning for a
  fully automatic delineation of the left atrium and pulmonary veins from late
  {Gadolinium-Enhanced} cardiac {MRI} images.
\newblock In \emph{Conf Proc IEEE Eng Med Biol Soc.} vol. 2018, 1123--1127.
\newblock \doi{10.1109/EMBC.2018.8512550}
\bibAnnoteFile{Yang_2018_EMBC}

\bibitem[{Yang et~al.(2018{\natexlab{b}})Yang, Zhuang, Khan, Haldar, Nyktari,
  Li et~al.}]{Yang_2018_MedPhy}
Yang, G., Zhuang, X., Khan, H., Haldar, S., Nyktari, E., Li, L., et~al.
  (2018{\natexlab{b}}).
\newblock Fully automatic segmentation and objective assessment of atrial scars
  for long-standing persistent atrial fibrillation patients using late
  gadolinium-enhanced {MRI}.
\newblock \emph{Medical physics} 45, 1562--1576.
\newblock \doi{10.1002/mp.12832}
\bibAnnoteFile{Yang_2018_MedPhy}

\bibitem[{Yang et~al.(2017{\natexlab{a}})Yang, Zhuang, Khan, Haldar, Nyktari,
  Ye et~al.}]{Yang_2017_MIUA}
Yang, G., Zhuang, X., Khan, H., Haldar, S., Nyktari, E., Ye, X., et~al.
  (2017{\natexlab{a}}).
\newblock Segmenting atrial fibrosis from late {Gadolinium-Enhanced} cardiac
  {MRI} by {Deep-Learned} features with stacked sparse {Auto-Encoders}.
\newblock In \emph{Medical Image Understanding and Analysis} (Springer
  International Publishing), 195--206.
\newblock \doi{10.1007/978-3-319-60964-5\_17}
\bibAnnoteFile{Yang_2017_MIUA}

\bibitem[{Yang et~al.(2016)Yang, Sun, Li, Wang, and Xu}]{Yang_2016_MICCAI}
Yang, H., Sun, J., Li, H., Wang, L., and Xu, Z. (2016).
\newblock Deep fusion net for multi-atlas segmentation: Application to cardiac
  {MR} images.
\newblock In \emph{Medical Image Computing and Computer Assisted Intervention}
  (Springer International Publishing), 521--528.
\newblock \doi{10.1007/978-3-319-46723-8\_60}
\bibAnnoteFile{Yang_2016_MICCAI}

\bibitem[{Yang et~al.(2017{\natexlab{b}})Yang, Bian, Yu, Ni, and
  Heng}]{Yang_2017b_MMWHS}
Yang, X., Bian, C., Yu, L., Ni, D., and Heng, P.-A. (2017{\natexlab{b}}).
\newblock {3D} convolutional networks for fully automatic fine-grained whole
  heart partition.
\newblock In \emph{International Workshop on Statistical Atlases and
  Computational Models of the Heart} (Springer), 181--189
\bibAnnoteFile{Yang_2017b_MMWHS}

\bibitem[{Yang et~al.(2017{\natexlab{c}})Yang, Bian, Yu, Ni, and
  Heng}]{Yang_2017_STACOM}
Yang, X., Bian, C., Yu, L., Ni, D., and Heng, P.-A. (2017{\natexlab{c}}).
\newblock {Class-Balanced} deep neural network for automatic ventricular
  structure segmentation.
\newblock In \emph{Statistical Atlases and Computational Models of the Heart.
  {ACDC} and {MMWHS} Challenges} (Springer International Publishing), 152--160.
\newblock \doi{10.1007/978-3-319-75541-0\_16}
\bibAnnoteFile{Yang_2017_STACOM}

\bibitem[{Yang et~al.(2017{\natexlab{d}})Yang, Bian, Yu, Ni, and
  Heng}]{Yang_2017c_MMWHS}
Yang, X., Bian, C., Yu, L., Ni, D., and Heng, P.-A. (2017{\natexlab{d}}).
\newblock Hybrid loss guided convolutional networks for whole heart parsing.
\newblock In \emph{International Workshop on Statistical Atlases and
  Computational Models of the Heart} (Springer), 215--223
\bibAnnoteFile{Yang_2017c_MMWHS}

\bibitem[{Ye et~al.(2019)Ye, Wang, Zhang, and Wang}]{Ye_2019_Access}
Ye, C., Wang, W., Zhang, S., and Wang, K. (2019).
\newblock Multi-depth fusion network for whole-heart {CT} image segmentation.
\newblock \emph{IEEE Access} 7, 23421--23429
\bibAnnoteFile{Ye_2019_Access}

\bibitem[{Yu and Koltun(2016)}]{Yu_2015_ICLR}
Yu, F. and Koltun, V. (2016).
\newblock {Multi-Scale} context aggregation by dilated convolutions.
\newblock In \emph{International Conference on Learning Representations}. 1--13
\bibAnnoteFile{Yu_2015_ICLR}

\bibitem[{Yu et~al.(2017{\natexlab{a}})Yu, Cheng, Dou, Yang, Chen, Qin
  et~al.}]{Yu_2017_MICCAI}
Yu, L., Cheng, J.-Z., Dou, Q., Yang, X., Chen, H., Qin, J., et~al.
  (2017{\natexlab{a}}).
\newblock Automatic {3D} cardiovascular {MR} segmentation with
  {Densely-Connected} volumetric {ConvNets}.
\newblock In \emph{Medical Image Computing and Computer Assisted Intervention}
  (Springer International Publishing), 287--295.
\newblock \doi{10.1007/978-3-319-66185-8\_33}
\bibAnnoteFile{Yu_2017_MICCAI}

\bibitem[{Yu et~al.(2017{\natexlab{b}})Yu, Guo, Wang, Yu, and
  Chen}]{Yu_2017_BiomedEng}
Yu, L., Guo, Y., Wang, Y., Yu, J., and Chen, P. (2017{\natexlab{b}}).
\newblock Segmentation of fetal left ventricle in echocardiographic sequences
  based on dynamic convolutional neural networks.
\newblock \emph{IEEE Trans. Biomed. Eng.} 64, 1886--1895
\bibAnnoteFile{Yu_2017_BiomedEng}

\bibitem[{Yu et~al.(2019)Yu, Wang, Li, Fu, and Heng}]{Yu_2019_MICCAI}
Yu, L., Wang, S., Li, X., Fu, C.-W., and Heng, P.-A. (2019).
\newblock Uncertainty-aware self-ensembling model for semi-supervised {3D} left
  atrium segmentation.
\newblock In \emph{Medical Image Computing and Computer Assisted Intervention}.
  605--613
\bibAnnoteFile{Yu_2019_MICCAI}

\bibitem[{Yue et~al.(2019)Yue, Luo, Ye, Xu, and Zhuang}]{Yue_2019_MICCAI}
Yue, Q., Luo, X., Ye, Q., Xu, L., and Zhuang, X. (2019).
\newblock Cardiac segmentation from {LGE} {MRI} using deep neural network
  incorporating shape and spatial priors.
\newblock In \emph{Medical Image Computing and Computer Assisted Intervention}.
  559--567
\bibAnnoteFile{Yue_2019_MICCAI}

\bibitem[{Zabihollahy et~al.(2018)Zabihollahy, White, and
  Ukwatta}]{Zabihollahy_2018_MI}
Zabihollahy, F., White, J.~A., and Ukwatta, E. (2018).
\newblock Myocardial scar segmentation from magnetic resonance images using
  convolutional neural network.
\newblock In \emph{Medical Imaging 2018: {Computer-Aided} Diagnosis}
  (International Society for Optics and Photonics), vol. 10575, 105752Z.
\newblock \doi{10.1117/12.2293518}
\bibAnnoteFile{Zabihollahy_2018_MI}

\bibitem[{Zhang(2010)}]{Zhang_2010_Thesis}
Zhang, D.~P. (2010).
\newblock \emph{Coronary artery segmentation and motion modelling}.
\newblock Ph.D. thesis, Imperial College London
\bibAnnoteFile{Zhang_2010_Thesis}

\bibitem[{Zhang et~al.(2019{\natexlab{a}})Zhang, Du, Liu, Hou, Zhao, and
  Ding}]{Zhang_2019_Access}
Zhang, J., Du, J., Liu, H., Hou, X., Zhao, Y., and Ding, M.
  (2019{\natexlab{a}}).
\newblock {LU-NET}: An improved {U-Net} for ventricular segmentation.
\newblock \emph{IEEE Access} 7, 92539--92546.
\newblock \doi{10.1109/ACCESS.2019.2925060}
\bibAnnoteFile{Zhang_2019_Access}

\bibitem[{Zhang et~al.(2018{\natexlab{a}})Zhang, Gajjala, Agrawal, Tison,
  Hallock, Beussink-Nelson et~al.}]{Zhang_2018_Circulation}
Zhang, J., Gajjala, S., Agrawal, P., Tison, G.~H., Hallock, L.~A.,
  Beussink-Nelson, L., et~al. (2018{\natexlab{a}}).
\newblock Fully automated echocardiogram interpretation in clinical practice:
  feasibility and diagnostic accuracy.
\newblock \emph{Circulation} 138, 1623--1635
\bibAnnoteFile{Zhang_2018_Circulation}

\bibitem[{Zhang et~al.(2018{\natexlab{b}})Zhang, Karanikolas, Ak{\c c}akaya,
  and Giannakis}]{Zhang_2018_ICASSP}
Zhang, L., Karanikolas, G.~V., Ak{\c c}akaya, M., and Giannakis, G.~B.
  (2018{\natexlab{b}}).
\newblock Fully automatic segmentation of the right ventricle via {Multi-Task}
  deep neural networks.
\newblock In \emph{{IEEE} International Conference on Acoustics, Speech and
  Signal Processing}. 6677--6681.
\newblock \doi{10.1109/ICASSP.2018.8461556}
\bibAnnoteFile{Zhang_2018_ICASSP}

\bibitem[{Zhang et~al.(2019{\natexlab{b}})Zhang, Wang, Yang, Sanford, Harmon,
  Turkbey et~al.}]{Zhang_2019_Arxiv}
Zhang, L., Wang, X., Yang, D., Sanford, T., Harmon, S., Turkbey, B., et~al.
  (2019{\natexlab{b}}).
\newblock When unseen domain generalization is unnecessary? rethinking data
  augmentation.
\newblock \emph{Arxiv Preprint} abs/1906.03347.
\newblock Available at \url{http://arxiv.org/abs/1906.03347} (Accessed
  September 1, 2019)
\bibAnnoteFile{Zhang_2019_Arxiv}

\bibitem[{Zhang et~al.(2019{\natexlab{c}})Zhang, Zhang, Du, Zhang, and
  Li}]{Zhang_2019_Computing}
Zhang, W., Zhang, J., Du, X., Zhang, Y., and Li, S. (2019{\natexlab{c}}).
\newblock An end-to-end joint learning framework of artery-specific coronary
  calcium scoring in non-contrast cardiac {CT}.
\newblock \emph{Computing} 101, 667--678
\bibAnnoteFile{Zhang_2019_Computing}

\bibitem[{Zhao et~al.(2019)Zhao, Balakrishnan, Durand, Guttag, and
  Dalca}]{Zhao_2019_CVPR}
Zhao, A., Balakrishnan, G., Durand, F., Guttag, J.~V., and Dalca, A.~V. (2019).
\newblock Data augmentation using learned transformations for one-shot medical
  image segmentation.
\newblock In \emph{Conference on Computer Vision and Pattern Recognition}.
  8543--8553
\bibAnnoteFile{Zhao_2019_CVPR}

\bibitem[{Zhao(2018)}]{LASC_2018}
[Dataset] Zhao, X.~Z., J. (2018).
\newblock 2018 left atrial segmentation challenge dataset.
\newblock \url{http://atriaseg2018.cardiacatlas.org/}
\bibAnnoteFile{LASC_2018}

\bibitem[{Zheng et~al.(2019)Zheng, Delingette, and Ayache}]{Zheng_2019_MedIA}
Zheng, Q., Delingette, H., and Ayache, N. (2019).
\newblock Explainable cardiac pathology classification on cine {MRI} with
  motion characterization by semi-supervised learning of apparent flow.
\newblock \emph{Medical Image Analysis} 56, 80--95.
\newblock \doi{10.1016/j.media.2019.06.001}
\bibAnnoteFile{Zheng_2019_MedIA}

\bibitem[{Zheng et~al.(2018)Zheng, Delingette, Duchateau, and
  Ayache}]{Zheng_2018_TMI}
Zheng, Q., Delingette, H., Duchateau, N., and Ayache, N. (2018).
\newblock {3-D} consistent and robust segmentation of cardiac images by deep
  learning with spatial propagation.
\newblock \emph{IEEE Transactions on Medical Imaging} 37, 2137--2148.
\newblock \doi{10.1109/TMI.2018.2820742}
\bibAnnoteFile{Zheng_2018_TMI}

\bibitem[{Zheng et~al.(2008)Zheng, Barbu, Georgescu, Scheuering, and
  Comaniciu}]{Zheng_2008_TMI}
Zheng, Y., Barbu, A., Georgescu, B., Scheuering, M., and Comaniciu, D. (2008).
\newblock Four-chamber heart modeling and automatic segmentation for {3-D}
  cardiac {CT} volumes using marginal space learning and steerable features.
\newblock \emph{IEEE Transactions on Medical Imaging} 27, 1668--1681
\bibAnnoteFile{Zheng_2008_TMI}

\bibitem[{Zhou et~al.(2019)Zhou, Deng, and Wu}]{Zhou_2019_Arxiv}
Zhou, L., Deng, W., and Wu, X. (2019).
\newblock Robust image segmentation quality assessment without ground truth.
\newblock \emph{Arxiv Preprint} abs/1903.08773.
\newblock Available at \url{http://arxiv.org/abs/1903.08773} (Accessed
  September 1, 2019)
\bibAnnoteFile{Zhou_2019_Arxiv}

\bibitem[{Zhou and Yang(2019)}]{Zhou_2018_RAL}
Zhou, X.-Y. and Yang, G.-Z. (2019).
\newblock Normalization in training {U-Net} for {2D} biomedical semantic
  segmentation.
\newblock \emph{IEEE Robotics and Automation Letters} PP, 1--1.
\newblock \doi{10.1109/LRA.2019.2896518}
\bibAnnoteFile{Zhou_2018_RAL}

\bibitem[{Zhuang et~al.(2019)Zhuang, Li, Payer, Stern, Urschler, Heinrich
  et~al.}]{Zhuang_2019_MedIA}
Zhuang, X., Li, L., Payer, C., Stern, D., Urschler, M., Heinrich, M.~P., et~al.
  (2019).
\newblock Evaluation of algorithms for {Multi-Modality} whole heart
  segmentation: An {Open-Access} grand challenge.
\newblock \emph{Medical Image Analysis} 58, 101537.
\newblock \doi{https://doi.org/10.1016/j.media.2019.101537}.
\newblock \url{http://www.sdspeople.fudan.edu.cn/zhuangxiahai/0/mmwhs17/}
\bibAnnoteFile{Zhuang_2019_MedIA}

\bibitem[{Zhuang et~al.(2010)Zhuang, Rhode, Razavi, Hawkes, and
  Ourselin}]{Zhuang_2010_TMI}
Zhuang, X., Rhode, K.~S., Razavi, R.~S., Hawkes, D.~J., and Ourselin, S.
  (2010).
\newblock A registration-based propagation framework for automatic whole heart
  segmentation of cardiac {MRI}.
\newblock \emph{IEEE Transactions on Medical Imaging} 29, 1612--1625.
\newblock \doi{10.1109/TMI.2010.2047112}
\bibAnnoteFile{Zhuang_2010_TMI}

\bibitem[{Zotti et~al.(2017)Zotti, Luo, Lalande, Humbert, and
  Jodoin}]{Zotti_2017_STACOM}
Zotti, C., Luo, Z., Lalande, A., Humbert, O., and Jodoin, P.-M. (2017).
\newblock {GridNet} with automatic shape prior registration for automatic {MRI}
  cardiac segmentation.
\newblock In \emph{International Workshop on Statistical Atlases and
  Computational Models of the Heart}. 73--81
\bibAnnoteFile{Zotti_2017_STACOM}

\bibitem[{Zotti et~al.(2019)Zotti, Luo, Lalande, and Jodoin}]{Zotti_2019_JBHI}
Zotti, C., Luo, Z., Lalande, A., and Jodoin, P.-M. (2019).
\newblock Convolutional neural network with shape prior applied to cardiac
  {MRI} segmentation.
\newblock \emph{IEEE Journal of Biomedical and Health Informatics} 23,
  1119--1128.
\newblock \doi{10.1109/JBHI.2018.2865450}
\bibAnnoteFile{Zotti_2019_JBHI}

\bibitem[{Zreik et~al.(2016)Zreik, Leiner, De~Vos, Van~Hamersvelt, Viergever,
  and Isgum}]{Zreik_2016_ISBI_CT}
Zreik, M., Leiner, T., De~Vos, B.~D., Van~Hamersvelt, R.~W., Viergever, M.~A.,
  and Isgum, I. (2016).
\newblock Automatic segmentation of the left ventricle in cardiac {CT}
  angiography using convolutional neural networks.
\newblock In \emph{International Symposium on Biomedical Imaging}. 40--43
\bibAnnoteFile{Zreik_2016_ISBI_CT}

\bibitem[{Zreik et~al.(2018{\natexlab{a}})Zreik, Lessmann, van Hamersvelt,
  Wolterink, Voskuil, Viergever et~al.}]{Zreik_2018_MedIA}
Zreik, M., Lessmann, N., van Hamersvelt, R.~W., Wolterink, J.~M., Voskuil, M.,
  Viergever, M.~A., et~al. (2018{\natexlab{a}}).
\newblock Deep learning analysis of the myocardium in coronary {CT} angiography
  for identification of patients with functionally significant coronary artery
  stenosis.
\newblock \emph{Medical Image Analysis} 44, 72--85
\bibAnnoteFile{Zreik_2018_MedIA}

\bibitem[{Zreik et~al.(2018{\natexlab{b}})Zreik, van Hamersvelt, Wolterink,
  Leiner, Viergever, and I{\v{s}}gum}]{Zreik_2018_TMI}
Zreik, M., van Hamersvelt, R.~W., Wolterink, J.~M., Leiner, T., Viergever,
  M.~A., and I{\v{s}}gum, I. (2018{\natexlab{b}}).
\newblock A recurrent {CNN} for automatic detection and classification of
  coronary artery plaque and stenosis in coronary {CT} angiography.
\newblock \emph{IEEE Transactions on Medical Imaging} 38, 1588--1598.
\newblock \doi{10.1109/TMI.2018.2883807}
\bibAnnoteFile{Zreik_2018_TMI}

\end{thebibliography}
\clearpage


%

 \end{document}